\documentclass[aps,prb,twocolumn,amsmath,amssymb,superscriptaddress,longbibliography]{revtex4-2}
\usepackage[breaklinks,colorlinks,bookmarks=false,citecolor=blue,linkcolor=red,urlcolor=blue]{hyperref}
\usepackage{blindtext,enumitem,graphicx,dcolumn,color,xcolor,amsmath,braket,bm,changes,chemformula,cleveref,times}

\usepackage{float}
\makeatletter
\let\newfloat\newfloat@ltx
\makeatother

\usepackage{algorithm}
\usepackage{algpseudocode}

\graphicspath{{figures_prb/}}
\usepackage{blindtext}
\usepackage{changes}
\usepackage{braket}
\usepackage{amssymb}

\crefname{equation}{Eq.}{Eqs.}
\Crefname{equation}{Eq.}{Eqs.}
\crefname{figure}{Fig.}{Figs.}
\Crefname{figure}{Fig.}{Figs.}
\crefname{section}{Sec.}{Secs.}
\Crefname{section}{Sec.}{Secs.}
\crefname{algorithm}{Alg.}{Algs.}
\Crefname{algorithm}{Alg.}{Algs.}

\begin{document}
\title{Spectroscopy and complex-time correlations using minimally entangled typical thermal states}

\author{Zhenjiu Wang}
\email{wangzj@lzu.edu.cn}
\affiliation{Arnold Sommerfeld Center for Theoretical Physics,
University of Munich, Theresienstr. 37, 80333 Munich, Germany}
\affiliation{Max Planck Institute for the Physics of Complex Systems, N\"othnitzer Strasse 38, Dresden 01187, Germany}
\affiliation{Lanzhou Center for Theoretical Physics, Key Laboratory of Quantum Theory and Applications of MoE, Key Laboratory of Theoretical Physics of Gansu Province, and School of Physical Science and Technology, Lanzhou University, Lanzhou, Gansu 730000, China.}

\author{Paul McClarty}
\affiliation{Max Planck Institute for the Physics of Complex Systems, N\"othnitzer Strasse 38, Dresden 01187, Germany}
\affiliation{Laboratoire Léon Brillouin, CEA, CNRS, Université Paris-Saclay, CEA Saclay, 91191 Gif-sur-Yvette, France}

\author{Dobromila Dankova}
\affiliation{Max Planck Institute for the Physics of Complex Systems, N\"othnitzer Strasse 38, Dresden 01187, Germany}

\author{Andreas Honecker}
%\email{Andreas.Honecker@cyu.fr}
\affiliation{Laboratoire de Physique Th\'eorique et Mod\'elisation, CNRS UMR 8089, CY Cergy Paris Universit\'e, 95302 Cergy-Pontoise, France}%%

\author{Alexander Wietek}
\email{awietek@pks.mpg.de}
\affiliation{Max Planck Institute for the Physics of Complex Systems, N\"othnitzer Strasse 38, Dresden 01187, Germany}

\begin{abstract}
Tensor network states have enjoyed great success at capturing aspects of strong correlation physics. However, obtaining dynamical correlators at non-zero temperatures is generically hard even using these methods. Here, we introduce a practical approach to computing such correlators using minimally entangled typical thermal states (METTS). While our primary method directly computes dynamical correlators of physical operators in real time, we propose extensions where correlations are evaluated in the complex-time plane. The imaginary time component bounds the rate of entanglement growth and strongly alleviates the computational difficulty allowing the study of larger system sizes. To extract the physical correlator one must take the limit of purely real-time evolution. We present two routes to obtaining this information (i) via an \textit{analytic} correlation function in complex time combined with a stochastic analytic continuation method to obtain the real-time limit and (ii) a \textit{Hermitian} correlation function that asymptotically captures the desired correlation function quantitatively without requiring effort of numerical analytic continuation. We show that these numerical techniques capture the finite-temperature dynamics of the Shastry-Sutherland model $-$ a model of interacting spin one-half in two dimensions.
%Obtaining real frequency spectral functions using methods based on matrix-product-states is generically hard at finite temperature. 
%Based on a minimally entangled typical thermal states approach, we provide a practical method of calculating the spectrum via  the correlation function of physical operators in the complex time plane.  
%Dynamical correlation functions are computed via time-evolution in the basis of matrix product state where hybridizing real and imaginary time improves the numerical efficiency: imaginary time evolution plays the role of dissipation and the speed of entanglement growth is bounded.  
%Although the real frequency spectrum is exactly defined only in the limit of purely real time, we introduce two alternative definitions of complex time correlation function, both of which help extract the real frequency  information efficiently:  an  \textit{analytic} correlation function, which is the nature extension from the  imaginary time axis to the complex plane, can be combined with stochastic analytic  continuation method to obtain the spectrum; whereas a \textit{Hermitian} correlation function, asymptotically captures the spectrum  quantitatively without requiring effort of numerical analytic continuation. 
\end{abstract}

\date{\today}

\maketitle
\tableofcontents

\section{Introduction}
Understanding the macroscopic behavior of quantum materials from microscopic descriptions of strongly interacting particles is one of the great challenges of contemporary physics. Developments in numerical simulations have given us powerful tools to gain insights beyond systems where an analytical treatment is still feasible. While several theoretical models in condensed matter physics are amenable to highly accurate quantum Monte Carlo (QMC) simulations~\cite{Gubernatis2016}, others like frustrated magnets or generic strongly interacting fermion systems present us with the \textit{sign problem} when tackled using QMC~\cite{Loh1990}, limiting the range of parameters and observables that can be studied. Tensor network methods, on the other hand, are not afflicted by a sign problem~\cite{Orus2019,Banuls2023}. Instead, these methods face difficulties whenever confronted with the problem of dealing with highly entangled states. The arguably most well-known tensor network method is the density matrix renormalization group (DMRG)~\cite{White1992,White1993,Schollwoeck2005} which, in its modern version, is based on matrix product states (MPS)~\cite{Schollwoeck2011}. Initially proposed as a method to study static properties of ground states, several extensions thereof have enabled the study of static observables at finite temperature~\cite{PhysRevB.56.5061,
PhysRevLett.82.3855,
Feiguin2005,White2009,Stoudenmire2010,Wietek2021,Wietek2021b,Feng2022,Feng2023,Nocera2016}, dynamical spectral functions at zero temperature~\cite{PhysRevB.52.R9827,
Kuehner1999,
PhysRevLett.93.076401,
PhysRevB.83.161104,
PhysRevB.83.195115,
PhysRevB.85.205119,
PhysRevE.94.053308,
Gohlke2017,Verresen2019,Drescher2022}, and unitary dynamics in non-equilibrium settings~\cite{PhysRevLett.93.040502,
PhysRevLett.93.076401,
Daley_2004,
PhysRevB.70.121302,
Kennes2014,Essler2014,Karrasch2013b,Karrasch2014}. 

To fully build a bridge between microscopic models and their experimental response functions, simulations of dynamical correlators at finite temperature are necessary. This includes dynamical spin structure factors as probed in neutron scattering experiments~\cite{McClarty_2017,Muehlbauer2019}, and electronic Green's functions probed in angular resolved photoemission spectroscopy (ARPES)~\cite{Sobota2021}. Moreover, transport coefficients can be understood within the Kubo linear response formalism. The computation of dynamical response functions at finite temperature in two spatial dimensions, however, still poses significant difficulties for tensor network methods. 

A well-known DMRG/MPS-based approach to compute time-dependent correlation functions is to perform real-time evolution. This technique has been applied at zero temperature~\cite{PhysRevLett.93.076401,Gohlke2017,Verresen2019}, but also approaches at finite temperature using purifications have been reported~\cite{Barthel2009,Karrasch2012,Karrasch2013,Barthel2013,Tiegel2014,Jansen2022} as well as an interesting approach to investigate high-temperature transport using dissipation-assisted operator evolution~\cite{Rakovsky2022}. For alternative approaches to finite-$T$ dynamics, see also Refs.~\cite{PhysRevB.71.241101,PhysRevB.80.205117}. 

The real-time evolution approach, however, suffers from the fact that the entanglement of generic quantum systems grows unbounded if evolved for long times. In practice, this limits the system size which can be simulated as well as the resolution of the spectral function which is attained. The computation of real-time correlators is also especially tricky in QMC simulations, where an even more intricate phase problem is encountered. Interestingly, however, much of many-body theory is built on the notion of imaginary-time correlators. While these correlations do not immediately allow for the extraction of dynamical response functions to arbitrary precision at any real frequency, properties like excitation gaps can be reliably estimated~\cite{Suwa2015}. Within the framework of MPS techniques, the computation of imaginary-time evolution does not face the same problem of unbounded entanglement growth. In fact, at asymptotically long times imaginary time evolution (under mild assumptions) yields the many-body ground state, whose entanglement for most physically relevant systems is expected to be small as compared to genuine quantum states with volume-law entanglement. Famously, ground states of locally interacting and gapped systems obey an area law~\cite{Hastings2007,Wolf2008}. Thus, imaginary-time evolution poses fewer challenges as compared to real-time evolution when applying MPS techniques, which allows one to study larger systems and a wider range of parameters.

In this paper, we introduce an approach to evaluating dynamical spectral functions based on the minimal entangled typical thermal states (METTS)~\cite{White2009,Stoudenmire2010,Wietek2021,Wietek2021b,Feng2022,Feng2023} technique to simulate systems at nonzero temperature. Moreover, we propose to use a mixed approach where correlators are evaluated in fully complex time, i.e., having both a real-time and imaginary-time component. We present efficient algorithms to obtain the dynamical spectrum of physical operators at nonzero temperature,  by computing correlation functions along a certain angle in the complex time plane. Such an approach has originally been proposed in Ref.~\cite{Krilov2001} for QMC simulations but has until now not been widely adopted. Recently, however, complex time correlation functions have been revisited in the context of MPS ground state simulations and applied to quantum impurity models~\cite{Cao2023,Grundner2023}. 

As we will show, complex-time correlation functions can be evaluated using well-known MPS time evolution algorithms such as the time-dependent variational principle (TDVP)~\cite{Haegeman2011,Haegeman2016}. Our aim is to combine the advantages of computing imaginary-time correlations and real-time correlations. The imaginary contribution of time evolution dissipates the large entanglement entropy while the real-time part of the correlation function benefits the efficiency of obtaining real frequency spectral function.

We compare two approaches that generalize time evolution to complex times. The first approach generalizes the imaginary time correlator that is well-discussed in the QMC community to the complex plane in an analytic way, such that the spectral functions in real frequency can be obtained following a numerical analytical continuation method.  In the second approach, we introduce a \textit{Hermitian} complex correlation function, which can be implemented analogously to the real-time correlation function.

This publication also serves as a companion paper to Ref.~\cite{Wang2024a} by the authors, whose goal is to explain the anomalous thermal broadening observed in \NoCaseChange{SrCu}$_2($BO$_3)_2$ as modeled by the Shastry-Sutherland model~\cite{ShastrySutherland1981,miyahara2003}. At low temperatures, a sharp band of triplon excitations is observed by inelastic neutron scattering experiments~\cite{kageyama2000,gaulin2004,zayed2014} which rapidly vanishes when the temperature is only raised to a fraction of the gap. From perturbation theory, the triplon gap can be computed as,
\begin{equation}
    \Delta = J_{D}\left[ 1 - (J/{J_D})^2 - \frac{1}{2}(J/{J_D})^3 - \frac{1}{8}(J/{J_D})^4 \right].
\end{equation}
For the coupling ratio $J/J_D=0.63$ relevant for \NoCaseChange{SrCu}$_2($BO$_3)_2$ used here, we obtain $\Delta\approx 0.458 J_D$. To genuinely understand these observations, dynamical spin structure factors at nonzero temperatures need to be evaluated in a controlled way.

\begin{figure}
    \centering
    \includegraphics[width=\columnwidth]{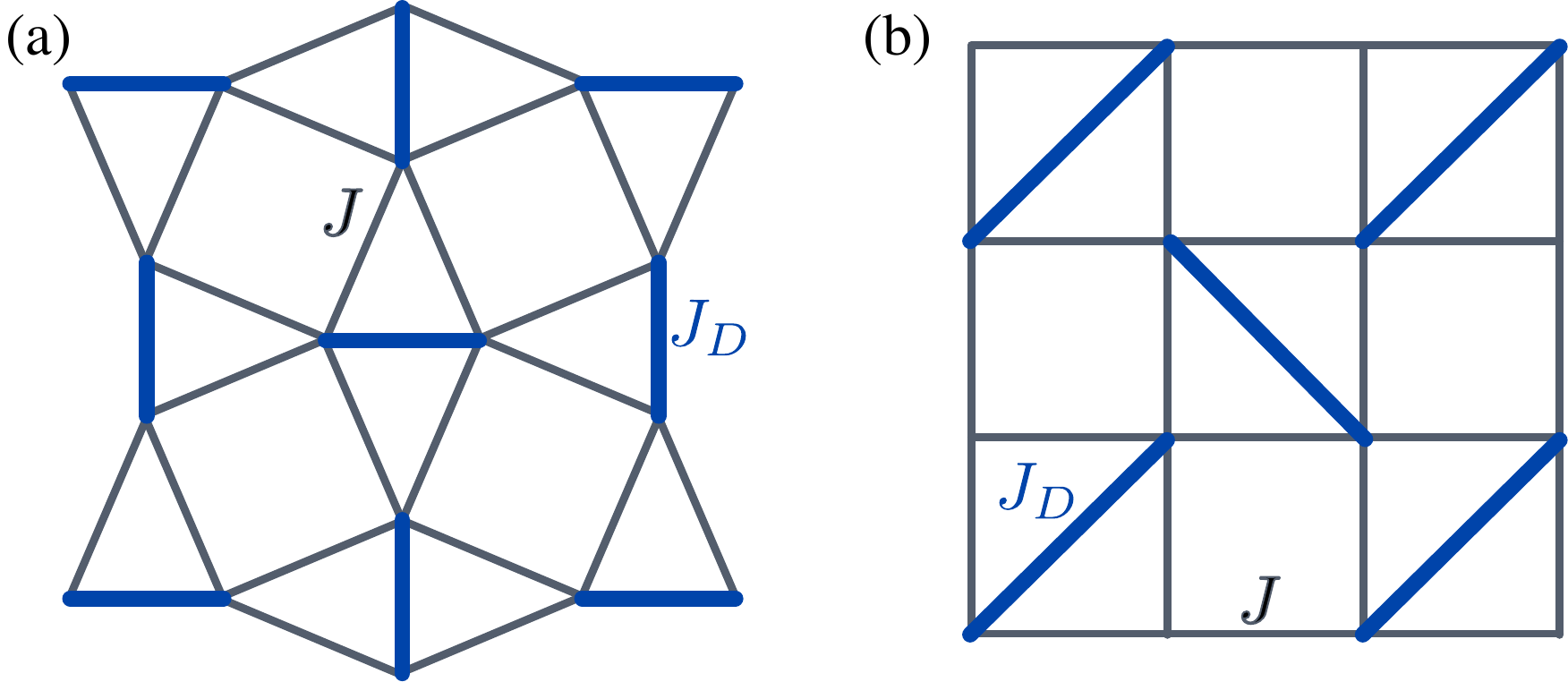}
    \caption{Sketch of the Shastry-Sutherland lattice geometry. The geometry of copper atoms in SrCu$_2($BO$_3)_2$ shown in (a) can be continuously deformed into a square lattice with a $2\times 2$ unit cell with the additional diagonal bonds as shown in (b). The bold blue bonds are referred to as dimers, $J$ denotes the interdimer couplings marked with grey lines and $J_D$ denotes the intradimer couplings marked with bold blue lines.}
    \label{fig:model}
\end{figure}

The outline of our paper is as follows. We begin with  \cref{sec:SS_Ham} introducing the Shastry-Sutherland model we use to benchmark our methods. In \cref{sec:METTS}, we review the conventional thermal METTS algorithm of simulating static observables at nonzero temperatures. The dynamical METTS algorithm of obtaining spectral function via real-time correlations is introduced in Sec.~\ref{sec:real_time}. As both these algorithms are based on efficient matrix-product state time-evolution algorithms, we discuss the time-evolution methods employed in \cref{sec:TDVP}. Details on the Fourier analysis of time-dependent correlation functions are discussed in \cref{sec:spectral}. To validate our approach, we perform a comparison of the dynamical METTS algorithm to spectral functions obtained via the Lehmann representation using full exact diagonalization and DMRG in \cref{sec:ED}.  

In the second part of our publication, we discuss how the computational cost from real-time evolution can be eased by performing genuinely complex-time evolution. In \cref{Sec:two_def}, we introduce two definitions of complex time correlation functions. One of these, that we call the 
\textit{analytic} correlation function is an interpolation between the real-time correlation function and the imaginary-time Matsubara Green's functions often employed in many-body physics. We discuss an efficient algorithm evaluating this correlation function using METTS in \cref{sec:analyticcorr}. From these correlation functions, the real-time spectral function can be obtained via analytic continuation. We describe our approach employing stochastic analytic continuation in  \cref{sec:SAC}. A second interesting kind of complex time correlation, we call the \textit{Hermitian} correlation function is introduced in \cref{sec:hermitiancorr}. We show that this correlation can be efficiently evaluated with bounded entanglement while being a good approximation to the real-time correlation function. Finally, we give a summary and perspective in Sec.~\ref{sec:outlook}.

\section{The Shastry-Sutherland model and \NoCaseChange{SrCu}$_2($BO$_3)_2$}\label{sec:SS_Ham}
The Shastry-Sutherland model \cite{ShastrySutherland1981,miyahara2003} is chosen to showcase and benchmark our numerical method. The model is given by, 
\begin{equation}
H = J_D\sum_{\langle i,j \rangle} \bm{S}_i\cdot \bm{S}_j + J\sum_{\langle\langle i,j \rangle\rangle} \bm{S}_i\cdot \bm{S}_j.
\end{equation}
where $\bm{S}_i=(S^x_i, S^y_i, S^z_i)$ denotes the spin-$1/2$ operators, $J_D$ denotes the intradimer coupling and $J$ denotes the interdimer coupling on the Shastry-Sutherland lattice, see \cref{fig:model}. 

The Shastry-Sutherland model \cite{ShastrySutherland1981} provides a classic instance of magnetic frustration in two dimensions. The magnetic moments interact via Heisenberg exchange couplings on bonds organized into triangular units (Fig.~\ref{fig:model}). When $J = 0$, the lattice breaks up into pairs of interacting spins so that the ground state is a tiling of singlets on the $J_D$ bonds and the elementary excitations are decoupled triplets. Remarkably, the singlet tiling is an exact eigenstate of the fully interacting model (finite $J$) and is the ground state for $J/J_D \lesssim 0.65$. In this paper, we focus on this regime of the model which is also relevant to the material \NoCaseChange{SrCu}$_2($BO$_3)_2$ \cite{Kageyama1999,miyahara2003}.
%\textcolor{red}{PM: minimally cite here???}.
In the material, the magnetic copper ions are organized into essentially decoupled layers each forming a Shastry-Sutherland lattice with couplings well approximated by $J/J_D \approx 0.63$
\cite{miyahara69thermodynamic,miyahara2003,wietek2019}. The singlet ground state of the model has been verified experimentally. We are interested in the dynamics of the model which has also been studied extensively. The decoupled triplets at $J = 0$ are coupled at finite $J/J_D$ but magnetic frustration almost perfectly localizes these modes leading to an almost flat band of triply degenerate {\it triplons} that appear in the dynamical spin-spin correlator. At finite temperature, in the material, the triplon modes have been observed to broaden dramatically \cite{kageyama2000,Lemmens2000,gaulin2004,zayed2014, wulferding2021}. It is this feature on which we focus in the following. 

We consider cylindrical geometries of length $L$ and width $W$ with periodic boundary conditions in the short $y$-direction and open boundary conditions in the longer $x$-direction. In this work, we focus on width $W=4$ and $W=6$ cylinders with length $L=16$. A peculiarity when using open boundary conditions for the Shastry-Sutherland model is that there exist spins that are not coupled by any intradimer coupling $J_D$ and only with interdimer couplings $J$. Flipping such a dangling spin would constitute an artificial boundary mode and, therefore, we remove interdimer couplings that couple to these boundary dangling spins.

We are interested in the dynamical spin structure factor given by
\begin{equation}
\label{eq:dynspinstructure}
    S^z(\bm{q}, \omega) =
\int_{-\infty}^{\infty} dt \; e^{i\omega t } \langle S^z(-\bm{q}, t) S^z(\bm{q}, 0) \rangle,
\end{equation}
where $S^z(\bm{q}, t) = e^{iHt}S^z(\bm{q}) e^{-iHt}$ and,
\begin{equation}
\label{eq:spinkspace}
S^z(\bm{q}) = \frac{1}{\sqrt{N}}\sum_{i=1}^N e^{i\bm{q} \cdot \bm{r}_i} S^z_i, 
\end{equation}
where $\bm{r}_i$ denotes the position of site $i$ and $N$ denotes the number of lattice sites. The position of the lattice sites $\bm{r}_i$ as well as the unit cell and basis sites are considered in the square lattice geometry shown in \cref{fig:model}(b). he Shastry-Sutherland model features a flat band of triplon excitations, which stretches throughout the entire Brillouin zone. At low temperatures we expect similar physical phenomena to occur regardless of the exact momentum. In this manuscript we focus on two momenta, $\bm{q}=(\pi,0)$ and $\bm{q} = (\pi/2,\pi/2)$ to illustrate the physics, but expect similar behavior at different values of $\bm{q}$.

To reduce the boundary effects of the cylinder we use a modified definition,
\begin{equation}
\label{eq:spinkspacewindow}
\tilde{S}^z(\bm{q}) = \frac{1}{\sqrt{N}}\sum_{i=1}^N w(\bm{r}_i) e^{i\bm{q} \cdot \bm{r}_i} S^z_i, 
\end{equation}
where $w(\bm{r}) = w(x, y)$ denotes a windowing function. We choose a Tukey window of width $a=4$ as defined by,
\begin{align}
&w(x, y) = \\ \nonumber 
&\begin{cases}
 \frac{1}{2}\left\{1 + \cos\left(\frac{\pi}{a} (x - a)\right)\right\}   &\text{ if } 0 \leq x \leq a\\
 \frac{1}{2}\left\{1 + \cos\left(\frac{\pi}{a} (x - (L - 1) + a)\right)\right\}  &\text{ if } x \geq (L-1) - a\\
   1 &\text{ otherwise.  }  \\
\end{cases}
\end{align}
Explicitly, for $L=16$ and $a=4$ this means $w(0,y) = w(L-1, y) = 0$, $w(1,y) = w(L-2, y) \approx 0.14644$, $w(2,y) = w(L-3, y) =0.5$, 
$w(3,y) = w(L-4, y) \approx 0.85355$, and
$w(x,y)=1$ else.
For notational simplicity, $S(\bm{q}, \omega)$ will, in the following, refer to the spectral function employing the windowed definition of $\tilde{S}^z(\bm{q})$ instead of $S^z(\bm{q})$, whenever cylindrical boundary conditions are used.

\section{Dynamical METTS algorithm}
After first reviewing the thermal METTS algorithm we introduce the dynamical METTS algorithm. We discuss computational details such as the appropriate choice of MPS time evolution algorithms and windowing for Fourier transformation of data simulated up to a finite time. The algorithm is then validated against results from exact diagonalization and DMRG.

\subsection{\label{sec:METTS}Review of the thermal METTS algorithm}

The METTS algorithm~\cite{White2009,Stoudenmire2010} computes thermal expectation values of any operator $\mathcal{O}$ by sampling over expectation values of certain pure states. Generically, we can write a thermal expectation value as,
\begin{align} 
\langle \mathcal{O} \rangle_\beta &= \frac{1}{\mathcal{Z}}\text{Tr}[e^{-\beta H}\mathcal{O}] \\
&= \frac{1}{\mathcal{Z}} \sum_i \langle \sigma_i | e^{-\beta H /2} \mathcal{O} e^{-\beta H / 2} | \sigma_i \rangle \\
&= \frac{1}{\mathcal{Z}} \sum_i p_i \langle \psi_i | \mathcal{O} | \psi_i \rangle,
\end{align}
where $\beta = 1/(k_{\text{B}} T)$ (henceforth $k_{\text{B}}=1$) denotes the inverse temperature,  $\mathcal{Z} = \text{Tr}[e^{-\beta H}]$ the partition function, and $\braket{\ldots}_\beta= \frac{1}{\mathcal{Z}} \text{Tr}\left[ e^{-\beta H} \ldots \right]$ denotes a thermal expectation value. The sum is performed over the computational basis of product states,
\begin{equation}
| \sigma_i \rangle = | \sigma_i^1 \rangle | \sigma_i^2 \rangle  \;\ldots \;| \sigma_i^N \rangle,   
\end{equation}
and we define the METTS state,
\begin{equation}
\label{eq:mettsstate}
|\psi_i\rangle = \frac{1}{\sqrt{p_i}} e^{-\beta H / 2}| \sigma_i \rangle,
\end{equation}
where $p_i=\langle \sigma_i | e^{-\beta H}| \sigma_i \rangle \geq 0$ is a (non-negative) weight. Notice that the partition function can also be written as $\mathcal{Z}=\sum_i p_i$. The METTS states are normalized, i.e., $|\langle \psi_i | \psi_i \rangle|^2 = 1$. To sample the METTS states $|\psi_i\rangle$ with probability $p_i / \mathcal{Z}$ a Markov chain is constructed where the transition probability is given by,
\begin{equation}
T_{i\rightarrow j} = |\langle \psi_i | \sigma_j \rangle |^2.
\end{equation}
The original works on METTS~\cite{White2009,Stoudenmire2010} proved that this transition probability fulfills the detailed balance conditions,
\begin{equation}
    p_i T_{i\rightarrow j} = p_j T_{j\rightarrow i},
\end{equation}
and, therefore, the stationary distribution of this Markov chain is given by the probabilities $p_i$.
As a Markov chain Monte Carlo simulation, thermalization and autocorrelation effects have to be taken into account to estimate averages and standard errors. In the following, we denote by $R$ the number of random METTS states sampled. In our calculations, we choose $R=100$ due to our computational budget, from which we obtain error estimates by analyzing thermalization and autocorrelation times as in standard error analysis of Markov chain Monte Carlo methods. The METTS algorithm is particularly well-suited to tensor network simulations describing states of low entanglement. Due to the fact that we sample with non-negative probabilities $p_i$ the Monte Carlo sampling itself is not accompanied by a \textit{sign problem}, as often encountered in QMC simulations. The computational complexity is rather transferred to the entanglement of the METTS states $\ket{\psi_i}$.

\subsection{Dynamical METTS algorithm in real-time}\label{sec:real_time}

%\begin{figure}[t]
%    \centering
%    \includegraphics[width=\columnwidth]{sofqt_sofqw_comparison_W_6.pdf}
%    \caption{METTS samples (in color) for the time-dependent correlation function $S_i(\bm{q},t)$(a,c) and the dynamical structure 
%    factor $S_i(\bm{q},\omega)$(b,d) of the Shastry-Sutherland model at $J/J_D=0.63$ at temperatures $T/J_D=0.1$ (a,b) and $T/J_D=0.2$(c,d) on the $16\times 6$ lattice. The black solid lines and the surrounding gray-shaded region show the (error) estimate from averaging these samples for $S(\bm{q},t)$ and $S(\bm{q},\omega)$ respectively.}
%    \label{fig:samples}
%\end{figure}

\begin{figure}[t!]
    \centering
    \includegraphics[width=0.9\columnwidth]{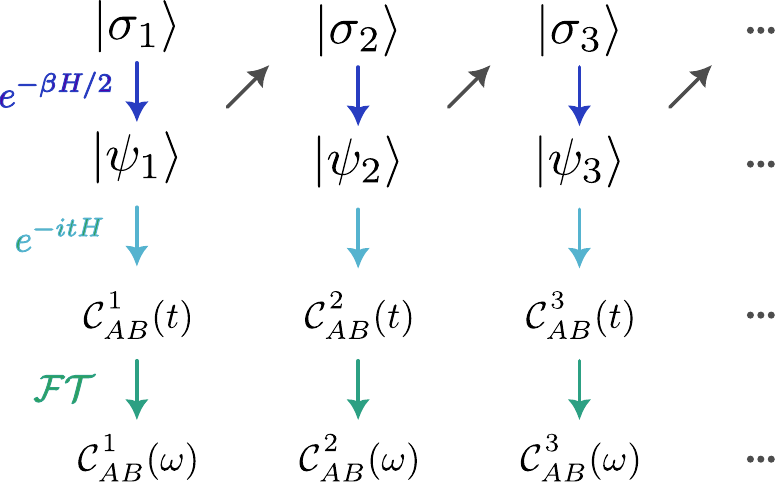}
    \caption{Illustration of the dynamical METTS algorithm. A sequence of product states $|\sigma_i\rangle$ and METTS  $|\psi_i\rangle$ is sampled according to the thermal METTS algorithm. For every METTS the time-dependent correlation function $\mathcal{C}_{AB}^{i}(t)$ as in \cref{eq:mettscorr}
    is computed. Product states $| \sigma_{i+1}\rangle$ are sampled with transition probability $T_{i\rightarrow i+1}=|\langle \psi_i | \sigma_{i+1}\rangle |$. From the individual METTS correlation functions $\mathcal{C}_{AB}^{i}(t)$ the spectral function $\mathcal{C}_{AB}(\omega)$ is computed via a Fourier transform and subsequent averaging.}
    \label{fig:illustration}
\end{figure}

To study dynamical structure factors we are interested in the \textit{conventional} time-dependent correlation function $\mathcal{C}_{AB}(t)$ given by,
\begin{equation}
\mathcal{C}_{AB}(t) = \langle A(t)B \rangle_\beta = \langle e^{iHt}A e^{-iHt} B\rangle_\beta,
\end{equation}
where $A$ and $B$ denote observables subject to investigation. For the dynamical spin structure factor considered in this work, we choose,
\begin{equation}
    A^\dagger = B = S^z(\bm{q}),
\end{equation}
as defined in \cref{eq:spinkspace,eq:spinkspacewindow}. A dynamical spectral function is then given as the Fourier transform,
\begin{equation}\label{Eq:real_time_to_frequency}
\mathcal{C}_{AB}(\omega) = \int_{-\infty}^{\infty} dt \;e^{i\omega t } \mathcal{C}_{AB}(t).
\end{equation}
We expand the corresponding expectation values in the basis of METTS states,
\begin{equation}
\mathcal{C}_{AB}(t) = \frac{1}{\mathcal{Z}}  \sum_i p_i \langle \psi_i |  e^{iHt}A e^{-iHt} B | \psi_i \rangle,
\end{equation}
and define the states,
\begin{align}
&|v_i(t)\rangle = e^{-iHt} B |\psi_i \rangle \label{eq:auxstatevi},\\
&|w_i(t)\rangle = e^{-iHt} |\psi_i \rangle. \label{eq:auxstatewi}
\end{align}
Notice, while the state $\ket{w_i(t)}$ is normalized, we generically have, 
\begin{equation}
\braket{v_i(t)|v_i(t)} = \text{constant} \neq 1.
\end{equation} 
We introduce the time-dependent correlation function of an individual METTS state as,
\begin{align}
\label{eq:mettscorr}
\mathcal{C}_{AB}^{i}(t) 
&\equiv \langle \psi_i |  e^{iHt}A e^{-iHt} B | \psi_i \rangle \\
&= \langle w_i(t) |  A  | v_i(t) \rangle.
\end{align}
Hence, the time-dependent correlation function and the dynamical spectral function are expressed as,
\begin{align}
\mathcal{C}_{AB}(t) = \frac{1}{\mathcal{Z}}  \sum_i p_i \mathcal{C}_{AB}^{i}(t),\\
\quad \mathcal{C}_{AB}(\omega) = \frac{1}{\mathcal{Z}}  \sum_i p_i \mathcal{C}_{AB}^{i}(\omega),
\end{align}

\begin{algorithm}
   \begin{algorithmic}
   \State Choose a random initial product state $| \sigma_1 \rangle = | \sigma_1^1 \rangle \cdot \ldots|\sigma_1^N \rangle$
   \For{$i = 1, \ldots , R$}
        \State 1 Compute the METTS state $|\psi_i\rangle = \frac{1}{\sqrt{p_i}} e^{-\beta H /2}| \sigma_i \rangle$
        \State Compute $|v_i(0)\rangle = B |\psi_i\rangle$ and set
        $|w_i(0)\rangle = |\psi_i\rangle$
        \State Set $t=0$  
        \While{$t \leq \Omega$}
            \State Compute $\mathcal{C}^i(t) = \langle w_i(t) |  A  | v_i(t) \rangle$
            \State Compute $|v_i(t+\delta t)\rangle = e^{-iH\delta t}|v_i(t)\rangle$ 
            \State Compute $|w_i(t+\delta t)\rangle = e^{-iH\delta t}|w_i(t)\rangle$ 
            \State $t \rightarrow t + \delta t$
        \EndWhile
        \State Store correlator $\mathcal{C}^i(t)$ to disk
        \State Pick next product state $|\sigma_{i+1} \rangle$
        with probability $|\langle \psi_i | \sigma_{i+1} \rangle |^2$
   \EndFor
   \end{algorithmic}
 \caption{Dynamical METTS for correlators $\mathcal{C}(t)$}
 \label{alg:dynmetts}
\end{algorithm}

where, 
\begin{equation}
\label{eq:mettspectral}
\mathcal{C}_{AB}^{i}(\omega) =
\int_{-\infty}^{\infty} dt \; e^{i\omega t } \mathcal{C}_{AB}^{i}(t).
\end{equation}
This suggests the following algorithm to compute dynamical correlation functions $\mathcal{C}_{AB}^{i}(t)$ and $\mathcal{C}_{AB}^{i}(\omega)$. We sample METTS states $\ket{\psi_i}$ by sampling from the Markov chain with transition probability $T_{i\rightarrow j} = |\langle \psi_i | \sigma_j \rangle |^2$, just as in the thermal METTS algorithm. 
For every METTS state $\ket{\psi_i}$ we  compute $ |v_i(0)\rangle = B |\psi_i \rangle $ and set $ |w_i(0)\rangle = |\psi_i \rangle $ at the beginning.

The next step is to 
perform a real-time evolution to compute  $|v_i(t)\rangle$ and $|w_i(t)\rangle$ as in \cref{eq:auxstatevi,eq:auxstatewi}. Notice that the METTS states $ | \psi_i \rangle$ are not eigenstates of the Hamiltonian $H$, such that the time-evolved states $|v_i(t)\rangle$ and $|w_i(t)\rangle$ have to be explicitly computed. Naturally, this time evolution can only be carried out until a maximum time $\Omega$. Measurements of $\mathcal{C}_{AB}^{i}(t)$ are only performed at discrete time steps which are a multiple of an elementary step size $\delta_t$.

\begin{figure}[t!]
    \includegraphics[width=\columnwidth]{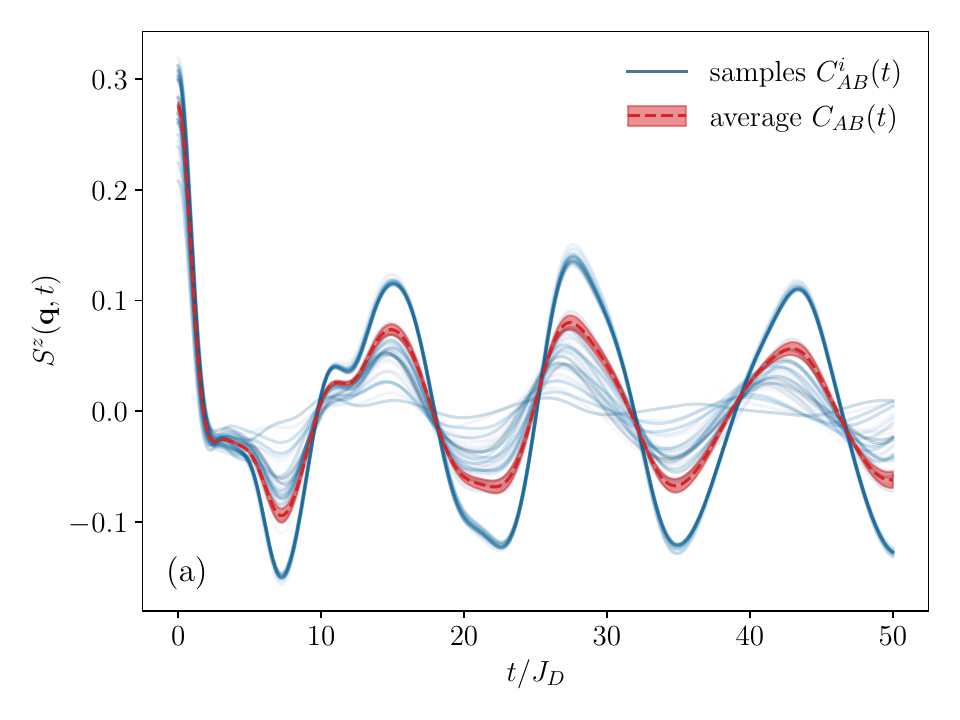}\\
    \includegraphics[width=\columnwidth]{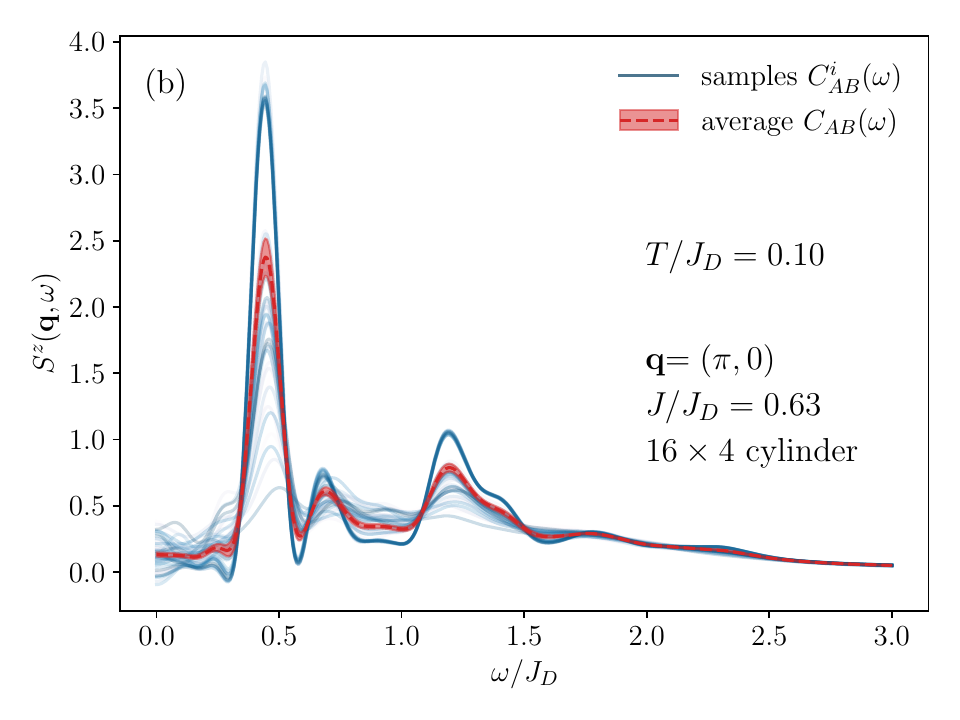}\\
    \caption{Exemplary METTS samples (blue lines) of (a) the correlation functions $\mathcal{C}_{AB}^{i}(t)$ and (b) the spectral functions $\mathcal{C}_{AB}^{i}(\omega)$ of the Shastry-Sutherland model on a $16\times 4$ cylinder for $J/J_D=0.63$ and $T/J_D=0.1$. We investigate the spin correlations and structure factors with $A^\dagger=B=S^z(\bm{q})$ as in \cref{eq:spinkspace,eq:spinkspacewindow} with $\bm{q} = (\pi, 0)$. The estimated averages $\mathcal{C}_{AB}(t)$ and $\mathcal{C}_{AB}(\omega)$ with a statistical error estimate from $R=100$ samples are shown as the red line with the shaded light red region.}
    \label{fig:samples}
\end{figure}

For convenience, we introduce the shorthand notations,
\begin{align}
    &C(t) = C_{AB}(t), \quad &C(\omega) = C_{AB}(\omega), \\
    &C^i(t) = C^i_{AB}(t), \quad &C^i(\omega) = C^i_{AB}(\omega).
\end{align}
The algorithm laid out above is summarized in \cref{alg:dynmetts}, and illustrated graphically in Fig.~\ref{fig:illustration}. Using the Shastry-Sutherland model as a concrete example, we show the corresponding correlation function $\mathcal{C}_{AB}^{i}(t)$ and $\mathcal{C}_{AB}^{i}(\omega)$ for several sampled METTS states as well as their average value in Fig.~\ref{fig:samples}. At the given temperature $T/J_D=0.1$ we observe a ``bunching'' of correlation functions $\mathcal{C}_{AB}^{i}(t)$ around the $T=0$ correlation function whereas several other random realizations disperse, illustrating the effect of thermal fluctuations.

The real-time approach of measuring spectral functions at finite temperature is rigorous as in the limit of long time domain and sufficient sampling of METTS states $\ket{\psi_i}$. 
However,  a limitation of this algorithm is naturally the unbounded growth of entanglement entropy in real-time evolution. In particular, at higher temperatures entanglement growth can hinder accurate calculations at given computational resources. As shown in Fig.~\ref{fig:entanglement_samples} below, the bipartite entanglement entropy $S_{\text{vN}}$ of time evolution based on METTS states of the Shastry-Sutherland model at $T/J_D=0.2$ grows much more quickly on average than the one at $T/J_D=0.1$.  Meanwhile, pronounced fluctuations of the entropy after time evolution exist among various sampled METTS states.  

In the limit of infinitely long simulation time $\Omega\rightarrow \infty$ and infinitely accurate calculations of the entangled states $\ket{\psi_i}$, $\ket{v_i(t)}$, and $\ket{w_i(t)}$, this algorithm is exact assuming ergodicity of the METTS Markov chain. A finite time horizon $\Omega$ introduces an artificial broadening to the spectral function on the order of $\propto 1 / \Omega$. The accuracy in computing time-evolutions of MPS states can be well controlled by modern algorithms described in the next section.

\subsection{Time-evolution using matrix product states}\label{sec:TDVP}
\label{sec:timeevo}

The key computational bottleneck in both the thermal as well as the dynamical METTS algorithm is the computation of either the imaginary-time evolution of the product states $\ket{\sigma_i}$ or the real-time evolutions $\ket{v_i(t)}$ and $\ket{w_i(t)}$ of the METTS state $\ket{\psi_i}$. Research on time-evolution techniques for matrix product states has received a considerable amount of attention over the years resulting in several practical methods that are in common use. In a recent review~\cite{Paeckel2019}, these methods have been thoroughly compared in various test scenarios. These thorough comparisons suggest that the time-dependent variational principle (TDVP) for MPS~\cite{Haegeman2011,Haegeman2016} often performs optimally, even though other approaches like the WII method~\cite{Zaletel2015} can perform equally well.

Similarly to DMRG, the TDVP method can be performed in a single-site (1TDVP) and a two-site version (2TDVP). While the 1TDVP method conserves energy exactly when performing a real-time evolution and is faster than 2TDVP, it does not allow the bond dimension to be increased without further extensions. 2TDVP on the other hand allows the MPS bond dimension to increase during the evolution. Both TDVP methods come with two control parameters. The first one is the time step $\tau$ which can be chosen to be remarkably large while only incurring a small error. In Ref.~\cite{Wietek2021} a detailed study was performed of the accuracy of TDVP depending on the time step size $\tau$.  There, a time step size between $\tau=0.1, \ldots, 0.2$ in units of the hopping amplitude $t$ of the Hubbard model was found to be optimal. The second control parameter is the cutoff $\varepsilon$ which is the analog of the DMRG truncation error which, here, is chosen to be $10^{-6}$. Interestingly, TDVP does not become more accurate in the limit $\tau\rightarrow 0$, as then more truncations are necessarily leading to an increased overall error. 

When performing long real-time evolutions we use 2TDVP to increase the MPS bond dimension until a maximal bond dimension $\chi_{\max}$ is encountered. Thereafter, 1TDVP is employed to further time-evolve the MPS. Results are then studied as a function of $\chi_{\max}$ to investigate convergence. A peculiarity of the TDVP algorithm is that upon evolving states with small bond dimensions (e.g.\ product states), a so-called projection error is encountered~\cite{Paeckel2019}. To avoid this error, the initial time evolution for the thermal METTS states should be performed by a different method with high accuracy. For this purpose, the time-evolving block decimation~\cite{Vidal2003}  algorithm has been applied previously~\cite{Wietek2021}. However, more recently a basis extension technique for 1TDVP has been proposed~\cite{Yang2020}, which can be applied easily without necessitating a Trotterization with possible swap gates. In our specific example, at very low temperatures, the projection error leads to a potential issue when evaluating the \textit{analytic} correlation function away from the real-time axis. 
This is due to the low entanglement property of the ground state of the Shastry-Sutherland model, and we document this discussion in Appendix~\ref{sec:error_TDVP}. This issue is specific to the dimer product ground state of the Shastry-Sutherland model and is not expected to occur in general for more entangled ground states. Our implementation is based on the ITensor library~\cite{ITensor} and the TDVP package of ITensor~\cite{ITensorTDVP}.

\subsection{Spectral analysis and windowing}
\label{sec:spectral}
Evaluating the METTS spectral functions $\mathcal{C}^{i}(\omega)$ via the Fourier transform \cref{eq:mettspectral} cannot be done exactly as we are only provided with data for $\mathcal{C}^{i}(t)$ on discrete times,
\begin{equation}
    t = 0, \;\delta_t,\; 2\delta_t,\; \ldots,\; (S-1)\delta_t,\; S \delta_t (= \Omega).
\end{equation}
In our simulations, we found that a thorough treatment of the Fourier transform is crucial to arrive at accurate spectral functions. A naive discrete Fourier transform of the form,
\begin{equation}
    \label{eq:fouriernaive}
    \mathcal{C}^{i}(\omega) \approx \delta_t\sum_{k=-S}^{S} e^{-i\omega k \delta_t} \mathcal{C}^{i}(k \delta_t),
\end{equation}
leads to artificial spectral content due to the abrupt end of the signal and cannot be used for accurate measurements of the spectral function. Instead, we employ a windowing function $w(k)$ to consider the windowed Fourier transform~\cite{Prabhu2018}, 
\begin{equation}
    \label{eq:fourierwindow}
    \mathcal{C}^i(\omega) \approx \delta_t\sum_{k=-S}^{S} e^{-i\omega k \delta_t} \mathcal{C}^i(k \delta_t) w(k).
\end{equation}
In the context of DMRG simulations of spectral functions at zero temperature the use of several distinct windowing functions has been reported, including Parzen windows~\cite{Barthel2009} and Gaussian windows~\cite{Drescher2022}. Here, we find that a Hann window of the form,
\begin{equation}
    \label{eq:hann}
    w(k) = 0.5 + 0.5 \cos\left(\frac{\pi k}{S}\right),
\end{equation} 
for $k=0, \ldots, S$ yields the most accurate results when compared to exact diagonalization data in \cref{fig:ed_comparison}. Hence, we adopt the Hann window as the windowing function in all our subsequent simulations. 

Introducing a windowing function reduces the integrated spectral weight. This reduction is commonly remedied by multiplying with a window correction factor~\cite{Prabhu2018}. We opt for the so-called energy correction,
\begin{equation}
    \Gamma = \sqrt{\frac{1}{2S+1}\sum_{k=-S}^{S} w(k)^2},
\end{equation}
conserving the total integrated spectral weight. The estimator for the correct infinite-time Fourier transform is given by,
\begin{equation}
\tilde{\mathcal{C}}^i(\omega) = \frac{1}{\Gamma}\;\mathcal{C}^i(\omega).
\end{equation}
In the following, we will use the notation $\mathcal{C}^i(\omega)$ implicitly for the energy-corrected spectral function $\tilde{\mathcal{C}}^i(\omega)$.

\subsection{Validation by Exact Diagonalization and DMRG}
\label{sec:ED}

\begin{figure}[t!]
    \centering
    \includegraphics[width=\columnwidth]{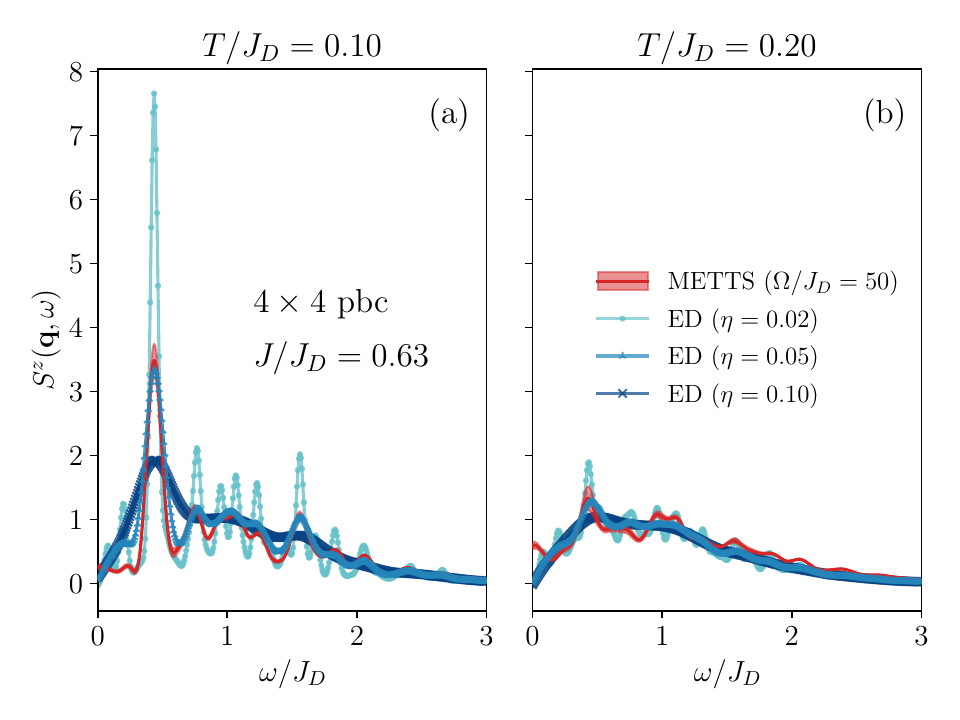}
    \caption{Comparison of results from dynamical METTS to the spectral function from exact diagonalization on a $4\times 4$ lattice with fully periodic boundary conditions at $J/J_D=0.63$ and $\bm{q}=(\pi/2, \pi/2)$ for temperatures $T/J_D=0.1$ (a) and $T/J_D=0.2$ (b). As the exact spectral function from ED is a sum of delta functions, we show Gaussian broadenings for widths $\eta=0.02, 0.05, 0.10$. For the dynamical METTS, we used $R=100$ samples with a real-time evolution up to $\Omega/J_D=50$. We observe close agreement between the METTS results and the ED results at Gaussian broadening $\eta=0.05$.}
    \label{fig:ed_comparison}
\end{figure}

To demonstrate the accuracy of our method we start by comparing to data on a $4\times 4$ simulation cluster with periodic boundary conditions, where exact diagonalization (ED) can be straightforwardly performed. In ED, we consider the Lehmann representation of the dynamical spectral function,
\begin{align}
\label{eq:lehmann}
    S^z(\bm{q}, \omega) =  \frac{2\pi}{\mathcal{Z}}\sum_{n,m}
    e^{-\beta E_n} |{\braket{m | S^z(\bm{q}) | n}}|^2 \delta(\omega - (E_n - E_m)),
\end{align}
where $\ket{n}$ and $\ket{m}$ denotes an eigenbasis with corresponding eigenvalues $E_n$ and $E_m$ of the Hamiltonian and $\delta$ denotes the Dirac delta function. The eigenvalues and eigenvectors are computed by a full diagonalization of the Hamiltonian matrix. Hence, both the positions of the poles $(\varepsilon_n - \varepsilon_m)$ as well as the weights are computed with machine precision. To analyze and compare the ED spectral function, it is customary to introduce a broadening of the spectral function by replacing the set of Dirac delta peaks with e.g.\ narrow Gaussian functions of width $\eta$, 
\begin{equation}
    \delta(\omega - p) \quad \longrightarrow \quad \frac{1}{\eta\sqrt{2\pi}} \exp\left[ -\frac{(\omega - p)^2}{2\eta^2}\right].
\end{equation}

A crucial difference to the proposed METTS method is the fact that the poles and weights are computed exactly and no time evolution is performed. A priori, it is not clear which precision can be expected when only simulating up to a maximal time $\Omega$ and also when only having access to the time-dependent correlator at discrete time steps which are a multiple of $\delta t$. Moreover, it is also not obvious what statistical error we can expect from sampling a finite number of METTS states.

To address these questions, we show results from ED as well as dynamical METTS simulations in \cref{fig:ed_comparison}. We focus on the parameter $J/J_D=0.63$ and $\bm{q}=(\pi/2, \pi/2)$ and consider two temperatures $T/J_D=0.10$ in (a) and $T/J_D=0.20$ (b). We sampled $1000$ METTS and performed a time-evolution up to a maximal time of $\Omega/ J_D = 50$ for every $10$-th METTS. By only considering a subset of all sampled METTS, we verified that the measurement time series has a negligible autocorrelation time. We also employed measurements in the $S^x$-basis, to further assure vanishing autocorrelation times, see e.g.~\cite{Stoudenmire2010,Wietek2021}. We employed 2TDVP as the time evolution algorithm with a cutoff $\varepsilon=10^{-6}$ without a constraint on the maximal bond dimension $\chi_{\max}$. While a $4 \times 4$ simulation cluster is arguably small, we would like to point out that the employed fully periodic boundary conditions are not without challenge for MPS methods, since long-range interactions are present. Hence, this benchmark also serves as a good first test of the quality of time-evolution algorithms employed to compute the METTS states as well as the real-time evolution. We apply the MPO representing $S^z(\bm{q})$ directly for a specific $q$ in our simulations instead of measuring correlations between pairs of sites. This is done in order to minimize the amount of measurements needed, and to allow for statistical self-averaging across the lattice during sampling. To compute the Fourier transform $S(\bm{q}, \omega)$ to frequency space, we compared different windowing functions and found the Hann window \cref{eq:hann} to give the most accurate results, cf.~\cref{sec:spectral} and \cref{sec:windowing}. For ED, we compare three different values of the broadening $\eta=0.02, 0.05, 0.10$. 

We report a close agreement between the ED and METTS simulations. While both the Gaussian broadening $\eta$ for the ED data as well as the finite time horizon $\Omega$ have a broadening effect on the exact spectrum \cref{eq:lehmann}, we observe that even small intricate structures of the spectrum (likely stemming from the small system size) are accurately captured by the METTS algorithm. In particular, the positions of the numerous side peaks are accurately reproduced, as well as the overall spectral amplitude. Choosing  $\Omega/ J_D = 50$ and applying the windowing procedure as explained in \cref{sec:spectral} yields a broadening of the spectral function comparable to a Gaussian broadening $\eta\approx 0.05$ in ED. Hence, we conclude that having a finite time horizon $\Omega$ as well as only a discrete set of measurements with time steps $\delta_t$ only leads to a small broadening of the spectrum. Discrepancies around the $\omega\rightarrow0$ limit stem from the finite simulation time-horizon $\Omega$. Moreover, we report that averaging over only $R=100$ METTS yields a satisfyingly small statistical error indicated by the shaded regions. 

\begin{figure}[t!]
    \centering
    \includegraphics[width=\columnwidth]{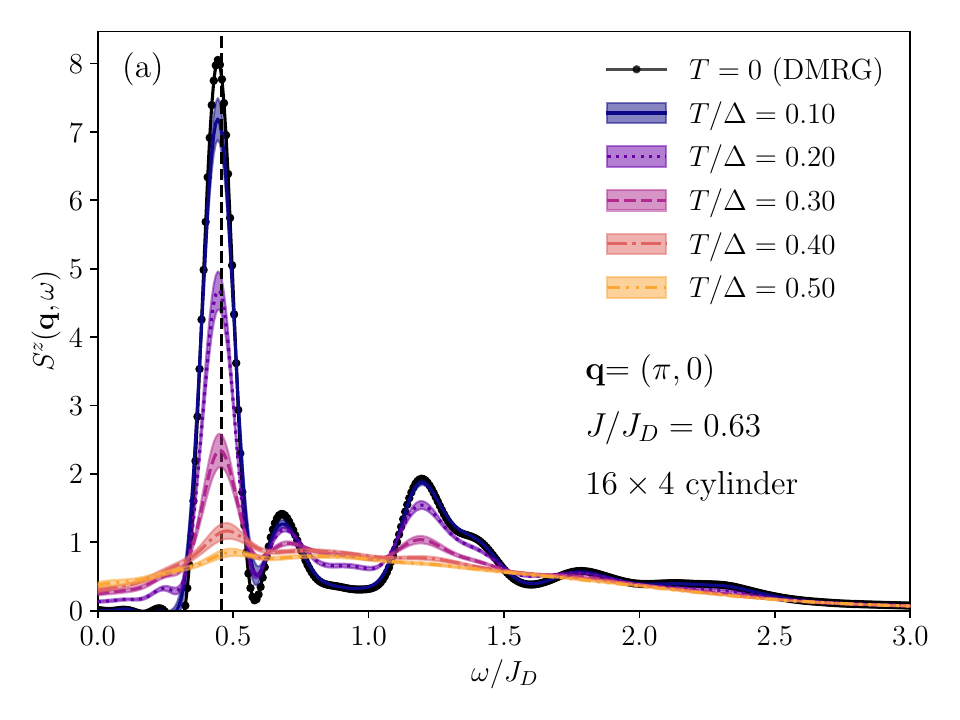}
        \includegraphics[width=\columnwidth]{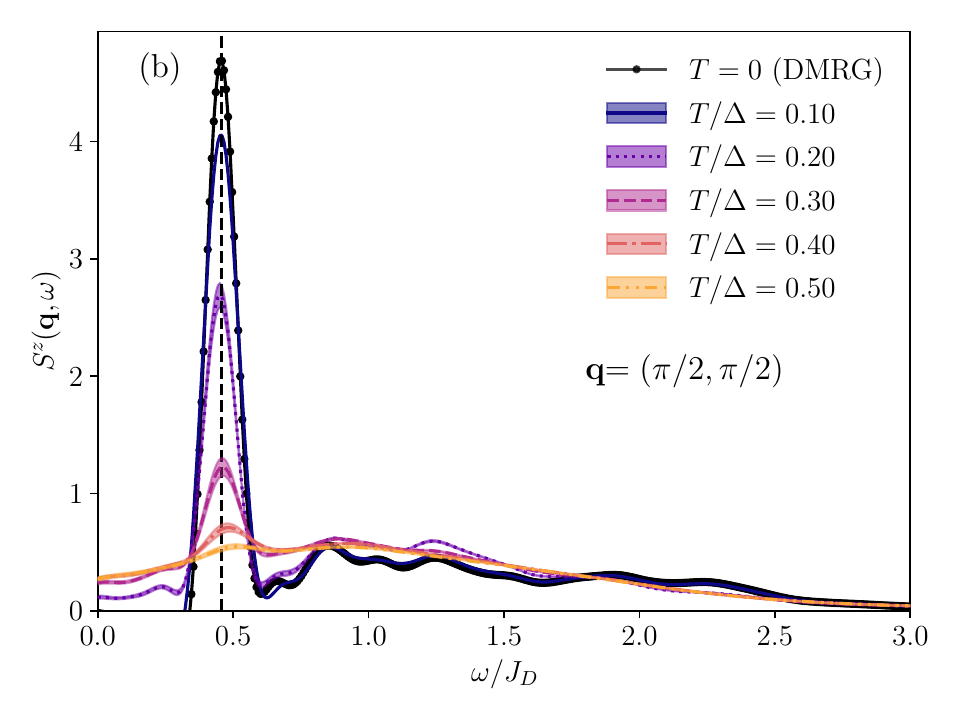}
    \caption{Dynamical spin structure factor $S^z(\bm{q}, \omega)$ of the Shastry-Sutherland model at $J/J_D=0.63$ on a $16\times4$ cylinder obtained using the real-time dynamical METTS algorithm at various temperatures in units of the triplon gap $\Delta=0.458 J_D$. We compare with $S^z(\bm{q}, \omega)$ at $T=0$ obtained using DMRG for momenta $\bm{q}=(\pi, 0)$ (a)
    and $\bm{q}=(\pi/2, \pi/2)$ (b). Shaded regions depict the standard error of the estimator obtained from $R=100$ METTS states with time evolution up to time $\Omega/J_D=50$. MPS time-evolution has been performed using the 2TDVP algorithm with $\chi_\text{max} = 2000$. We observe that even small features, as well as secondary and tertiary peaks, are well-resolved and converge towards the DMRG result for $T\rightarrow 0$.}
    \label{fig:widthfourresults}
\end{figure}

We now compare the results from simulations at nonzero temperature with ground state spectral functions from DMRG. This demonstrates the accuracy of the dynamical METTS algorithm in the limit of low temperatures. Results for spectral functions at different temperatures, including $T=0$ from DMRG, are shown in \cref{fig:widthfourresults}. We focus on the $16\times 4$ cylinder at $J/J_D=0.63$ and two wave vectors $\bm{q}= (\pi,0)$ and $\bm{q}=(\pi/2, \pi/2)$. Our maximal time horizon for the real-time evolution has been chosen as $\Omega/J_D=50$. Moreover, we collected $R=100$ independent METTS states for each temperature and observed that this allows sharp error estimates with a large signal-to-noise ratio, as apparent by the small error bars in \cref{fig:widthfourresults}. The time evolution has been performed using 2TDVP with maximal bond dimensions $\chi_{\max}=1000,1500,2000$.
Comparing different bond dimensions for different seeds leads us to conclude that for this particular system, the sample real-time correlation functions $C^i(t)$ are already well-converged at $\chi_{\max}=1000$. The spectral function from DMRG has similarly been obtained by a single time-evolution up to time $\Omega/J_D=50$. We observe a structured spectrum, where the dominant peak corresponds to single triplon excitations, but also secondary and tertiary peaks corresponding to multi-triplon excitations of the Shastry-Sutherland model. The good convergence towards the DMRG results for $T\rightarrow 0$, where even small features such as secondary and tertiary peaks are well-resolved, demonstrates the accuracy of the approach at low temperatures.

\begin{figure}
    \centering
    \includegraphics[width=\columnwidth]{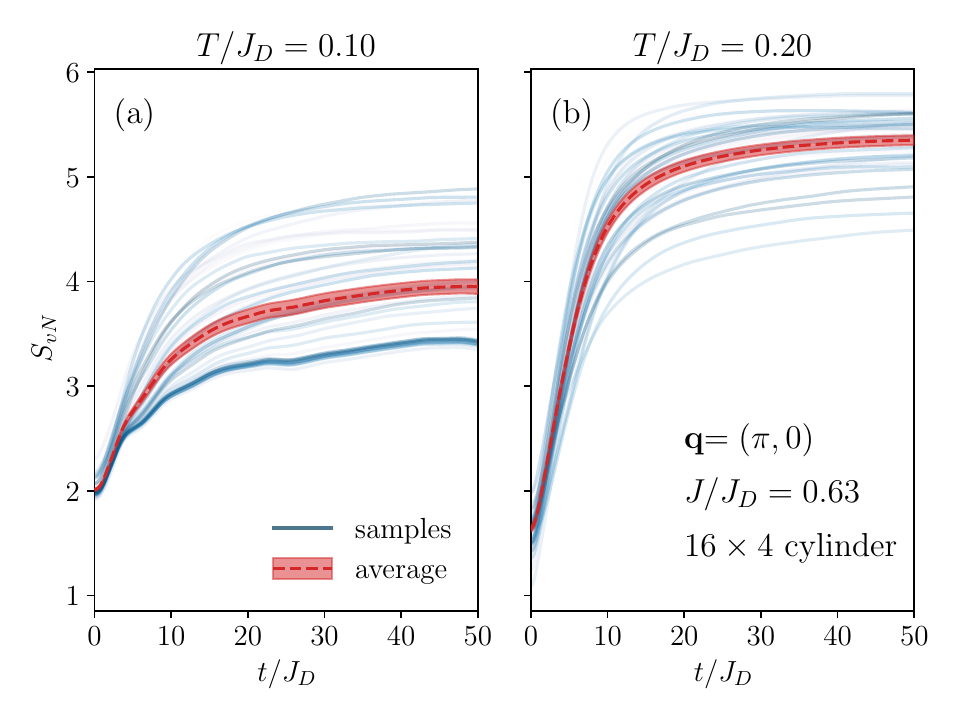}
    \caption{Von Neumann entanglement entropy $S_{\text{vN}}$ of time evolved METTS states $|v_i(t)\rangle$ from \cref{eq:auxstatevi}  on the $16\times4$ lattice for a bipartition in the center of the lattice. We choose $J/J_D=0.63$ and $\bm{q}= (\pi, 0)$. Increasing the temperature from $T/J_D=0.1$ (a) to $T/J_D=0.2$ (b) increases the average entanglement entropy, as shown by the red line with shaded region estimating the statistical error from $R=100$ samples.}
    \label{fig:entanglement_samples}
\end{figure}

A key quantity determining the accuracy of matrix product states is the von Neumann entanglement entropy,
\begin{equation}
    \label{eq:svn}
    \mathcal{S}_{\text{vN}}(\rho, A) = - \text{Tr}[\rho_A \log(\rho_A)],
\end{equation}
where the reduced density matrix with respect to a subsystem $A$ is given by $\rho_A = \text{Tr}_B [\rho]$ ($B$ denotes the complement of $A$) and $\rho$ denotes an arbitrary density matrix. For pure states $\ket{\psi}$, the density matrix is given by 
\begin{equation}
\rho(\ket{\psi}) = \ket{\psi}\bra{\psi}.
\end{equation}
For any METTS state $\ket{\psi_i}$, the entanglement entropy in the limit $T\rightarrow\infty$ is vanishing, as then the METTS state is a product state. In the limit  $T\rightarrow 0$, the METTS states approach the ground state and its entanglement entropy is naturally bounded. We analyze the behavior of the time-evolved METTS states $\ket{v_i(t)}$ as in \cref{eq:auxstatevi} in the dynamical METTS algorithm. Results for the $16 \times 4$ cylinder with $J/J_D=0.63$ are shown in \cref{fig:entanglement_samples}. The data shown has been obtained using 2TDVP with a maximal bond dimension $\chi_{\max} = 2000$. We observe a significant increase in entanglement entropy for a bipartition in the center of the system over the simulated time horizon. Increasing temperature also leads to an increase in entanglement entropy of the time evolved states $\ket{v_i(t)}$ even though the original METTS states $\ket{\psi_i}$ have lower entanglement at a higher temperature. The behavior of the second kind of auxiliary states $\ket{w_i(t)}$ is similar.

Even though the growth of entanglement is sizable for this system, the time-dependent correlation functions $C^i(t)$ can still be converged with $\chi_{\max}=1000,1500,2000$ on the width $W=4$ cylinder. This changes when considering cylinders of width $W=6$. There, we found that correlation functions even at a bond dimension $\chi_{\max}=2000$ can be not converged for several samples, especially when increasing temperature. Due to this reason, we introduce the complex-time correlation function as an approach to limit the growth of entanglement in the next section.

\section{Complex-time correlation functions}
Highly entangled quantum states typically require a large bond dimension to be efficiently compressed using a matrix-product state, which limits the system sizes that can reliably be studied. In the present case, we found the width $W=4$ cylinders to be amenable to the accurate real-time evolution algorithm, whereas $W=6$ is significantly more challenging and ultimately did not lead to converged results given our current computational resources. The main problem hindering such simulations is the entanglement growth with simulation time if performing real-time evolution, as shown in \cref{fig:entanglement_samples}. In the following sections, we discuss how this problem can be ameliorated, by performing complex-time evolution. 

% Here
 
\subsection{Two variants of complex-time correlation functions}
\label{Sec:two_def} 
Real-time evolution of MPS is computationally expensive due to unbounded entanglement growth for generic wave functions. Imaginary-time correlation functions, on the other hand, can be efficiently computed since for imaginary-time $\tau \rightarrow \infty$ every wavefunction $|\phi\rangle$ will converge to the ground state $|\phi\rangle = | \psi_0 \rangle$ whenever both wave functions have non-vanishing overlap $ |\langle \phi | \psi_0\rangle| \neq 0$, i.e., 
\begin{equation}
 \lim\limits_{\tau\rightarrow \infty}   e^{-\tau H} |\phi\rangle  \simeq | \psi_0 \rangle, 
\end{equation}
where $\simeq$ denotes equivalence up to a phase of the normalized quantum states. In the following, we will parameterize a line through the origin in the complex plane by the angle $\theta$,
\begin{equation}
\label{eq:zdef}
    z  =  t - i \tau  = |z|e^{  -i \theta  }.
\end{equation}
The negative sign when defining $\tau$ in \cref{eq:zdef} is owed to the fact that we aim to perform imaginary time evolutions only with negative exponents and this convention simplifies further notation significantly. For real-time evolution, we write a time-dependent correlation function in terms of the time-evolution operator $U(t)$,
\begin{equation}
\mathcal{C}_{AB}(t) = \braket{A(t) B}, \text{ where } A(t) = U(t)^\dagger A U(t),
\end{equation}
where $U(t) = \exp(-iHt)$ denotes the unitary time-evolution operator of the time-independent Hamiltonian $H$. $U$ being unitary rests both on the fact that $H$ is Hermitian, but also on the fact that $t$ is real. Hence, the complex time-evolution operator of the form,
\begin{equation}
    \mathcal{U}(z) = e^{-iHz},
\end{equation}
where $z \in \mathbf{C}$ now denotes a complex time, is generically not a unitary operator for $\text{Im}(z) \neq 0$,
\begin{equation}
\mathcal{U}(z)^{-1} \neq \mathcal{U}(z)^{\dagger}.
\end{equation}
There are two natural %\replaced{natural}{conceivable}
alternatives when defining complex time correlation functions,
\begin{align}
    \label{eq:analyticcorr}
    \mathcal{A}_{AB}(z) &= \braket{ \mathcal{U}(z)^{-1}A \mathcal{U}(z) B}_\beta = \braket{e^{iHz} A e^{-iHz} B}_\beta,\\
    \label{eq:hermitiancorr}
    \mathcal{H}_{AB}(z) &= \braket{ \mathcal{U}(z)^{\dagger}A \mathcal{U}(z) B}_\beta = \braket{ e^{iH\overline{z}} A e^{-iHz} B}_\beta.
\end{align}
Here, we denote by $\overline{z} = t+i\tau$ the complex conjugate of $z$. $\mathcal{A}(z)$ is an analytic function of $z$ and therefore we will refer to this quantity as the \textit{analytic} correlation function. Moreover, for purely imaginary times this correlation function is given by,
\begin{equation}\label{Eq:C_a_imaginary}
    \mathcal{A}_{AB}(-i\tau) = \braket{e^{H\tau} A e^{-H\tau} B}_\beta,
\end{equation}
which is the well-known imaginary-time correlation function used in the Matsubara formalism. 
%The analytic correlator features an interesting (quasi-)periodicity in imaginary time,
%\begin{equation}
%\mathcal{A}_{AB}(z - i\beta) = \mathcal{A}_{BA}(-z) =  \overline{\mathcal{A}_{B^\dagger A^\dagger}(z)},
%\end{equation}
%as can be readily derived.  For purely imaginary time $z=i\tau$, and real operators $A$ and $B$ with $A^\dagger = B$ this corresponds to the usual periodicity in imaginary time, as exhibited by e.g.\ Matsubara Green's functions.

One notices that evaluating $ \langle \psi_i | e^{ H \tau } $ from the left side of Eq.~(\ref{Eq:C_a_imaginary}) necessarily requires a time evolution process with \textit{positive} time $\tau$.  
Generically,   
computing $e^{+H\tau} |\psi\rangle$ as might occur when evaluating $\mathcal{A}_{AB}(z)$ will ultimately lead to numerical instability. 
Clearly, this instability issue exists not only for imaginary time evolution but also for generic complex time. 
As we will demonstrate in \cref{sec:analyticcorr}, this issue in the evaluation of $\mathcal{A}_{AB}(z)$ can be circumvented within a METTS algorithm in an efficient way.  

The Hermitian correlator $\mathcal{H}_{AB}(z)$ as in \cref{eq:hermitiancorr}, on the other hand, has only recently been investigated in the context of MPS and impurity models~\cite{Grundner2023} and has several intriguing properties which are discussed \cref{sec:hermitiancorr}. We choose the name \textit{Hermitian}, since whenever $A$ is a Hermitian operator, i.e., $A^\dagger = A$, the time-evolved operator $\mathcal{U}(z)^{\dagger} A \mathcal{U}(z)$ is also Hermitian. This property is shared with the real-time correlation function, but not with the analytic correlation function, where in general $\mathcal{A}_{AB}(z)^\star \neq \mathcal{A}_{BA}(z)$.

There are computational advantages in evaluating this time-dependent correlation function. For convenience, we again introduce the shorthand notations,
\begin{align}
    \mathcal{A}(z) = \mathcal{A}_{AB}(z),\\ 
    \mathcal{H}(z)=\mathcal{H}_{AB}(z).
\end{align}

\begin{figure}
    \centering
    \includegraphics[width=\columnwidth]{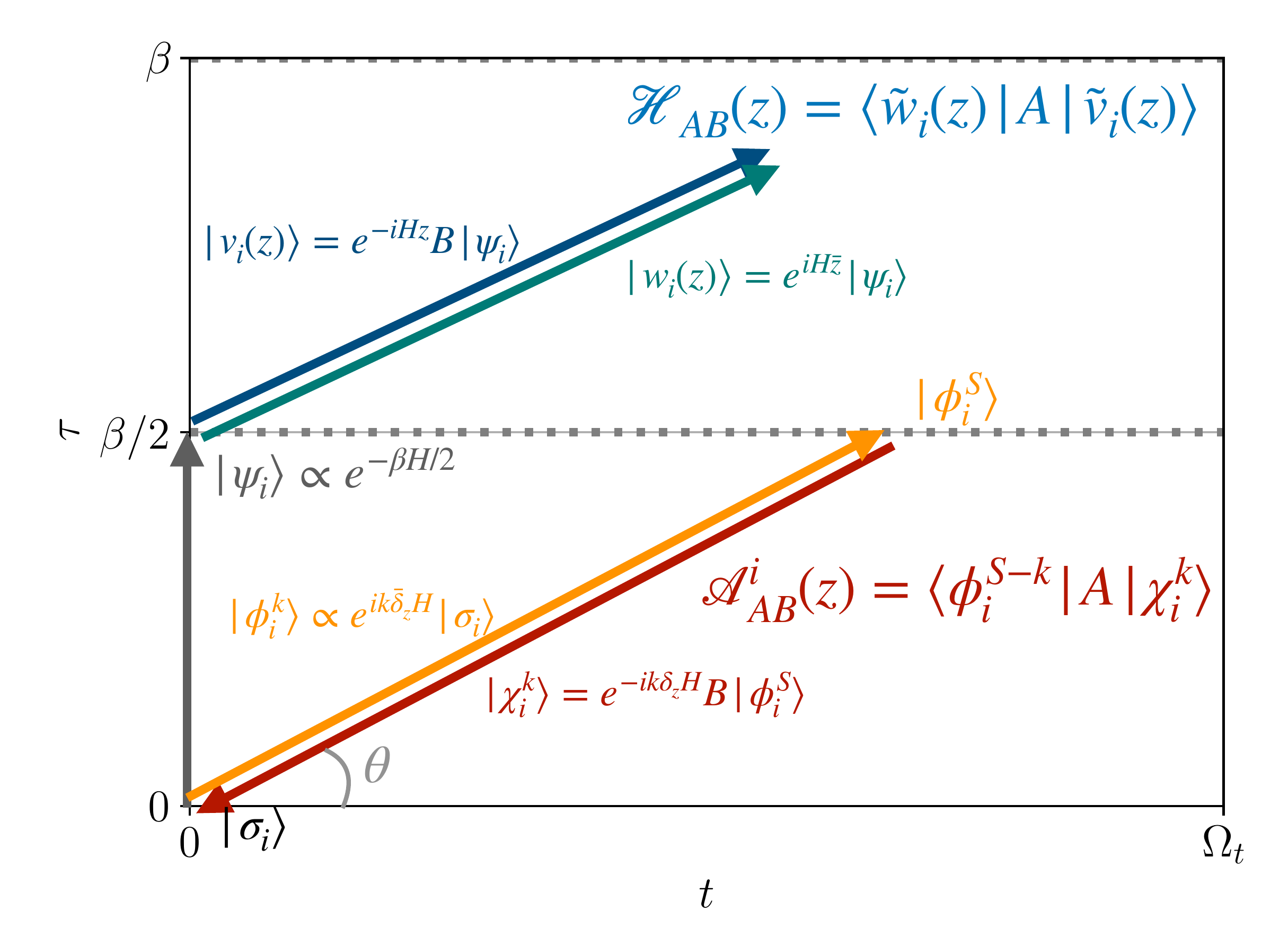}
    \caption{Sketch of the complex contours on which the analytical correlation function $\mathcal{A}(z)$ and the Hermitian correlation functions $\mathcal{H}(z)$ are evaluated. For the analytical correlation function, the maximal imaginary time is chosen to be $\Omega_\tau = \beta / 2$ in order to allow for an efficient and stable numerical algorithm.}
    \label{fig:omega_sketch}
\end{figure}

\subsection{Analytic complex-time correlation functions}
\label{sec:analyticcorr}   
We propose an efficient method for evaluating the 
\textit{analytic} complex time correlation function $\mathcal{A}(z)$.  
Its definition as in \cref{eq:analyticcorr} is expressed as, 
\begin{align} 
\label{Eq:OO_corr}
   \mathcal{A}(z) &= \braket{e^{iHz} A e^{-iHz} B}_\beta \nonumber \\
   &= \frac{1}{\mathcal{Z}} \sum_i p_i  \langle \psi_i |  
e^{ i z H }  A e^{ - i z H }  B    
| \psi_i \rangle   
\end{align} 
into the basis of METTS $\ket{\psi_i}$. We choose to evaluate the analytical correlator on the contour $z=|z|e^{-i\theta}$ for,
\begin{equation}
    |z| \in [ 0, \Omega ], \quad \text{Re}(z) \geq 0,
\end{equation}
where $\Omega$ denotes a final time horizon. We define,
\begin{equation}
    \Omega_t = \Omega \cos \theta, \quad \Omega_\tau = \Omega \sin \theta, \quad \Omega_z = \Omega_t - i \Omega_\tau.
\end{equation}
Thus, the real and imaginary time components take values, 
\begin{equation} 
t \in [ 0, \Omega_t] \subseteq \mathbb{R},  \quad  \tau  \in [ 0, \Omega_\tau ]\subseteq \mathbb{R}.
\end{equation}
As shown further below, when choosing
\begin{equation}
    \Omega_\tau = \frac{ \beta }{ 2 }
\end{equation}
the analytic correlator $\mathcal{A}(z)$ can be evaluated without ever having to perform an imaginary time evolution with positive time $e^{+\tau H} \ket{\psi}$. For actual computations, we discretize the complex time $z$ into $S$ steps with step size $\delta_z =  \Omega_z / S$, i.e.
\begin{equation}
    z = 0, \; \delta_z, \; 2 \delta_z, \; \ldots, \;(S-1) \delta_z, \; S \delta_z (= \Omega_z),
\end{equation}
where,
\begin{equation}
    \delta_z = \delta_t - i\delta_\tau = |\delta_z|e^{  -i \theta  }.
\end{equation}
The complex contour and the quantities $\Omega_z$, $\Omega_t$, $\Omega_\tau$ for which we evaluate $\mathcal{A}(z)$ are shown in \cref{fig:omega_sketch}.  

A naive approach to evaluating $\mathcal{A}(z)$ would be to compute intermediate states,
\begin{align}
\label{eq:timeevomettscomplexwrong}
&|v_i(z)\rangle = e^{-iHz} B |\psi_i \rangle = e^{-iHt} e^{-\tau H}B |\psi_i\rangle \\
&|w_i(z)\rangle = e^{-iH\overline{z}} |\psi_i \rangle = e^{-iHt} e^{+\tau H} |\psi_i\rangle,
\end{align}  
with $z=t-i\tau$ similar to the real-time correlation algorithm \cref{alg:dynmetts}. However, this approach is plagued by severe numerical instability when computing the state $|w_i(t)\rangle$.
Computing a positive exponential $e^{+\tau H} |\psi_i \rangle$ is numerically highly unstable and prohibits most practical calculations. This is due to the fact that the initial METTS state is a low-energy state. High-energy contributions are truncated during the time evolution but would need to be resolved accurately when performing positive time evolution causing severe numerical instabilities.
  
When restricting our time horizon to $\Omega_\tau = \beta / 2$ we can circumvent ever having to compute positive real exponential in the context of the METTS algorithm. Thus, given we are only interested in simulating complex times with $\Omega_\tau < \beta / 2$ we propose that the analytic correlator can be evaluated in a numerically stable way. This is done by absorbing part of the positive time evolution into the negative time evolution performed for the METTS states. Notice that a smaller complex angle $\theta$ allows a longer real-time component $\Omega_t = \Omega \cos \theta$ while still fulfilling $\Omega_\tau = \Omega\sin \theta < \beta / 2$.

The key idea is that the positive imaginary time evolution can be absorbed by the negative imaginary time evolution performed when calculating an individual METTS $\ket{\psi_i}$. For this, we rewrite \cref{Eq:OO_corr} as,   
\begin{align}
\label{Eq:Redefine}
   \mathcal{A}&(z) = \braket{e^{ i z H } A  e^{ - i z H }  B}_\beta = \nonumber \\
 &= \braket{e^{ -i \Omega_t H}e^{ i z H } A  e^{ - i z H }  Be^{ i \Omega_t H}}_\beta \nonumber \\ 
  &= \frac{1}{\mathcal{Z}} \sum_i p_i  \langle \psi_i | e^{ -i \Omega_t H} 
e^{ i z H } A  e^{ - i z H }  B   e^{  i \Omega_t H}  | \psi_i \rangle \nonumber  \\ 
 &= \frac{1}{\mathcal{Z}} \sum_i p_i \; \mathcal{A}^i(z),
\end{align}
where we defined the sample analytic correlation function as,
\begin{equation}
\label{eq:analyticsample}
\mathcal{A}^i(z) \equiv  \langle \psi_i | e^{ -i \Omega_t H } 
e^{ i z H } A e^{ - i z H } B   e^{  i \Omega_t H} | \psi_i \rangle.
\end{equation}
Introducing the factor $e^{\pm i \Omega_tH}$ as in \cref{eq:analyticsample} allows to rewrite $\mathcal{A}^i(z)$ in a simple form,
\begin{align}
 \mathcal{A}^i(z) &=  \langle \psi_i | e^{ -i \Omega_t H } 
e^{ i z H } A e^{ - i z H } B   e^{  i \Omega_t H} | \psi_i \rangle   \nonumber  \\ 
%&= \frac{1}{p_i} \langle \sigma_i  |   
%   e^{ -\beta H/2} e^{-i \Omega_t H } e^{ i z H }   
%  A   e^{ - i z H }   B e^{i \Omega_t H} e^{ -\beta H/2} |\sigma_i \rangle \\ 
%&=  \frac{1}{p_i} \langle \sigma_i  |   
%   e^{ -\Omega_\tau H} e^{-i \Omega_t H } e^{ i z H }   
%  A   e^{ - i z H }   B e^{i \Omega_t H} e^{ -\Omega_\tau H} |\sigma_i \rangle \\ 
%&=  \frac{1}{p_i} \langle \sigma_i  |   
%   e^{ -i\Omega_z H} e^{ i z H }   
%  A   e^{ - i z H }   B e^{i \overline{\Omega}_z H} |\sigma_i \rangle \\ 
&= \frac{1}{p_i} \langle \sigma_i  |   
   e^{ - i (\Omega_z - z) H }  A   e^{ - i z H }   B e^{i \overline{\Omega}_z H}|\sigma_i \rangle.
   \label{eq:analyticcorrderiv1}
\end{align}
For $k=0, \ldots, S$ we introduce the auxiliary computational states $\ket{\phi_i^k}$ and $\ket{\chi_i^k}$ defined as,
\begin{align}
\ket{\phi_i^k} 
&\equiv \frac{1}{\sqrt{p_i}} e^{i k \overline{\delta}_z H} \ket{\sigma_i} &= \frac{1}{\sqrt{p_i}} e^{i k {\delta_t} H} e^{- k {\delta_\tau} H} \ket{\sigma_i}, \label{eq:analyticphi} \\
\ket{\chi_i^k} &\equiv e^{-i k \delta_z H} B \ket{\phi_i^S} &= e^{-i k \delta_t H} e^{- k \delta_\tau H} B \ket{\phi_i^S} .\label{eq:analyticchi} 
\end{align}

Importantly, we see that these states can all be computed by performing a negative imaginary time evolution, i.e., computing $e^{-\tau H} \ket{\psi}$ for some state $\ket{\psi}$. This insight is the main reason why $\mathcal{A}^i(z)$ can be computed efficiently. Notice that, 
\begin{align}
    \bra{\phi_i^{S-k}} &= \frac{1}{\sqrt{p_i}}\bra{\sigma_i}e^{-i(\Omega_z - k\delta_z)H}, \\
    \ket{\phi_i^S} &= \frac{1}{\sqrt{p_i}} e^{i \overline{\Omega}_z H}|\sigma_i \rangle.
\end{align}
Using a discretized $z = k \delta_z$ we can  rewrite \cref{eq:analyticcorrderiv1} as,
\begin{equation}
  \mathcal{A}^i(k\delta_z) =\bra{\phi^{S-k}_i} A \ket{\chi_i^{k}}.
\end{equation}

\begin{algorithm}[t!]
  \caption{Dynamical METTS for analytic correlator $\mathcal{A}(z)$}
   \begin{algorithmic}[1]
   \State Choose a random initial product state $| \sigma_1 \rangle = | \sigma_1^1 \rangle \cdot \ldots| \sigma_1^N \rangle$
    \For{$i = 1, \ldots , R$} 
  \State Compute $|\psi_i\rangle = \frac{1}{\sqrt{p_i}} e^{-\beta H/2}| \sigma_i \rangle$ and $p_i=\braket{\sigma_i | e^{-\beta H} |\sigma_i}$    
  \State Set $\ket{\phi_i^0} = \frac{1}{\sqrt{p_i}}\ket{\sigma_i}$
  \For{k\; = \;1, \ldots, S}
      \State Compute $\ket{\phi_i^k} = e^{i\delta_t H} e^{-\delta_\tau H}\ket{\phi_i^{k-1}}$
      \State Store $\ket{\phi_i^k}$ to disk for later use
  \EndFor 
  \For{k\;=\;0, \ldots, S}
  \If{$k = 0$}
    \State Compute $| \chi_i^0 \rangle = B  |\phi_i^S \rangle$ 
  \Else
    \State Compute $\ket{\chi_i^k} = e^{-i\delta_t H} e^{-\delta_\tau H}\ket{\chi_i^{k-1}}$
  \EndIf
      \State Read $\ket{\phi_i^{S-k}}$ from disk
      \State Evaluate $\mathcal{A}^i(k\delta_z) =\bra{\phi^{S-k}_i} A \ket{\chi_i^{k}}$
  \EndFor
  \State Pick next product state $|\sigma_{i+1} \rangle$
   with probability $|\langle \psi_i | \sigma_{i+1} \rangle |^2$  
  \EndFor     
\end{algorithmic}
\label{alg:dynmetts_ana} 
\end{algorithm}

To evaluate these expectation values for all discretized values of $z= k \delta_z$ ($k=0, \ldots, S$), we compute the states $\ket{\phi_i^k}$ for $k=0, \ldots, S$ successively by complex time evolution and store all intermediate states to disk for later use. Then we compute the state $ \ket{\chi^0_i} = B \ket{\psi_i^S}$ from which we successively compute $ \ket{\chi^k_i}$ for $k=1, \ldots, S$ and thereby evaluate the desired correlation functions $\mathcal{A}^i(z) = \bra{\phi^{S-k}_i} A \ket{\chi_i^{k}}$. $\mathcal{A}(z)$ is then obtained by averaging over $\mathcal{A}^i(z)$. We summarize the described algorithm in \cref{alg:dynmetts_ana}.

\cref{alg:dynmetts_ana} is in several aspects similar to \cref{alg:dynmetts}. Both algorithms require computing the METTS $\ket{\psi_i}$ and then time-evolves two further states to compute time-dependent correlators. Moreover, the basic thermal METTS Markov chain is constructed in both cases. A difference in \cref{alg:dynmetts_ana} is the necessity of storing the auxiliary states $\ket{\phi_i^k}$. On present-day computers, this typically neither imposes severe time nor memory constraints. However, in the limit of small angle $\theta$ the memory requirement can be considerable due to the large number of discrete time steps and requires corresponding storage capacities.

\begin{figure}[t!]
    \centering
    \includegraphics[width=\columnwidth]{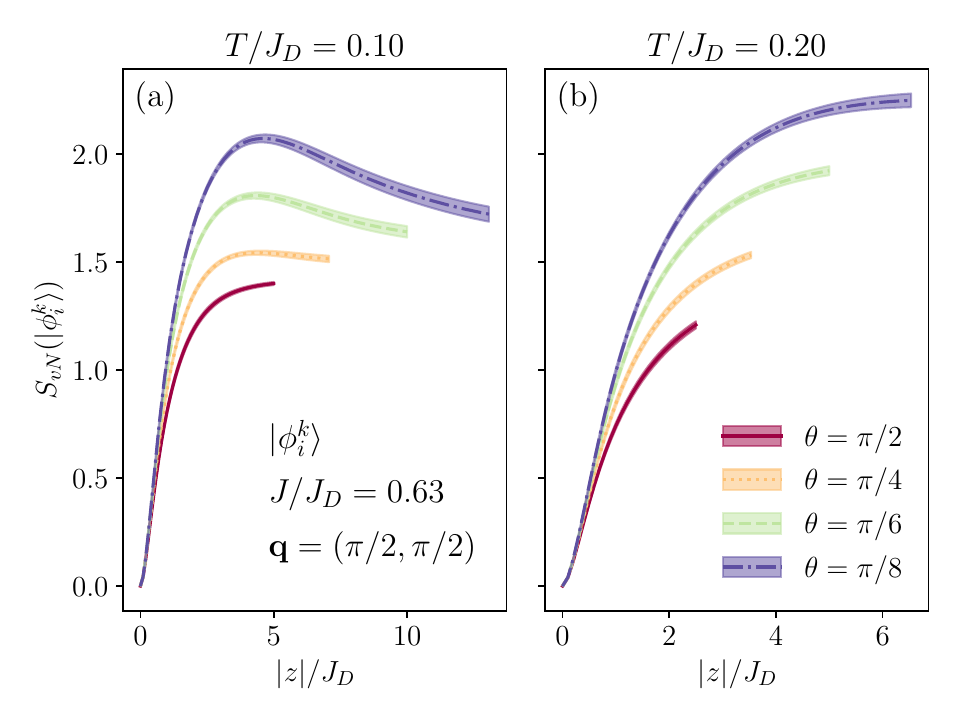}
    \caption{Average bipartite von Neumann entanglement entropy $S_{\text{vN}}$ of the auxiliary state $\ket{\phi_i^k}$ \cref{eq:analyticphi} for evaluating the analytical correlation
    function $\mathcal{A}(z)$. We consider the $16\times 4$ Shastry-Sutherland lattice for different values of the complex angle $\theta$ and $J/J_D=0.63$ and compare results for temperatures (a) $T/J_D=0.1$ and (b) $T/J_D=0.2$. The average is computed from $R=100$ METTS states and the shaded region shows the estimated standard deviation of the mean.}
    \label{fig:entanglement_step1}
\end{figure}

\begin{figure}[t!]
    \centering
    \includegraphics[width=\columnwidth]{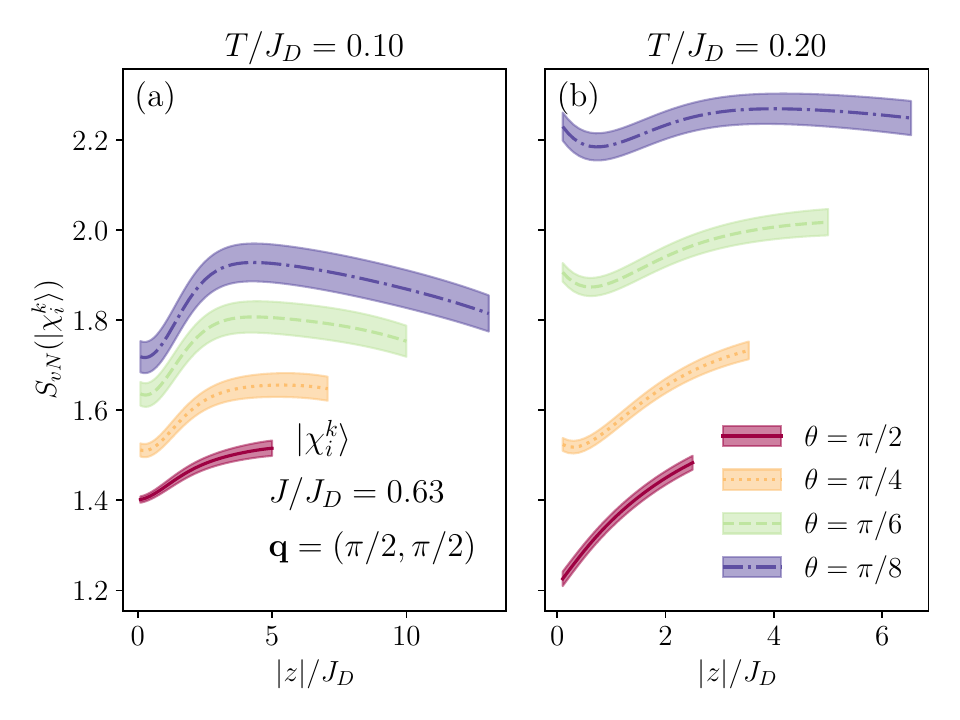}
    \caption{Same as Fig.~\ref{fig:entanglement_step1}, but for the auxiliary state $\ket{\chi_i^k}$ \cref{eq:analyticchi}.}
    \label{fig:entanglement_step2}
\end{figure}

The key quantity determining the efficiency of the algorithm is the bipartite von-Neumann entanglement entropy $S_{\text{vN}}$ as in \cref{eq:svn} of the MPS $\ket{\phi_i^k}$ and $\ket{\chi_i^k}$, shown in Figs.~\ref{fig:entanglement_step1} and \ref{fig:entanglement_step2}, respectively. The entanglement is considered for a bipartition at the center of the system. We evaluated the analytic correlator for the dynamical spin structure factor $S^z(\bm{q}, \omega)$ as in \cref{eq:dynspinstructure} for the $16 \times 4$ Shastry-Sutherland model at $J/J_D=0.63$ at temperatures $T/J_D=0.1$ and $T/J_D=0.2$. The auxiliary state $\ket{\phi_i^0} = \ket{\sigma_i}$ is a classical product state such that the entropy in \cref{fig:entanglement_step1} starts from zero. Slower entanglement growth is observed for larger values of $\theta$ as expected. For $\theta=\pi/2$ we are performing fully imaginary time evolution, which yields the slowest entanglement growth. Meanwhile,  fluctuations of $S_{\text{vN}}$ among the sampled states during METTS dynamics are rather small, as indicated by the shaded regions around the mean.  The absolute values of $S_{\text{vN}}$ are comparable for both $\ket{\phi_i^k}$ and $\ket{\chi_i^k}$ and grow with increasing temperature. We observe a significant reduction in entanglement entropy when comparing the values to the entanglement of the states $|v_i(t)\rangle$ \cref{eq:auxstatevi} as shown in \cref{fig:entanglement_samples}. While we observed maximal values of $S_{\text{vN}}\approx 5.3$ for $T/J_D=0.2$ for $|v_i(t)\rangle$ in \cref{fig:entanglement_samples}, we only find maximal values of $S_{\text{vN}}\approx 2.3$ for $\theta=\pi / 8$ in \cref{fig:entanglement_step1,fig:entanglement_step2}. Naturally, the chosen time horizon $\Omega$ is significantly shorter when evaluating the analytical correlator. However, this demonstrates that $\mathcal{A}(z)$ can be evaluated at significantly reduced cost at non-zero values of $\theta$. We want to point out that setting a value of $\theta$ determines the value of $\Omega$ due to the constraint $\Omega_\tau = \beta / 2$ we impose for the analytical correlator. Hence, the maximal complex times $|z|$ differ for different values of $\theta$. In our simulations, the maximal computed entanglement entropy exhibits a plateau due to simulating at finite bond dimension.

%Finally, we remind the readers of a possible systematic error in this approach.
%One must apply Eq.~\ref{Eq:Redefine} carefully since 
%the periodic boundary condition along time direction 
%for a thermal density matrix   
%is restored only after sufficiently long METTS sampling steps. In other words, the projector $ | \psi_i \rangle   \langle \psi_i | $ itself does not commute with the  propagator $ e^{ i \Omega_t H} $.  Let us denote the number of sampled  METTS configurations as $N_{M}$. 
%Although undoubtedly this systematic error can be well controlled in large $N_M$ limit, the question is  whether or not 
%the $ N_M $ scaling of this error can beat the one of statistical error from METTS sampling (i.e., $\sigma \propto 1/\sqrt{ R}$).  
%This issue may become crucial due to the stochastic analytic continuation (SAC) method that we will introduce in the next section: one necessarily requires the potential systematic errors 
%to be much smaller than the statistical one of the correlation function. 

\subsection{Stochastic analytic continuation}
\label{sec:SAC}  

To retrieve the spectral function from $\mathcal{A}(z)$ we perform analytic continuation, for which various approaches have been proposed and tested in the context of quantum Monte Carlo simulations~\cite{Silver_1990,Sandvik_SAC_98,beach2004identifying,SHAO20231}. Although mostly used for performing continuation from imaginary time correlation functions, these approaches are often also applicable for the analytic correlator $\mathcal{A}(z)$. Generally, inverse Laplacian approaches from imaginary time correlation functions lead to a biased result where the details of the efficiency are case-dependent. In most cases, dominant peaks capturing well-defined quasiparticles can be resolved satisfactorily. However, for a spectrum with contributions from both a low energy peak and a higher energy continuum, numerical continuation faces difficulties in quantitatively resolving the information at high frequency.  On the other hand,  we propose the numerical analytic continuation to be more efficient via correlation functions defined on the complex time case. % since real-time contributions of the correlator, even for short periods of time,  carry information about the high-frequency spectrum.  
At first sight, this natural extension would improve the performance of conventional imaginary-time-based numerical analytic continuation approaches since Green's functions that we use as input are closer to the desired real frequency angle. Most importantly, the imaginary time Green's function is dominated by the low-energy spectrum of the system.  Instead, the high-energy part of the spectrum contributes significantly to the real-time correlator. Hence analytic continuations in our general consideration, with real-time contributions of the correlator, even for short periods, may give an improvement in resolving the spectrum.

We now document our approach to stochastic analytic continuation (SAC) 
based on correlation functions in the complex time plane. In the following, we choose the observables to be $A^\dagger = B \equiv \mathcal{O}$.
The aim is to obtain the finite temperature spectral function,
\begin{equation}
\begin{aligned} 
   S( \omega )  \equiv  \frac{2\pi}{\mathcal{Z}} \sum_{ m, n } 
    e^{ - \beta E_m } | \langle  n | \mathcal{O} |  m \rangle |^2 
   \delta ( \omega - (E_m - E_n))    \end{aligned},
\end{equation}
where $|m\rangle$ ($|n\rangle $) is the $m$-th ($n$-th) eigenvector of the Hamiltonian with energy $E_m$ ($E_n$) cf.~\cref{eq:lehmann}. 

%As mentioned above, we perform continuation from the analytic correlation function  $\mathcal{A}(z)$ obtained from METTS simulation as in \cref{alg:dynmetts_ana},   
%\begin{equation}
%\begin{aligned}
% \mathcal{C}^{\text{a}}(z) & \equiv \langle O^{\dagger} (z) O (0) %\rangle 
%\end{aligned}    
%\end{equation}   
%where $ \langle O^{\dagger} (z) O (0) \rangle $ is defined in Eq.~\ref{Eq:def_two_corr} and the way of measuring this correlation function has been demonstrated in Sec.~\ref{Sec:CTE}. 

The correlation function is related to the spectral function via the following Laplacian transformation, 
\begin{equation}
\begin{aligned}
 \mathcal{A}  (z) & = \int_{ -\infty }^{\infty} d \omega  e^{ - z \omega }  S ( \omega ) \frac{1}{ 2 \pi} 
 \\  
   & = \int_0^{\infty}  d \omega  S(\omega) ( e^{ - z \omega } + e^{ - (\beta - z ) \omega } )  \frac{1}{2 \pi}   \\ 
   & \equiv \int_0^{\infty}  d \omega  A(\omega) K( z, \omega)   ,
\end{aligned}
\end{equation}
where $S(-\omega) = e^{ -\beta \omega } S(\omega) $ and we defined $A(\omega) \equiv S(\omega) (1+e^{-\beta \omega}) /2 $ and  
the integration kernel as,
\begin{equation} 
K( z, \omega) \equiv ( e^{- z \omega}  + e^{-(\beta - z )\omega }) (1+e^{-\beta\omega} )^{-1} \pi^{-1},
\end{equation}
which is by definition also a complex function. We parameterize $A(\omega)$ as a sum of delta functions, 
\begin{equation}\label{def:A_delta}
  A (\omega) = \sum_i^N A_i \delta (\omega - \omega_i ).
\end{equation}
During the stochastic sampling process, the testing Green's function is,       
\begin{equation} 
  \mathcal{A}^S(z) = \int_0^{\infty}  d \omega  A(\omega) K( z, \omega) .
\end{equation} 

We run stochastic sampling by shifting the positions and amplitudes of the delta functions in \cref{def:A_delta},  
  based on the following probability distribution,     
\begin{equation}
   P( A ) \propto e^{ - \frac{ \chi^2 (A)}{ \alpha } }  ,
\end{equation} 
where $\alpha$ denotes the fictitious temperature, and $ \chi^2 (A) $ is the $ \chi^2 $ that measures the deviation of correlation function based on any given $A(\omega)$ during stochastic sampling,  with the physical one that is observed from METTS calculation.  
In the current case,   
fluctuations from both real and imaginary parts of the correlation function are included in the definition of $ \chi^2 $,  
\begin{equation} 
\begin{aligned}\label{Eq:def_chi}
   \chi^2 \equiv  \sum_{ k = 1  }^{ S } 
   \sum_{ k' = 1 }^{ S }  
  &  \left({C^R_{ k  k' }}^{-1} 
 \text{Re}( \mathcal{A}^S (z_k ) - \overline{\mathcal{A}} (z_{k'} ) )^2  \right.\\ 
 + & \left.{C^I_{ k  k' }}^{-1} 
 \text{Im}( \mathcal{A}^S (z_k ) - \overline{\mathcal{A}} (z_{k'} ) )^2  \right),
\end{aligned} 
\end{equation}        
where $\overline{\mathcal{A}}(z_k)$ denotes the mean value of the analytic correlator at the $k$-th time step $z_k = k \delta_z$.  %Here  $ [ 1, N_t ] $ is the time domain of correlator used in our SAC.
%and it seems natural to take $ N_{ \text{start}} = 1$ such that the full range correlator is considered. 
The  covariance matrices corresponding to the real and imaginary parts of the correlator are defined respectively,     
%\begin{equation}
%    C_{ i j } \equiv  \frac{1}{ N_{ \text{Metts} } (N_{ \text{Metts} } ) } \sum_{ n =1 }^{N_{ \text{Metts} } } 
%( \mathcal{C}^n_{\text{a}}(z_i ) - \overline{\mathcal{C}}_{\text{a}}(z_i ) ) 
%( \mathcal{C}^n_{\text{a}}(z_j ) - \overline{\mathcal{C}}_{\text{a}}(z_j ) )   
%\end{equation}
\begin{equation} 
\begin{aligned} 
    C^R_{ k k' } \equiv & \frac{1}{ R ( R - 1  ) }  \\ 
   & \sum_{ i =1 }^{  R  }  
  \text{Re}( \mathcal{A}^i (z_k ) - \overline{\mathcal{A}} (z_k ) ) 
  \text{Re}( \mathcal{A}^i  (z_{k'} ) - \overline{\mathcal{A}}  (z_{k'} ) )   \\ 
    C^I_{ k k' } \equiv & \frac{1}{  R 
    (  R - 1  ) }  \\ 
   & \sum_{ i =1 }^{  R  }  
  \text{Im}( \mathcal{A}^i (z_k ) - \overline{\mathcal{A}} (z_k ) ) 
  \text{Im}( \mathcal{A}^i (z_{k'} ) - \overline{\mathcal{A}}(z_{k'} ) )   ,
\end{aligned} 
\end{equation}
where $\mathcal{A}^i (z_k )$ is the correlation function measured from the $i$-th METTS state. % and $ R $  denotes the total number of sampled Metts states used for measurement. 

%The optimal determination of temperature $\Theta$ has been argued in previous articles.  Here, we take $\Theta$ as the one  
%where the average $\chi^2$ is roughly the summation of $\chi^2_{min}$ and  
%number of time slices we have. Here is $ \chi^2_{min} $ is the minimal value of $\chi^2$. 
%Up to now, the main difference between current approach and the conventional SAC method based on imaginary time correlation function is the complexity of time $z$, $G(z)$ and the integration kernel $K(z, \omega)$. 

So far, the major difference between the current setup and the standard analytic continuation approach  based on imaginary time correlation functions is the   
complexity of time $z$, correlation function $G(z)$, and the integration kernel $K(z, \omega)$. 
The determination of the optimal fictitious sampling temperature $\alpha$ is a crucial issue in SAC methods.   
It relies on the dependence of the average $ \chi^2 $ on temperature $\alpha$, which is case-dependent.     
The crossover regime where $\chi^2$ gradually saturates to the minimum value of 
chi-squared $ \chi^2_{\text{min}}$ is usually argued to be the optimal choice. 
On the other hand,  
in the case that the time domain is far from the imaginary axis, 
we observe a mild decay of $ \chi^2 $ as $\alpha$ is decreased from $1$.   
Here we test two approaches to evaluating the spectrum.  The first method
is to perform an annealing process sequentially from  
 $\alpha=1$ to low temperature and pick up the  optimal sampling temperature $\alpha=\alpha_c$ as the one where the corresponding $\chi^2$ satisfies, 
\begin{equation}
    \chi^2 = \chi^2_{\text{min}} + \sqrt{ 2 S }
\end{equation}  
 and we collect the averaged spectrum $\alpha_c$. The second approach is to average the sampled spectrum at a temperature list $ \ln (1/ \alpha) = 0, 0.1, 0.2, ... 6.0$ via annealing. 
 Note that this is a weighted average and the weight is proportional to the slope of $\chi^2$-$\alpha$ curve.  Here, the formal approach is a simple generalization of the one used in Ref.~\cite{SHAO20231} to the complex plane, whereas the latter method, which was also frequently used in imaginary time cases, follows Ref.~\cite{beach2004identifying}. 
 
%So we choose to perform a list of stochastic sampling processes  at $  \ln (1/ \alpha) = 0, 0.1, 0.2, ... 3.0$, and average their output of $S(\omega)$ to get the final spectral function.   

%The optimal determination of sampling temperature $\Theta$ has been argued in previous articles.   Following Ref.~\cite{SHAO20231},    
%we prepare the simulation with an annealing procedure such that the minimum value of 
%chi-squared $ \chi^2_{\text{min}}$ is approximately determined.  Then we restart an annealing process at the second step, such that $\Theta$ is lowered gradually until the following condition is satisfied  
%\begin{equation}
%  \chi^2 = \chi^2_{\text{min}} + N_{ d o f } 
%\end{equation}
%where $N_{d o f}$ is the number of independent degrees of freedom.  

\begin{figure}[t!]
    \centering
    \includegraphics[width=\columnwidth]{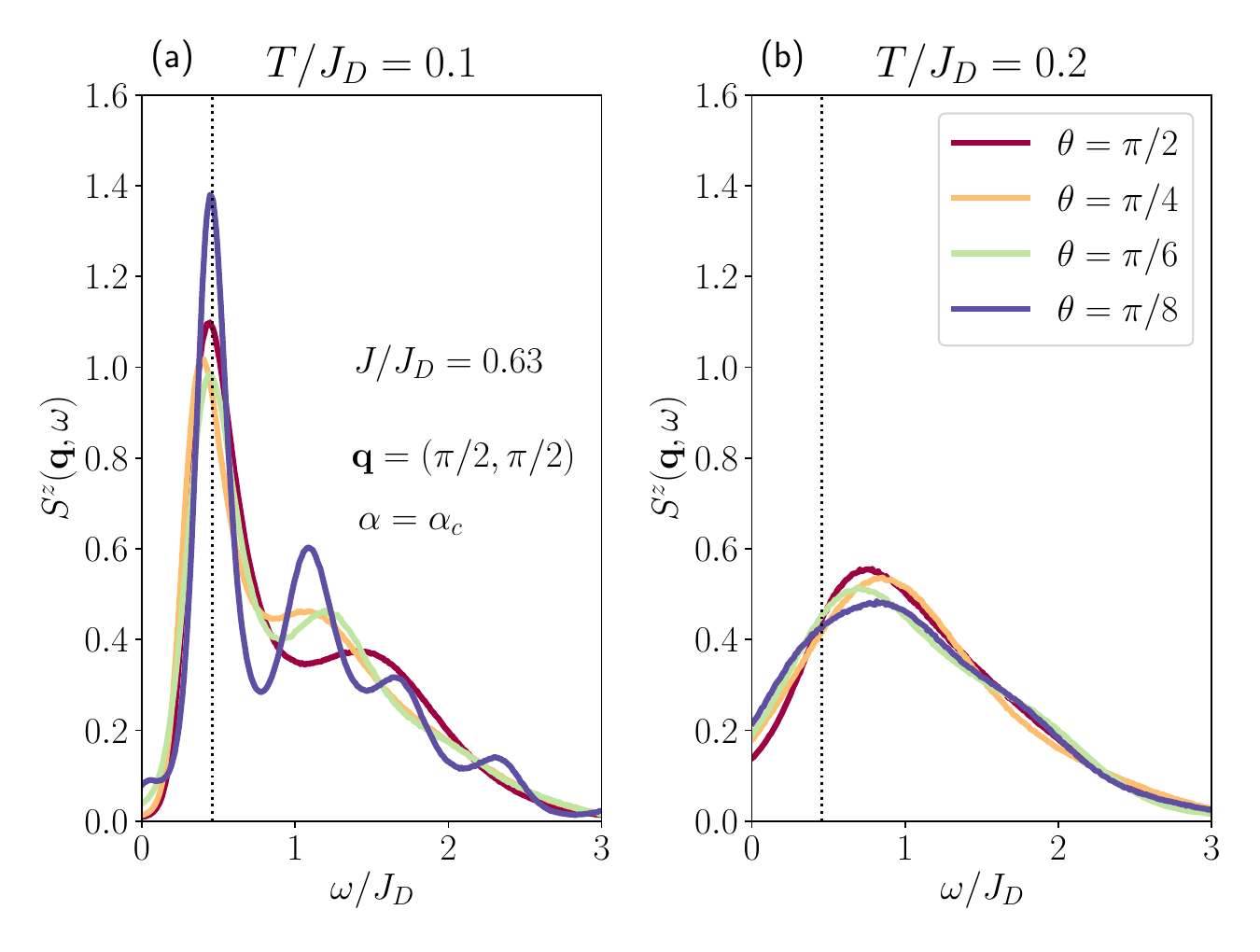}
    \caption{Dynamical structure factors $ {S}(\bm{q}, \omega)$ from stochastic analytic continuation based on    
    analytic correlation functions $ S(\bm{q}, z)$ at several complex angles.   
    Here the spectrum in obtained from sampling at 
    $ \alpha = \alpha_c $. Data is shown for the $16\times 4$ lattice with $J/J_D=0.63$ and $\bm{q}=(\pi/2, \pi/2)$. We compare between temperatures $T/J_D=0.1$ (a) and $T/J_D=0.2$ (b).
    }
    \label{fig:thetas_analytic_alpha_c}
\end{figure}

\begin{figure}[t!]
    \centering
    \includegraphics[width=\columnwidth]{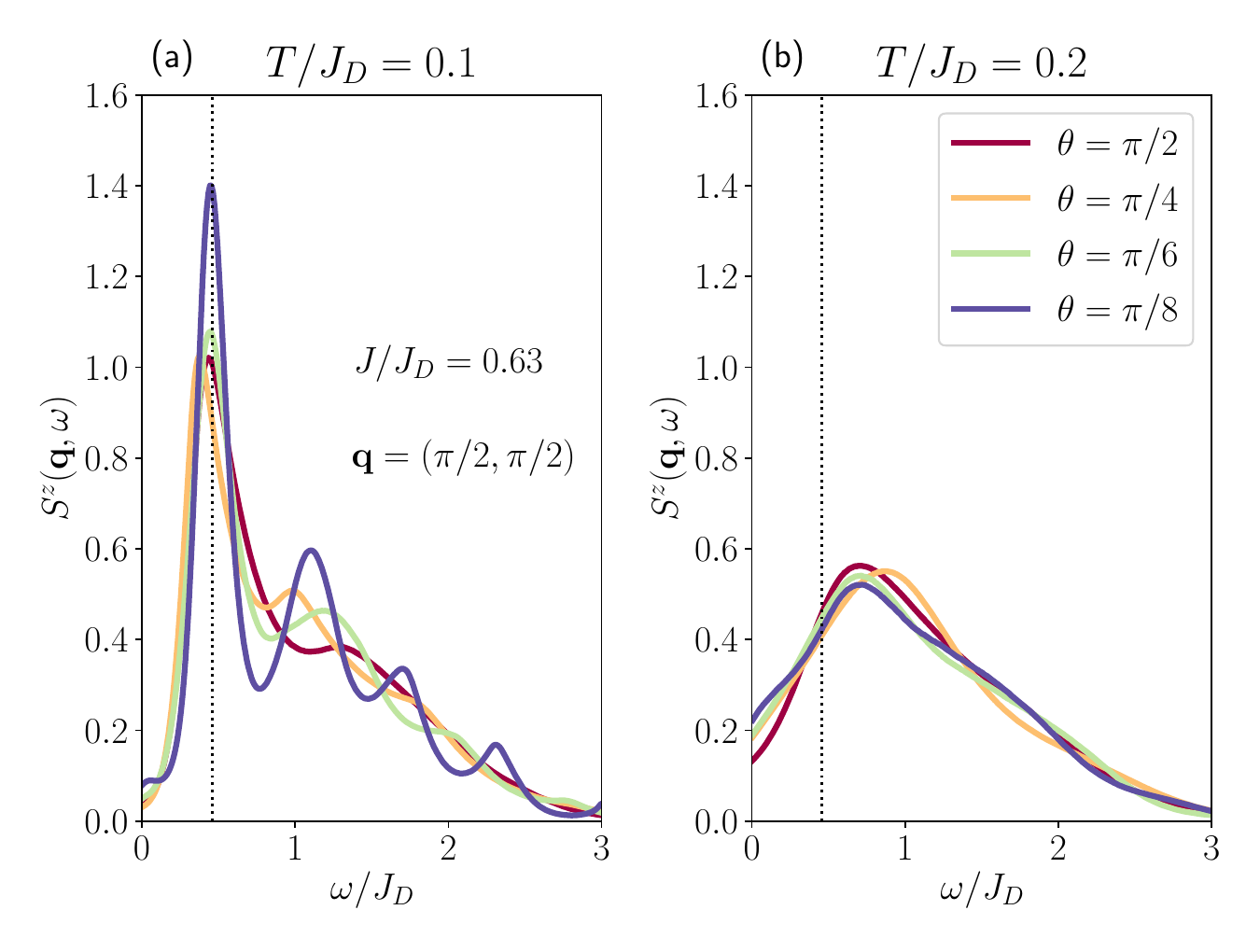}
    \caption{ Same as Fig.~\ref{fig:thetas_analytic_alpha_c},  where 
    the spectrum is obtained via   
    weighted  averaging the collected histogram from $ \log \alpha = 0 $ to $ \log \alpha = -6 $. 
    }
    \label{fig:thetas_analytic}
\end{figure}

We show results of analytic continuation at $T/J_D=0.1$ and $0.2$ for several complex angles in Figs.~\ref{fig:thetas_analytic_alpha_c} and 
\ref{fig:thetas_analytic}.    
First, the spectrum obtained via the sampling at $\alpha_c$ does not differ from the one via weighted averaging the data from high to low temperature. Clearly, the position of the lowest triplon mode at $T/J_D=0.1$ is constant with both perturbation theory and the one from the \textit{Hermitian} correlation function  (at $\theta \longrightarrow 0$ limit) as shown in  Fig.~\ref{fig:thetas}.  Meanwhile, the second triplon peak at a higher frequency, which is not clearly visible for the case of $\theta = \pi/2$, can be resolved better from complex angle calculations.   
%However,  width of  the lowest peak  is broadened dramatically at $\theta=\pi/2$ case, and it shrinks gradually as the complex angle is tuned towards the real time axis. 
On the other hand, we did not observe a significant $\theta$ dependence for the thermal broadened spectrum at $T/J_D=0.2$.  

The optimal angle for calculating the analytic correlation function is a question of balancing the increased computational cost at a smaller complex angle and the higher uncertainty in analytical continuation at a larger one. Certainly, among the numerical cost-accessible regions in the complex time plane, one may push the calculation to the smallest possible complex angle to acquire the most accurate spectral function.

%The only difference between current approach and the conventional SAC method that is well known in QMC community is that time $t$ in both correlation function and the kernel are taken as complex value. Although there is no mathematical prove to show the benefits, one could naively expect a better performance when the angle of time is closer to the real axis.  
%We perform a simple test of complex time SAC based on a non-interacting fermion system.  Small Gaussian noises is added artificially to the exact complex time correlation function to test the SAC approach. 
%Fig.~ shows the dynamical charge structure factor for free fermions on the half filled square lattice at momentum $ \ $ spinless fermion o

\subsection{Hermitian complex-time correlation functions}
\label{sec:hermitiancorr}

\begin{figure}[t!]
    \centering
    \includegraphics[width=\columnwidth]{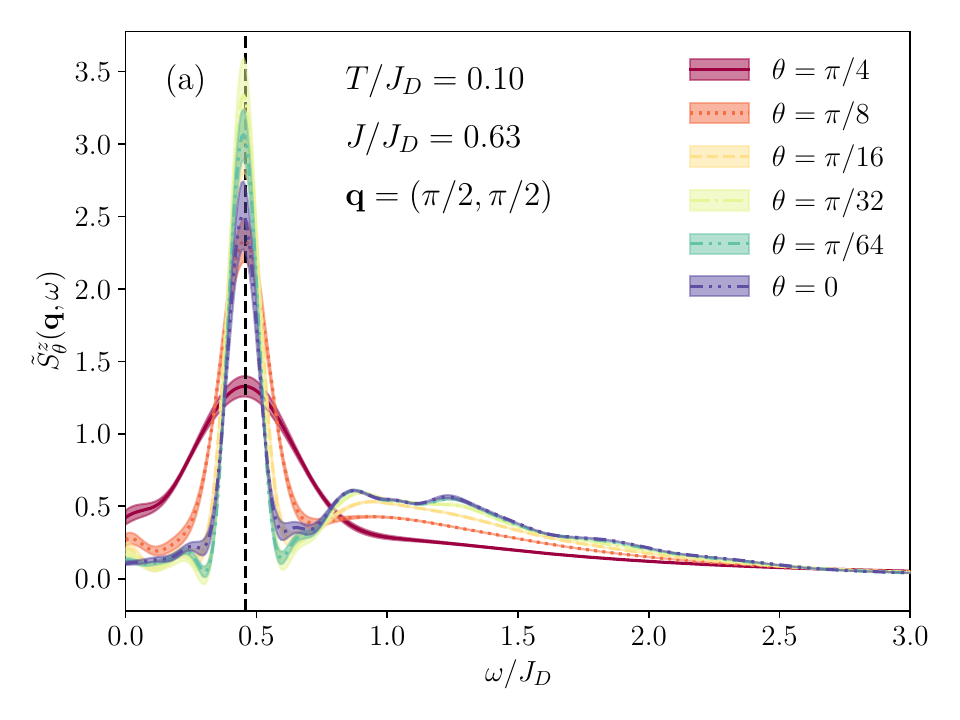}\\
    \includegraphics[width=\columnwidth]{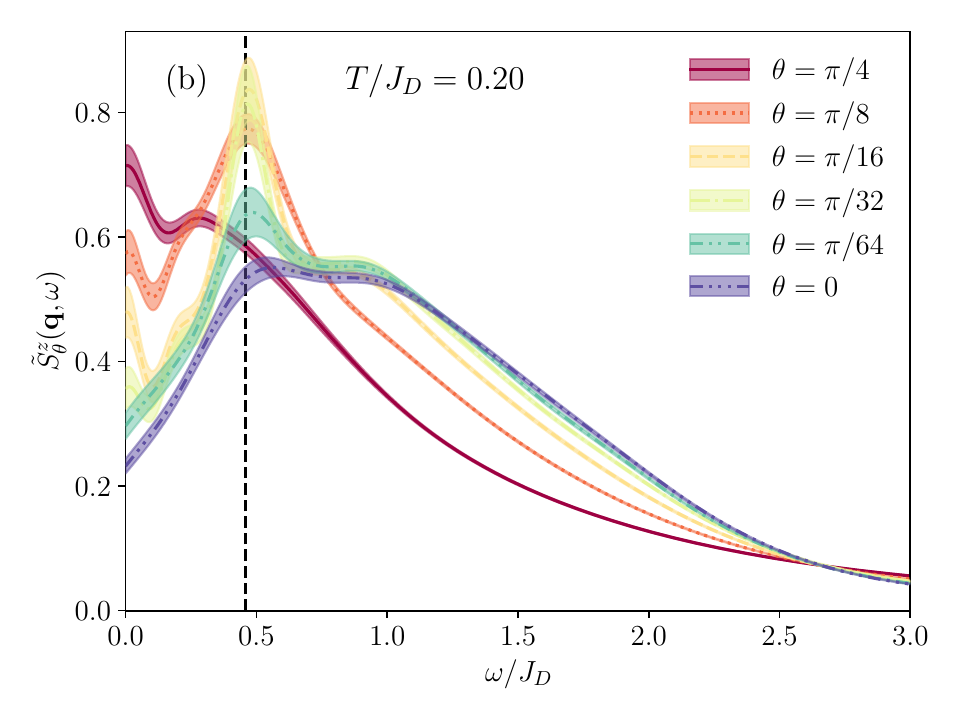}
    \caption{Hermitian dynamical structure factors $\tilde{S}^z_\theta(\bm{q}, \omega)$ from complex-time correlation functions $\tilde{S}^z_\theta(\bm{q}, t)$ at different complex angles $\theta$. Data is shown for the $16\times 4$ lattice with $J/J_D=0.63$ and $\bm{q}=(\pi/2, \pi/2)$. We compare between temperatures $T/J_D=0.1$ (a) and $T/J_D=0.2$ (b). We observe that several key features of the exact spectral function $S(\bm{q}, \omega)$ ($\theta=0$) are already captured well by sizable values of $\theta$. Larger values of $\theta$ significantly reduce the computational cost. Time evolution has been performed up to time $\Omega/J_D=50$ for $R=100$ samples.} 
    \label{fig:thetas}
\end{figure}

We now focus our attention on the second possible generalization of correlation functions to complex time, the Hermitian correlation function as defined in \cref{eq:hermitiancorr}, which expanded in the METTS basis yields,
\begin{align}
\mathcal{H}_{AB}(z) &= \braket{e^{iH\overline{z}} A e^{-iHz} B} \\
&=\frac{1}{\mathcal{Z}} \sum_i p_i  \langle \psi_i |   e^{iH\overline{z}} A e^{-iHz} B | \psi_i \rangle \\
&= \frac{1}{\mathcal{Z}} \sum_i p_i \mathcal{H}_{AB}^i(z),
\end{align}
where we define the sample Hermitian correlation function as,
\begin{equation}
\label{eq:samplehermcorr}
\mathcal{H}^{i}_{AB}(z) 
\equiv \langle \psi_i | e^{iH\overline{z}} A e^{-iHz} B | \psi_i \rangle .
\end{equation}
To evaluate this correlation function we can proceed analogously to the conventional time-dependent correlation function algorithm in \cref{alg:dynmetts}, and define the states,
\begin{align}
&|v_i(z)\rangle = e^{-iHz} B |\psi_i \rangle = e^{-H\tau} e^{-iHt} B |\psi_i \rangle, \\
&|w_i(z)\rangle = e^{iH\overline{z}} |\psi_i \rangle = e^{-H\tau} e^{-iHt} |\psi_i \rangle.
\end{align}
Using these states, we evaluate the sample Hermitian correlation functions as, 
\begin{equation}
\label{eq:samplehermcorr}
\mathcal{H}^{i}_{AB}(z) 
= \langle w_i(z) |  A  | v_i(z) \rangle,
\end{equation}
using \cref{alg:dynmetts} adjusted for the complex time evolution along the contour $z = |z|e^{-i\theta}$ shown in \cref{fig:omega_sketch}. A key difference to the dynamical METTS algorithm \cref{alg:dynmetts} is the fact that the time evolution operators $e^{-iHz}$ and $e^{iH\overline{z}}$ are not norm preserving. However, we could reinforce a normalization by setting, 
\begin{align}
n_v^2 &\equiv \braket{v_i(z) | v_i(z)} = \bra{\psi_i} B^\dagger e^{-2\tau H} B \ket{\psi_i}, \\
n_w^2 &\equiv \braket{w_i(z) | w_i(z)} = \bra{\psi_i} e^{-2 \tau H} 
\ket{\psi_i},\\
c^2_B &\equiv \braket{\psi_i| B^\dagger B |\psi_i}
\end{align}
and introducing the modified states,
\begin{align}
\label{eq:vstatesmodz}
&|\tilde{v}_i(z)\rangle = c_B\;|\tilde{v}_i(z)\rangle / n_v = c_B\; e^{-iHt} e^{-H\tau}  B |\psi_i \rangle / n_v, \\
&|\tilde{w}_i(z)\rangle = |\tilde{w}_i(z)\rangle / n_w = e^{-H\tau} e^{-iHt} e^{-H\tau}  |\psi_i \rangle / n_w.
\end{align}
Finally, we consider the modified sample Hermitian correlation functions,
\begin{equation}
\tilde{\mathcal{H}}^{i}_{AB}(z) 
= \langle \tilde{w}_i(z) |  A  | \tilde{v}_i(z) \rangle,
\end{equation}
giving rise to a modified Hermitian correlation function
\begin{equation}
    \tilde{\mathcal{H}}_{AB}(z) =  \frac{1}{\mathcal{Z}} \sum_i p_i \tilde{\mathcal{H}}^{i}_{AB}(z).
\end{equation}
Our strategy is to directly view $\tilde{\mathcal{H}}_{AB}(z)$ as an approximation of $\mathcal{C}_{AB}(t)$, 
\begin{equation}
    \mathcal{C}_{AB}(t) \approx \tilde{\mathcal{H}}_{AB}(t - it\tan \theta).
\end{equation}
We find that $\tilde{\mathcal{H}}_{AB}(z)$ is a significantly better approximation to $\mathcal{C}_{AB}(t)$ than the bare Hermitian correlator $\mathcal{H}_{AB}(z)$, which motivates the modified definition. 

The negative signs in the real exponentials $e^{-H\tau}$ have significantly favorable consequences on the stability of computing $e^{-H\tau} |\psi\rangle$, where $|\psi\rangle$ denotes an arbitrary wave function. Moreover, evaluating imaginary-time evolutions $e^{-H\tau} |\psi\rangle$ with a negative prefactor is well-suited for MPS methods. Whenever the overlap with the groundstate $\ket{\psi_0}$ of $H$ is nonzero, i.e., $\braket{\psi_0|\psi} \neq 0$, we have 
\begin{equation}
    \lim_{\tau\rightarrow\infty} e^{-H\tau} |\psi\rangle \simeq \ket{\psi_0}.
\end{equation}
Since ground states of local Hamiltonians are subject to bounds on their entanglement, we also expect states of the form $e^{-H\tau} |\psi\rangle$ to have manageable entanglement, whenever 
$|\psi\rangle$ is weakly entangled.

We show Hermitian spectral functions $S^z_\theta(\bm{q}, \omega)$ for the $16\times 4$ Shastry-Sutherland lattice at $J/J_D=0.63$ and temperatures $T/J_D = 0.1, 0.2$ in \cref{fig:thetas}. Remarkably, for $T/J_D=0.1$ the position of the dominant peak at $ \omega /J_D \approx 0.458 $ is always obtained consistently even for sizable values of the complex angle $\theta$.  Meanwhile, the height and width of the dominant peak are well captured as long as $\theta \leq \pi/16$. Moreover, even features like secondary peaks at higher temperatures are captured correctly for not too large values of the complex angle $\theta$. Similarly, the overall shape of the exact spectral function with $\theta=0$ at $T/J_D=0.2$ in Fig.~\ref{fig:thetas}(b) is generally well captured at larger values of $\theta$. However, the peak around the triplon gap for $\omega/J_D \approx 0.458$ is overestimated at finite angles of $\theta$.

\begin{figure}[t]
    \centering
    \includegraphics[width=\columnwidth]{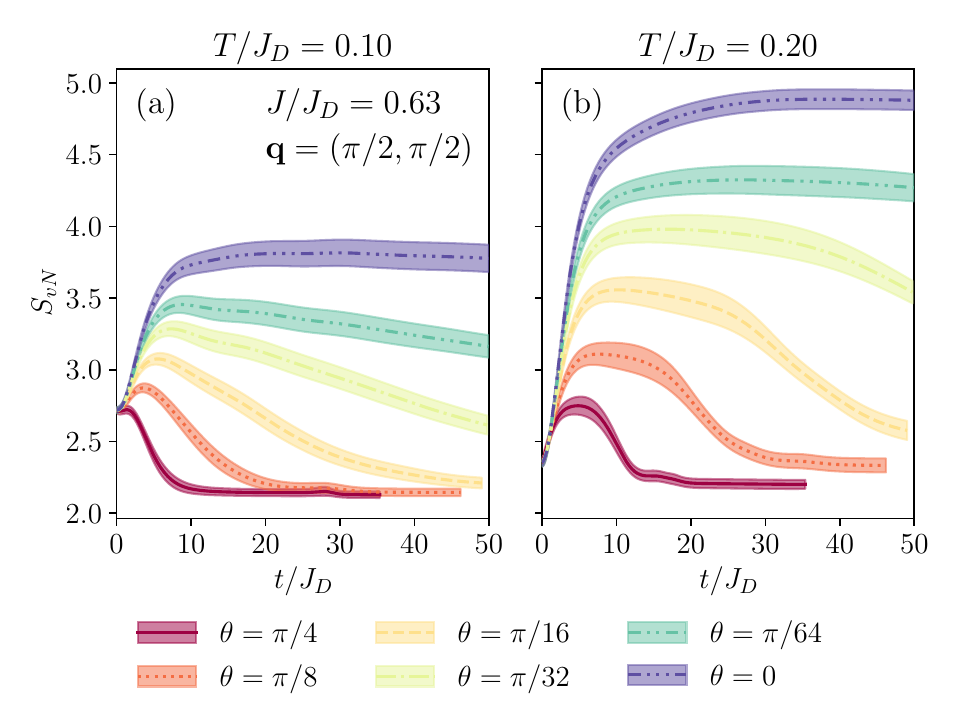}
    \caption{Average bipartite von Neumann entanglement entropy $S_{\text{vN}}$ of the states $\ket{\tilde{v}_i(z)}$ as in \cref{eq:vstatesmodz} for computing the Hermitian correlator \cref{eq:hermitiancorr}. Data on the $16\times 6$ Shastry-Sutherland lattice for different values of the complex angle $\theta$ at $J/J_D=0.63$ and $\bm{q}=(\pi/2, \pi/2)$. We compare results for temperatures $T/J_D=0.1$ (a) and $T/J_D=0.2$ (b). For larger values of $\theta$ we observe convergence of $S_{\text{vN}}$ towards the ground-state entanglement entropy.}
    \label{fig:entanglement_thetas}
\end{figure}

Our main motivation for employing the Hermitian correlation function is to limit the entanglement when time-evolving an MPS state. Thus, we show the average growth of the bipartite von Neumann entropy $S_{\text{vN}}$ of the states $\ket{\tilde{v}_i(z)}$ for different values of $\theta$ on a $16 \times 6$ cylinder in \cref{fig:entanglement_thetas}. We chose momenta $\bm{q}=(\pi/2, \pi/2)$, $ J/J_D =0.63$ and temperatures $T/J_D=0.1, 0.2$. As expected, for $\theta=0$ the entanglement is monotonically growing as a function of time. While the initial entanglement of the METTS state at $T/J_D=0.1$ in (a) is larger than at $T/J_D=0.2$ in (b), the entanglement grows more quickly when going to higher temperatures. As soon as $\theta>0$ we observe an initial growth of the entanglement, followed by a decay caused by the imaginary-time evolution component when applying $e^{-\tau H}$.
We see at nonzero values of $\theta$ that entanglement is quickly suppressed after attaining an intermediate maximum which grows when decreasing $\theta$. We thus conclude that a nonzero complex angle $\theta$ solves the issue of unbounded entanglement growth at the cost of introducing a controlled error when evaluating the spectral functions.

\section{Discussion and Perspectives}\label{sec:outlook}

We introduced a method to compute dynamical spectral functions at non-zero temperatures. Compared to previous algorithms using MPS techniques~\cite{Barthel2009,Karrasch2012,Karrasch2013,Barthel2013,Tiegel2014,Jansen2022}, our method does not rely on purification but employs METTS states. The entanglement of the MPS states used in current computation can be reduced, making simulations of larger cylinder geometries possible. By performing a series of analyses we propose that this method is capable of addressing questions in two-dimensional quantum magnets, exemplified by our study of the anomalous thermal broadening of SrCu$_2($BO$_3)_2$ and the Shastry-Sutherland model~\cite{Wang2024a}.

The real-time evolution of METTS states performed is the computational bottleneck due to the generically unbounded growth of entanglement. To alleviate the difficulties arising from large bond dimensions, we suggest that complex-time correlators can be used to lower the computational cost. The key idea is that a contribution from imaginary-time evolution with a negative exponent will ultimately evolve the METTS state to the ground state, for which entanglement bounds are known. While complex-time evolution had been proposed for QMC~\cite{Krilov2001} or MPS methods at $T=0$ applied to impurity models~\cite{Cao2023,Grundner2023}, we here propose the algorithms to employ this approach at non-zero temperatures using METTS with MPS. We considered two different ways of generalizing time-dependent correlation functions to complex time which we denote by the \textit{analytic}  and \textit{Hermitian} correlation functions. The analytic one is a natural generalization of the spectrum in Matsubara formalism to the complex plane, such that conventional numerical analytic continuation approaches can be easily generalized correspondingly. Most importantly, real-time contribution to the time-displaced correlator improves the quality of analytic continuation. Ref.~\cite{Cao2023} also provided an alternative method for performing the continuation from the complex plane, which could presumably also be applied to the present case. 

On the other hand, the definition of the \textit{Hermitian} correlator has the advantage of easy implementation and the imaginary time evolution dissipates the entanglement entropy dramatically. Meanwhile, the main structure of the real-frequency spectrum is well captured even in the presence of a systemic deviation due to imaginary contributions. Only recently, this version of a complex time correlator has been studied~\cite{Grundner2023} in the context of impurity models with MPS. Moreover, Ref.~\cite{Grundner2023} provides a method of extrapolating correlation functions from the complex plane to the real axis which demonstrably is capable of achieving high accuracy.

It is worth pointing out that the idea of introducing an imaginary-time component to reduce entanglement is similar in spirit to the dissipation-assisted operator evolution proposed in Ref.~\cite{Rakovsky2022}. The imaginary-time component acts as a natural ``dissipator''. One advantage of the present approach is that existing implementations of time-evolution algorithms, such as TDVP, can be readily employed to compute complex time evolution.

The SAC method based on the complex plane correlation function has still leftover possibilities to be improved in the future: one promising suggestion is to perform continuation via combined information of correlators at several angles along the complex plane.  Apart from the stochastic approach, an interesting direction is to combine the Nevanlinna analytic continuation~\cite{Nevanlinna_PRL} approach with the correlation functions obtained at certain METTS states. On the other hand, 
a common technique for computing dynamical spectral functions at zero temperatures by using real-time evolution with MPS, is to perform linear prediction in order to extend the time-series of correlation functions~\cite{White2008,Barthel2009,Wolf2015,Tian2021}. 
While it is reported that this works remarkably well for simple spectra with few poles, we have not been able to successfully apply this to the correlation functions computed at non-zero temperatures in this work. For generic METTS states we mostly find strong disagreement between the actual time-evolution and the linear prediction. This is likely due to the fact that the correlators
$\mathcal{C}_i(t)$ from METTS in general have a more complex structure than a superposition of a few sinusoidal waves. Thus, all the data we present is taken from actual time evolution. Recently, however, a more advanced method for performing linear prediction in the context of spectral functions with increased accuracy has been proposed~\cite{Tian2021}. It would be interesting to find out whether this method would be beneficial in the present context to extend the METTS time series. 

The scaling of the method is determined by the maximal MPS bond dimension $\chi$ which needs to be simulated and the MPO bond dimension $D$ which typically remains finite. To perform both the METTS sampling as well as the time-evolution of METTS states, the proposed method has the same scaling as DMRG, which is $\mathcal{O}(\chi^3 D + \chi^2 D^2)$, since we are using the TDVP time evolution method. The key novelty is, that due to the complex time evolution, the maximal necessary MPS bond dimension $\chi$ needed to resolve the time-evolved state is bounded, as opposed to real-time evolution, where the entanglement in general grows unbounded.

In summary, we demonstrated that the combination of METTS with either real or complex-time evolution can be applied to study spectral functions of two-dimensional quantum many-body systems using MPS. While the most accurate approach using real-time correlation functions is feasible on certain problem sizes, the complex-time correlation functions can be evaluated more efficiently. The reduction of computation cost can be tuned using the complex angle $\theta$, from highly accurate but less efficient at $\theta$ close to zero to highly efficient but less accurate at larger values of $\theta$. We, therefore, propose this method to serve as a bridge between the physics quantum lattice models and spectroscopy experiments performed on strongly correlated materials. 

\begin{acknowledgments}  
Z.W.\ thanks Anders Sandvik and Hui Shao for enlightening discussions on the stochastic analytic continuation method. A.W.\ is grateful to Siddharth Parameswaran, Roderich Moessner, Miles Stoudenmire, and Olivier Parcollet for insightful discussion. A.W.\ acknowledges support by the DFG through the Emmy Noether program (Grant No.\ 509755282). Z.W.\ was supported by the FP7/ERC Consolidator Grant QSIMCORR, No.\ 771891 and  by the Deutsche Forschungsgemeinschaft (DFG, German Research Foundation) under Germany's Excellence Strategy -- EXC-2111 -- 390814868.  
\end{acknowledgments}

\bibliography{main}

\begin{thebibliography}{81}%
\makeatletter
\providecommand \@ifxundefined [1]{%
 \@ifx{#1\undefined}
}%
\providecommand \@ifnum [1]{%
 \ifnum #1\expandafter \@firstoftwo
 \else \expandafter \@secondoftwo
 \fi
}%
\providecommand \@ifx [1]{%
 \ifx #1\expandafter \@firstoftwo
 \else \expandafter \@secondoftwo
 \fi
}%
\providecommand \natexlab [1]{#1}%
\providecommand \enquote  [1]{``#1''}%
\providecommand \bibnamefont  [1]{#1}%
\providecommand \bibfnamefont [1]{#1}%
\providecommand \citenamefont [1]{#1}%
\providecommand \href@noop [0]{\@secondoftwo}%
\providecommand \href [0]{\begingroup \@sanitize@url \@href}%
\providecommand \@href[1]{\@@startlink{#1}\@@href}%
\providecommand \@@href[1]{\endgroup#1\@@endlink}%
\providecommand \@sanitize@url [0]{\catcode `\\12\catcode `\$12\catcode
  `\&12\catcode `\#12\catcode `\^12\catcode `\_12\catcode `\%12\relax}%
\providecommand \@@startlink[1]{}%
\providecommand \@@endlink[0]{}%
\providecommand \url  [0]{\begingroup\@sanitize@url \@url }%
\providecommand \@url [1]{\endgroup\@href {#1}{\urlprefix }}%
\providecommand \urlprefix  [0]{URL }%
\providecommand \Eprint [0]{\href }%
\providecommand \doibase [0]{https://doi.org/}%
\providecommand \selectlanguage [0]{\@gobble}%
\providecommand \bibinfo  [0]{\@secondoftwo}%
\providecommand \bibfield  [0]{\@secondoftwo}%
\providecommand \translation [1]{[#1]}%
\providecommand \BibitemOpen [0]{}%
\providecommand \bibitemStop [0]{}%
\providecommand \bibitemNoStop [0]{.\EOS\space}%
\providecommand \EOS [0]{\spacefactor3000\relax}%
\providecommand \BibitemShut  [1]{\csname bibitem#1\endcsname}%
\let\auto@bib@innerbib\@empty
%</preamble>
\bibitem [{\citenamefont {Gubernatis}\ \emph {et~al.}(2016)\citenamefont
  {Gubernatis}, \citenamefont {Kawashima},\ and\ \citenamefont
  {Werner}}]{Gubernatis2016}%
  \BibitemOpen
  \bibfield  {author} {\bibinfo {author} {\bibfnamefont {J.}~\bibnamefont
  {Gubernatis}}, \bibinfo {author} {\bibfnamefont {N.}~\bibnamefont
  {Kawashima}},\ and\ \bibinfo {author} {\bibfnamefont {P.}~\bibnamefont
  {Werner}},\ }\href {https://doi.org/10.1017/cbo9780511902581} {\emph
  {\bibinfo {title} {Quantum {M}onte {C}arlo Methods: Algorithms for Lattice
  Models}}}\ (\bibinfo  {publisher} {Cambridge University Press},\ \bibinfo
  {year} {2016})\BibitemShut {NoStop}%
\bibitem [{\citenamefont {Loh}\ \emph {et~al.}(1990)\citenamefont {Loh},
  \citenamefont {Gubernatis}, \citenamefont {Scalettar}, \citenamefont {White},
  \citenamefont {Scalapino},\ and\ \citenamefont {Sugar}}]{Loh1990}%
  \BibitemOpen
  \bibfield  {author} {\bibinfo {author} {\bibfnamefont {E.~Y.}\ \bibnamefont
  {Loh}}, \bibinfo {author} {\bibfnamefont {J.~E.}\ \bibnamefont {Gubernatis}},
  \bibinfo {author} {\bibfnamefont {R.~T.}\ \bibnamefont {Scalettar}}, \bibinfo
  {author} {\bibfnamefont {S.~R.}\ \bibnamefont {White}}, \bibinfo {author}
  {\bibfnamefont {D.~J.}\ \bibnamefont {Scalapino}},\ and\ \bibinfo {author}
  {\bibfnamefont {R.~L.}\ \bibnamefont {Sugar}},\ }\bibfield  {title} {\bibinfo
  {title} {Sign problem in the numerical simulation of many-electron systems},\
  }\href {https://doi.org/10.1103/PhysRevB.41.9301} {\bibfield  {journal}
  {\bibinfo  {journal} {Phys. Rev. B}\ }\textbf {\bibinfo {volume} {41}},\
  \bibinfo {pages} {9301} (\bibinfo {year} {1990})}\BibitemShut {NoStop}%
\bibitem [{\citenamefont {Orús}(2019)}]{Orus2019}%
  \BibitemOpen
  \bibfield  {author} {\bibinfo {author} {\bibfnamefont {R.}~\bibnamefont
  {Orús}},\ }\bibfield  {title} {\bibinfo {title} {Tensor networks for complex
  quantum systems},\ }\href {https://doi.org/10.1038/s42254-019-0086-7}
  {\bibfield  {journal} {\bibinfo  {journal} {Nat. Rev. Phys.}\ }\textbf
  {\bibinfo {volume} {1}},\ \bibinfo {pages} {538} (\bibinfo {year}
  {2019})}\BibitemShut {NoStop}%
\bibitem [{\citenamefont {Ba\~{n}uls}(2023)}]{Banuls2023}%
  \BibitemOpen
  \bibfield  {author} {\bibinfo {author} {\bibfnamefont {M.~C.}\ \bibnamefont
  {Ba\~{n}uls}},\ }\bibfield  {title} {\bibinfo {title} {Tensor network
  algorithms: A route map},\ }\href
  {https://doi.org/10.1146/annurev-conmatphys-040721-022705} {\bibfield
  {journal} {\bibinfo  {journal} {Annu. Rev. Condens}\ }\textbf {\bibinfo
  {volume} {14}},\ \bibinfo {pages} {173} (\bibinfo {year} {2023})}\BibitemShut
  {NoStop}%
\bibitem [{\citenamefont {White}(1992)}]{White1992}%
  \BibitemOpen
  \bibfield  {author} {\bibinfo {author} {\bibfnamefont {S.~R.}\ \bibnamefont
  {White}},\ }\bibfield  {title} {\bibinfo {title} {{Density matrix formulation
  for quantum renormalization groups}},\ }\href
  {https://doi.org/10.1103/PhysRevLett.69.2863} {\bibfield  {journal} {\bibinfo
   {journal} {Phys. Rev. Lett.}\ }\textbf {\bibinfo {volume} {69}},\ \bibinfo
  {pages} {2863} (\bibinfo {year} {1992})}\BibitemShut {NoStop}%
\bibitem [{\citenamefont {White}(1993)}]{White1993}%
  \BibitemOpen
  \bibfield  {author} {\bibinfo {author} {\bibfnamefont {S.~R.}\ \bibnamefont
  {White}},\ }\bibfield  {title} {\bibinfo {title} {{Density-matrix algorithms
  for quantum renormalization groups}},\ }\href
  {https://doi.org/10.1103/PhysRevB.48.10345} {\bibfield  {journal} {\bibinfo
  {journal} {Phys. Rev. B}\ }\textbf {\bibinfo {volume} {48}},\ \bibinfo
  {pages} {10345} (\bibinfo {year} {1993})}\BibitemShut {NoStop}%
\bibitem [{\citenamefont {Schollw\"ock}(2005)}]{Schollwoeck2005}%
  \BibitemOpen
  \bibfield  {author} {\bibinfo {author} {\bibfnamefont {U.}~\bibnamefont
  {Schollw\"ock}},\ }\bibfield  {title} {\bibinfo {title} {The density-matrix
  renormalization group},\ }\href {https://doi.org/10.1103/RevModPhys.77.259}
  {\bibfield  {journal} {\bibinfo  {journal} {Rev. Mod. Phys.}\ }\textbf
  {\bibinfo {volume} {77}},\ \bibinfo {pages} {259} (\bibinfo {year}
  {2005})}\BibitemShut {NoStop}%
\bibitem [{\citenamefont {Schollwöck}(2011)}]{Schollwoeck2011}%
  \BibitemOpen
  \bibfield  {author} {\bibinfo {author} {\bibfnamefont {U.}~\bibnamefont
  {Schollwöck}},\ }\bibfield  {title} {\bibinfo {title} {The density-matrix
  renormalization group in the age of matrix product states},\ }\href
  {https://doi.org/https://doi.org/10.1016/j.aop.2010.09.012} {\bibfield
  {journal} {\bibinfo  {journal} {Ann. Phys. (N.Y.)}\ }\textbf {\bibinfo
  {volume} {326}},\ \bibinfo {pages} {96} (\bibinfo {year} {2011})}\BibitemShut
  {NoStop}%
\bibitem [{\citenamefont {Wang}\ and\ \citenamefont
  {Xiang}(1997)}]{PhysRevB.56.5061}%
  \BibitemOpen
  \bibfield  {author} {\bibinfo {author} {\bibfnamefont {X.}~\bibnamefont
  {Wang}}\ and\ \bibinfo {author} {\bibfnamefont {T.}~\bibnamefont {Xiang}},\
  }\bibfield  {title} {\bibinfo {title} {Transfer-matrix density-matrix
  renormalization-group theory for thermodynamics of one-dimensional quantum
  systems},\ }\href {https://doi.org/10.1103/PhysRevB.56.5061} {\bibfield
  {journal} {\bibinfo  {journal} {Phys. Rev. B}\ }\textbf {\bibinfo {volume}
  {56}},\ \bibinfo {pages} {5061} (\bibinfo {year} {1997})}\BibitemShut
  {NoStop}%
\bibitem [{\citenamefont {Ammon}\ \emph {et~al.}(1999)\citenamefont {Ammon},
  \citenamefont {Troyer}, \citenamefont {Rice},\ and\ \citenamefont
  {Shibata}}]{PhysRevLett.82.3855}%
  \BibitemOpen
  \bibfield  {author} {\bibinfo {author} {\bibfnamefont {B.}~\bibnamefont
  {Ammon}}, \bibinfo {author} {\bibfnamefont {M.}~\bibnamefont {Troyer}},
  \bibinfo {author} {\bibfnamefont {T.~M.}\ \bibnamefont {Rice}},\ and\
  \bibinfo {author} {\bibfnamefont {N.}~\bibnamefont {Shibata}},\ }\bibfield
  {title} {\bibinfo {title} {Thermodynamics of the {$t$-$J$} ladder: A stable
  finite-temperature density matrix renormalization group calculation},\ }\href
  {https://doi.org/10.1103/PhysRevLett.82.3855} {\bibfield  {journal} {\bibinfo
   {journal} {Phys. Rev. Lett.}\ }\textbf {\bibinfo {volume} {82}},\ \bibinfo
  {pages} {3855} (\bibinfo {year} {1999})}\BibitemShut {NoStop}%
\bibitem [{\citenamefont {Feiguin}\ and\ \citenamefont
  {White}(2005)}]{Feiguin2005}%
  \BibitemOpen
  \bibfield  {author} {\bibinfo {author} {\bibfnamefont {A.~E.}\ \bibnamefont
  {Feiguin}}\ and\ \bibinfo {author} {\bibfnamefont {S.~R.}\ \bibnamefont
  {White}},\ }\bibfield  {title} {\bibinfo {title} {Finite-temperature density
  matrix renormalization using an enlarged {H}ilbert space},\ }\href
  {https://doi.org/10.1103/PhysRevB.72.220401} {\bibfield  {journal} {\bibinfo
  {journal} {Phys. Rev. B}\ }\textbf {\bibinfo {volume} {72}},\ \bibinfo
  {pages} {220401} (\bibinfo {year} {2005})}\BibitemShut {NoStop}%
\bibitem [{\citenamefont {White}(2009)}]{White2009}%
  \BibitemOpen
  \bibfield  {author} {\bibinfo {author} {\bibfnamefont {S.~R.}\ \bibnamefont
  {White}},\ }\bibfield  {title} {\bibinfo {title} {{Minimally Entangled
  Typical Quantum States at Finite Temperature}},\ }\href
  {https://doi.org/10.1103/PhysRevLett.102.190601} {\bibfield  {journal}
  {\bibinfo  {journal} {Phys. Rev. Lett.}\ }\textbf {\bibinfo {volume} {102}},\
  \bibinfo {pages} {190601} (\bibinfo {year} {2009})}\BibitemShut {NoStop}%
\bibitem [{\citenamefont {Stoudenmire}\ and\ \citenamefont
  {White}(2010)}]{Stoudenmire2010}%
  \BibitemOpen
  \bibfield  {author} {\bibinfo {author} {\bibfnamefont {E.~M.}\ \bibnamefont
  {Stoudenmire}}\ and\ \bibinfo {author} {\bibfnamefont {S.~R.}\ \bibnamefont
  {White}},\ }\bibfield  {title} {\bibinfo {title} {{Minimally entangled
  typical thermal state algorithms}},\ }\href
  {https://doi.org/10.1088/1367-2630/12/5/055026} {\bibfield  {journal}
  {\bibinfo  {journal} {New J. Phys.}\ }\textbf {\bibinfo {volume} {12}},\
  \bibinfo {pages} {055026} (\bibinfo {year} {2010})}\BibitemShut {NoStop}%
\bibitem [{\citenamefont {Wietek}\ \emph
  {et~al.}(2021{\natexlab{a}})\citenamefont {Wietek}, \citenamefont {He},
  \citenamefont {White}, \citenamefont {Georges},\ and\ \citenamefont
  {Stoudenmire}}]{Wietek2021}%
  \BibitemOpen
  \bibfield  {author} {\bibinfo {author} {\bibfnamefont {A.}~\bibnamefont
  {Wietek}}, \bibinfo {author} {\bibfnamefont {Y.-Y.}\ \bibnamefont {He}},
  \bibinfo {author} {\bibfnamefont {S.~R.}\ \bibnamefont {White}}, \bibinfo
  {author} {\bibfnamefont {A.}~\bibnamefont {Georges}},\ and\ \bibinfo {author}
  {\bibfnamefont {E.~M.}\ \bibnamefont {Stoudenmire}},\ }\bibfield  {title}
  {\bibinfo {title} {Stripes, antiferromagnetism, and the pseudogap in the
  doped {H}ubbard model at finite temperature},\ }\href
  {https://doi.org/10.1103/PhysRevX.11.031007} {\bibfield  {journal} {\bibinfo
  {journal} {Phys. Rev. X}\ }\textbf {\bibinfo {volume} {11}},\ \bibinfo
  {pages} {031007} (\bibinfo {year} {2021}{\natexlab{a}})}\BibitemShut
  {NoStop}%
\bibitem [{\citenamefont {Wietek}\ \emph
  {et~al.}(2021{\natexlab{b}})\citenamefont {Wietek}, \citenamefont {Rossi},
  \citenamefont {\ifmmode~\check{S}\else \v{S}\fi{}imkovic}, \citenamefont
  {Klett}, \citenamefont {Hansmann}, \citenamefont {Ferrero}, \citenamefont
  {Stoudenmire}, \citenamefont {Sch\"afer},\ and\ \citenamefont
  {Georges}}]{Wietek2021b}%
  \BibitemOpen
  \bibfield  {author} {\bibinfo {author} {\bibfnamefont {A.}~\bibnamefont
  {Wietek}}, \bibinfo {author} {\bibfnamefont {R.}~\bibnamefont {Rossi}},
  \bibinfo {author} {\bibfnamefont {F.}~\bibnamefont {\ifmmode~\check{S}\else
  \v{S}\fi{}imkovic}}, \bibinfo {author} {\bibfnamefont {M.}~\bibnamefont
  {Klett}}, \bibinfo {author} {\bibfnamefont {P.}~\bibnamefont {Hansmann}},
  \bibinfo {author} {\bibfnamefont {M.}~\bibnamefont {Ferrero}}, \bibinfo
  {author} {\bibfnamefont {E.~M.}\ \bibnamefont {Stoudenmire}}, \bibinfo
  {author} {\bibfnamefont {T.}~\bibnamefont {Sch\"afer}},\ and\ \bibinfo
  {author} {\bibfnamefont {A.}~\bibnamefont {Georges}},\ }\bibfield  {title}
  {\bibinfo {title} {{M}ott insulating states with competing orders in the
  triangular lattice {H}ubbard model},\ }\href
  {https://doi.org/10.1103/PhysRevX.11.041013} {\bibfield  {journal} {\bibinfo
  {journal} {Phys. Rev. X}\ }\textbf {\bibinfo {volume} {11}},\ \bibinfo
  {pages} {041013} (\bibinfo {year} {2021}{\natexlab{b}})}\BibitemShut
  {NoStop}%
\bibitem [{\citenamefont {Feng}\ \emph {et~al.}(2022)\citenamefont {Feng},
  \citenamefont {Wietek}, \citenamefont {Stoudenmire},\ and\ \citenamefont
  {Singh}}]{Feng2022}%
  \BibitemOpen
  \bibfield  {author} {\bibinfo {author} {\bibfnamefont {C.}~\bibnamefont
  {Feng}}, \bibinfo {author} {\bibfnamefont {A.}~\bibnamefont {Wietek}},
  \bibinfo {author} {\bibfnamefont {E.~M.}\ \bibnamefont {Stoudenmire}},\ and\
  \bibinfo {author} {\bibfnamefont {R.~R.~P.}\ \bibnamefont {Singh}},\
  }\bibfield  {title} {\bibinfo {title} {Order, disorder, and monopole
  confinement in the spin-$\frac{1}{2}$ {XXZ} model on a pyrochlore tube},\
  }\href {https://doi.org/10.1103/PhysRevB.106.075135} {\bibfield  {journal}
  {\bibinfo  {journal} {Phys. Rev. B}\ }\textbf {\bibinfo {volume} {106}},\
  \bibinfo {pages} {075135} (\bibinfo {year} {2022})}\BibitemShut {NoStop}%
\bibitem [{\citenamefont {Feng}\ \emph {et~al.}(2023)\citenamefont {Feng},
  \citenamefont {Stoudenmire},\ and\ \citenamefont {Wietek}}]{Feng2023}%
  \BibitemOpen
  \bibfield  {author} {\bibinfo {author} {\bibfnamefont {C.}~\bibnamefont
  {Feng}}, \bibinfo {author} {\bibfnamefont {E.~M.}\ \bibnamefont
  {Stoudenmire}},\ and\ \bibinfo {author} {\bibfnamefont {A.}~\bibnamefont
  {Wietek}},\ }\bibfield  {title} {\bibinfo {title} {{B}ose-{E}instein
  condensation in honeycomb dimer magnets and
  {${\mathrm{Yb}}_{2}{\mathrm{Si}}_{2}{\mathrm{O}}_{7}$}},\ }\href
  {https://doi.org/10.1103/PhysRevB.107.205150} {\bibfield  {journal} {\bibinfo
   {journal} {Phys. Rev. B}\ }\textbf {\bibinfo {volume} {107}},\ \bibinfo
  {pages} {205150} (\bibinfo {year} {2023})}\BibitemShut {NoStop}%
\bibitem [{\citenamefont {Nocera}\ and\ \citenamefont
  {Alvarez}(2016{\natexlab{a}})}]{Nocera2016}%
  \BibitemOpen
  \bibfield  {author} {\bibinfo {author} {\bibfnamefont {A.}~\bibnamefont
  {Nocera}}\ and\ \bibinfo {author} {\bibfnamefont {G.}~\bibnamefont
  {Alvarez}},\ }\bibfield  {title} {\bibinfo {title} {Symmetry-conserving
  purification of quantum states within the density matrix renormalization
  group},\ }\href {https://doi.org/10.1103/PhysRevB.93.045137} {\bibfield
  {journal} {\bibinfo  {journal} {Phys. Rev. B}\ }\textbf {\bibinfo {volume}
  {93}},\ \bibinfo {pages} {045137} (\bibinfo {year}
  {2016}{\natexlab{a}})}\BibitemShut {NoStop}%
\bibitem [{\citenamefont {Hallberg}(1995)}]{PhysRevB.52.R9827}%
  \BibitemOpen
  \bibfield  {author} {\bibinfo {author} {\bibfnamefont {K.~A.}\ \bibnamefont
  {Hallberg}},\ }\bibfield  {title} {\bibinfo {title} {Density-matrix algorithm
  for the calculation of dynamical properties of low-dimensional systems},\
  }\href {https://doi.org/10.1103/PhysRevB.52.R9827} {\bibfield  {journal}
  {\bibinfo  {journal} {Phys. Rev. B}\ }\textbf {\bibinfo {volume} {52}},\
  \bibinfo {pages} {R9827} (\bibinfo {year} {1995})}\BibitemShut {NoStop}%
\bibitem [{\citenamefont {K\"uhner}\ and\ \citenamefont
  {White}(1999)}]{Kuehner1999}%
  \BibitemOpen
  \bibfield  {author} {\bibinfo {author} {\bibfnamefont {T.~D.}\ \bibnamefont
  {K\"uhner}}\ and\ \bibinfo {author} {\bibfnamefont {S.~R.}\ \bibnamefont
  {White}},\ }\bibfield  {title} {\bibinfo {title} {Dynamical correlation
  functions using the density matrix renormalization group},\ }\href
  {https://doi.org/10.1103/PhysRevB.60.335} {\bibfield  {journal} {\bibinfo
  {journal} {Phys. Rev. B}\ }\textbf {\bibinfo {volume} {60}},\ \bibinfo
  {pages} {335} (\bibinfo {year} {1999})}\BibitemShut {NoStop}%
\bibitem [{\citenamefont {White}\ and\ \citenamefont
  {Feiguin}(2004)}]{PhysRevLett.93.076401}%
  \BibitemOpen
  \bibfield  {author} {\bibinfo {author} {\bibfnamefont {S.~R.}\ \bibnamefont
  {White}}\ and\ \bibinfo {author} {\bibfnamefont {A.~E.}\ \bibnamefont
  {Feiguin}},\ }\bibfield  {title} {\bibinfo {title} {Real-time evolution using
  the density matrix renormalization group},\ }\href
  {https://doi.org/10.1103/PhysRevLett.93.076401} {\bibfield  {journal}
  {\bibinfo  {journal} {Phys. Rev. Lett.}\ }\textbf {\bibinfo {volume} {93}},\
  \bibinfo {pages} {076401} (\bibinfo {year} {2004})}\BibitemShut {NoStop}%
\bibitem [{\citenamefont {Dargel}\ \emph {et~al.}(2011)\citenamefont {Dargel},
  \citenamefont {Honecker}, \citenamefont {Peters}, \citenamefont {Noack},\
  and\ \citenamefont {Pruschke}}]{PhysRevB.83.161104}%
  \BibitemOpen
  \bibfield  {author} {\bibinfo {author} {\bibfnamefont {P.~E.}\ \bibnamefont
  {Dargel}}, \bibinfo {author} {\bibfnamefont {A.}~\bibnamefont {Honecker}},
  \bibinfo {author} {\bibfnamefont {R.}~\bibnamefont {Peters}}, \bibinfo
  {author} {\bibfnamefont {R.~M.}\ \bibnamefont {Noack}},\ and\ \bibinfo
  {author} {\bibfnamefont {T.}~\bibnamefont {Pruschke}},\ }\bibfield  {title}
  {\bibinfo {title} {Adaptive {L}anczos-vector method for dynamic properties
  within the density matrix renormalization group},\ }\href
  {https://doi.org/10.1103/PhysRevB.83.161104} {\bibfield  {journal} {\bibinfo
  {journal} {Phys. Rev. B}\ }\textbf {\bibinfo {volume} {83}},\ \bibinfo
  {pages} {161104} (\bibinfo {year} {2011})}\BibitemShut {NoStop}%
\bibitem [{\citenamefont {Holzner}\ \emph {et~al.}(2011)\citenamefont
  {Holzner}, \citenamefont {Weichselbaum}, \citenamefont {McCulloch},
  \citenamefont {Schollw\"ock},\ and\ \citenamefont {von
  Delft}}]{PhysRevB.83.195115}%
  \BibitemOpen
  \bibfield  {author} {\bibinfo {author} {\bibfnamefont {A.}~\bibnamefont
  {Holzner}}, \bibinfo {author} {\bibfnamefont {A.}~\bibnamefont
  {Weichselbaum}}, \bibinfo {author} {\bibfnamefont {I.~P.}\ \bibnamefont
  {McCulloch}}, \bibinfo {author} {\bibfnamefont {U.}~\bibnamefont
  {Schollw\"ock}},\ and\ \bibinfo {author} {\bibfnamefont {J.}~\bibnamefont
  {von Delft}},\ }\bibfield  {title} {\bibinfo {title} {Chebyshev matrix
  product state approach for spectral functions},\ }\href
  {https://doi.org/10.1103/PhysRevB.83.195115} {\bibfield  {journal} {\bibinfo
  {journal} {Phys. Rev. B}\ }\textbf {\bibinfo {volume} {83}},\ \bibinfo
  {pages} {195115} (\bibinfo {year} {2011})}\BibitemShut {NoStop}%
\bibitem [{\citenamefont {Dargel}\ \emph {et~al.}(2012)\citenamefont {Dargel},
  \citenamefont {W\"ollert}, \citenamefont {Honecker}, \citenamefont
  {McCulloch}, \citenamefont {Schollw\"ock},\ and\ \citenamefont
  {Pruschke}}]{PhysRevB.85.205119}%
  \BibitemOpen
  \bibfield  {author} {\bibinfo {author} {\bibfnamefont {P.~E.}\ \bibnamefont
  {Dargel}}, \bibinfo {author} {\bibfnamefont {A.}~\bibnamefont {W\"ollert}},
  \bibinfo {author} {\bibfnamefont {A.}~\bibnamefont {Honecker}}, \bibinfo
  {author} {\bibfnamefont {I.~P.}\ \bibnamefont {McCulloch}}, \bibinfo {author}
  {\bibfnamefont {U.}~\bibnamefont {Schollw\"ock}},\ and\ \bibinfo {author}
  {\bibfnamefont {T.}~\bibnamefont {Pruschke}},\ }\bibfield  {title} {\bibinfo
  {title} {Lanczos algorithm with matrix product states for dynamical
  correlation functions},\ }\href {https://doi.org/10.1103/PhysRevB.85.205119}
  {\bibfield  {journal} {\bibinfo  {journal} {Phys. Rev. B}\ }\textbf {\bibinfo
  {volume} {85}},\ \bibinfo {pages} {205119} (\bibinfo {year}
  {2012})}\BibitemShut {NoStop}%
\bibitem [{\citenamefont {Nocera}\ and\ \citenamefont
  {Alvarez}(2016{\natexlab{b}})}]{PhysRevE.94.053308}%
  \BibitemOpen
  \bibfield  {author} {\bibinfo {author} {\bibfnamefont {A.}~\bibnamefont
  {Nocera}}\ and\ \bibinfo {author} {\bibfnamefont {G.}~\bibnamefont
  {Alvarez}},\ }\bibfield  {title} {\bibinfo {title} {Spectral functions with
  the density matrix renormalization group: {K}rylov-space approach for
  correction vectors},\ }\href {https://doi.org/10.1103/PhysRevE.94.053308}
  {\bibfield  {journal} {\bibinfo  {journal} {Phys. Rev. E}\ }\textbf {\bibinfo
  {volume} {94}},\ \bibinfo {pages} {053308} (\bibinfo {year}
  {2016}{\natexlab{b}})}\BibitemShut {NoStop}%
\bibitem [{\citenamefont {Gohlke}\ \emph {et~al.}(2017)\citenamefont {Gohlke},
  \citenamefont {Verresen}, \citenamefont {Moessner},\ and\ \citenamefont
  {Pollmann}}]{Gohlke2017}%
  \BibitemOpen
  \bibfield  {author} {\bibinfo {author} {\bibfnamefont {M.}~\bibnamefont
  {Gohlke}}, \bibinfo {author} {\bibfnamefont {R.}~\bibnamefont {Verresen}},
  \bibinfo {author} {\bibfnamefont {R.}~\bibnamefont {Moessner}},\ and\
  \bibinfo {author} {\bibfnamefont {F.}~\bibnamefont {Pollmann}},\ }\bibfield
  {title} {\bibinfo {title} {{Dynamics of the {K}itaev-{H}eisenberg Model}},\
  }\href {https://doi.org/10.1103/PhysRevLett.119.157203} {\bibfield  {journal}
  {\bibinfo  {journal} {Phys. Rev. Lett.}\ }\textbf {\bibinfo {volume} {119}},\
  \bibinfo {pages} {157203} (\bibinfo {year} {2017})}\BibitemShut {NoStop}%
\bibitem [{\citenamefont {Verresen}\ \emph {et~al.}(2019)\citenamefont
  {Verresen}, \citenamefont {Moessner},\ and\ \citenamefont
  {Pollmann}}]{Verresen2019}%
  \BibitemOpen
  \bibfield  {author} {\bibinfo {author} {\bibfnamefont {R.}~\bibnamefont
  {Verresen}}, \bibinfo {author} {\bibfnamefont {R.}~\bibnamefont {Moessner}},\
  and\ \bibinfo {author} {\bibfnamefont {F.}~\bibnamefont {Pollmann}},\
  }\bibfield  {title} {\bibinfo {title} {Avoided quasiparticle decay from
  strong quantum interactions},\ }\href
  {https://doi.org/10.1038/s41567-019-0535-3} {\bibfield  {journal} {\bibinfo
  {journal} {Nat. Phys.}\ }\textbf {\bibinfo {volume} {15}},\ \bibinfo {pages}
  {750} (\bibinfo {year} {2019})}\BibitemShut {NoStop}%
\bibitem [{\citenamefont {Drescher}\ \emph {et~al.}(2023)\citenamefont
  {Drescher}, \citenamefont {Vanderstraeten}, \citenamefont {Moessner},\ and\
  \citenamefont {Pollmann}}]{Drescher2022}%
  \BibitemOpen
  \bibfield  {author} {\bibinfo {author} {\bibfnamefont {M.}~\bibnamefont
  {Drescher}}, \bibinfo {author} {\bibfnamefont {L.}~\bibnamefont
  {Vanderstraeten}}, \bibinfo {author} {\bibfnamefont {R.}~\bibnamefont
  {Moessner}},\ and\ \bibinfo {author} {\bibfnamefont {F.}~\bibnamefont
  {Pollmann}},\ }\bibfield  {title} {\bibinfo {title} {Dynamical signatures of
  symmetry-broken and liquid phases in an ${S} = \frac{1}{2}$ {H}eisenberg
  antiferromagnet on the triangular lattice},\ }\href
  {https://doi.org/10.1103/PhysRevB.108.L220401} {\bibfield  {journal}
  {\bibinfo  {journal} {Phys. Rev. B}\ }\textbf {\bibinfo {volume} {108}},\
  \bibinfo {pages} {L220401} (\bibinfo {year} {2023})}\BibitemShut {NoStop}%
\bibitem [{\citenamefont {Vidal}(2004)}]{PhysRevLett.93.040502}%
  \BibitemOpen
  \bibfield  {author} {\bibinfo {author} {\bibfnamefont {G.}~\bibnamefont
  {Vidal}},\ }\bibfield  {title} {\bibinfo {title} {Efficient simulation of
  one-dimensional quantum many-body systems},\ }\href
  {https://doi.org/10.1103/PhysRevLett.93.040502} {\bibfield  {journal}
  {\bibinfo  {journal} {Phys. Rev. Lett.}\ }\textbf {\bibinfo {volume} {93}},\
  \bibinfo {pages} {040502} (\bibinfo {year} {2004})}\BibitemShut {NoStop}%
\bibitem [{\citenamefont {Daley}\ \emph {et~al.}(2004)\citenamefont {Daley},
  \citenamefont {Kollath}, \citenamefont {Schollwöck},\ and\ \citenamefont
  {Vidal}}]{Daley_2004}%
  \BibitemOpen
  \bibfield  {author} {\bibinfo {author} {\bibfnamefont {A.~J.}\ \bibnamefont
  {Daley}}, \bibinfo {author} {\bibfnamefont {C.}~\bibnamefont {Kollath}},
  \bibinfo {author} {\bibfnamefont {U.}~\bibnamefont {Schollwöck}},\ and\
  \bibinfo {author} {\bibfnamefont {G.}~\bibnamefont {Vidal}},\ }\bibfield
  {title} {\bibinfo {title} {Time-dependent density-matrix
  renormalization-group using adaptive effective {H}ilbert spaces},\ }\href
  {https://doi.org/10.1088/1742-5468/2004/04/P04005} {\bibfield  {journal}
  {\bibinfo  {journal} {J. Stat. Mech.: Theor. Exp.}\ }\textbf {\bibinfo
  {volume} {2004}},\ \bibinfo {pages} {P04005} (\bibinfo {year}
  {2004})}\BibitemShut {NoStop}%
\bibitem [{\citenamefont {Schmitteckert}(2004)}]{PhysRevB.70.121302}%
  \BibitemOpen
  \bibfield  {author} {\bibinfo {author} {\bibfnamefont {P.}~\bibnamefont
  {Schmitteckert}},\ }\bibfield  {title} {\bibinfo {title} {Nonequilibrium
  electron transport using the density matrix renormalization group method},\
  }\href {https://doi.org/10.1103/PhysRevB.70.121302} {\bibfield  {journal}
  {\bibinfo  {journal} {Phys. Rev. B}\ }\textbf {\bibinfo {volume} {70}},\
  \bibinfo {pages} {121302} (\bibinfo {year} {2004})}\BibitemShut {NoStop}%
\bibitem [{\citenamefont {Kennes}\ \emph {et~al.}(2014)\citenamefont {Kennes},
  \citenamefont {Meden},\ and\ \citenamefont {Vasseur}}]{Kennes2014}%
  \BibitemOpen
  \bibfield  {author} {\bibinfo {author} {\bibfnamefont {D.~M.}\ \bibnamefont
  {Kennes}}, \bibinfo {author} {\bibfnamefont {V.}~\bibnamefont {Meden}},\ and\
  \bibinfo {author} {\bibfnamefont {R.}~\bibnamefont {Vasseur}},\ }\bibfield
  {title} {\bibinfo {title} {Universal quench dynamics of interacting quantum
  impurity systems},\ }\href {https://doi.org/10.1103/PhysRevB.90.115101}
  {\bibfield  {journal} {\bibinfo  {journal} {Phys. Rev. B}\ }\textbf {\bibinfo
  {volume} {90}},\ \bibinfo {pages} {115101} (\bibinfo {year}
  {2014})}\BibitemShut {NoStop}%
\bibitem [{\citenamefont {Essler}\ \emph {et~al.}(2014)\citenamefont {Essler},
  \citenamefont {Kehrein}, \citenamefont {Manmana},\ and\ \citenamefont
  {Robinson}}]{Essler2014}%
  \BibitemOpen
  \bibfield  {author} {\bibinfo {author} {\bibfnamefont {F.~H.~L.}\
  \bibnamefont {Essler}}, \bibinfo {author} {\bibfnamefont {S.}~\bibnamefont
  {Kehrein}}, \bibinfo {author} {\bibfnamefont {S.~R.}\ \bibnamefont
  {Manmana}},\ and\ \bibinfo {author} {\bibfnamefont {N.~J.}\ \bibnamefont
  {Robinson}},\ }\bibfield  {title} {\bibinfo {title} {Quench dynamics in a
  model with tuneable integrability breaking},\ }\href
  {https://doi.org/10.1103/PhysRevB.89.165104} {\bibfield  {journal} {\bibinfo
  {journal} {Phys. Rev. B}\ }\textbf {\bibinfo {volume} {89}},\ \bibinfo
  {pages} {165104} (\bibinfo {year} {2014})}\BibitemShut {NoStop}%
\bibitem [{\citenamefont {Karrasch}\ and\ \citenamefont
  {Schuricht}(2013)}]{Karrasch2013b}%
  \BibitemOpen
  \bibfield  {author} {\bibinfo {author} {\bibfnamefont {C.}~\bibnamefont
  {Karrasch}}\ and\ \bibinfo {author} {\bibfnamefont {D.}~\bibnamefont
  {Schuricht}},\ }\bibfield  {title} {\bibinfo {title} {Dynamical phase
  transitions after quenches in nonintegrable models},\ }\href
  {https://doi.org/10.1103/PhysRevB.87.195104} {\bibfield  {journal} {\bibinfo
  {journal} {Phys. Rev. B}\ }\textbf {\bibinfo {volume} {87}},\ \bibinfo
  {pages} {195104} (\bibinfo {year} {2013})}\BibitemShut {NoStop}%
\bibitem [{\citenamefont {Karrasch}\ \emph {et~al.}(2014)\citenamefont
  {Karrasch}, \citenamefont {Moore},\ and\ \citenamefont
  {Heidrich-Meisner}}]{Karrasch2014}%
  \BibitemOpen
  \bibfield  {author} {\bibinfo {author} {\bibfnamefont {C.}~\bibnamefont
  {Karrasch}}, \bibinfo {author} {\bibfnamefont {J.~E.}\ \bibnamefont
  {Moore}},\ and\ \bibinfo {author} {\bibfnamefont {F.}~\bibnamefont
  {Heidrich-Meisner}},\ }\bibfield  {title} {\bibinfo {title} {Real-time and
  real-space spin and energy dynamics in one-dimensional spin-$\frac{1}{2}$
  systems induced by local quantum quenches at finite temperatures},\ }\href
  {https://doi.org/10.1103/PhysRevB.89.075139} {\bibfield  {journal} {\bibinfo
  {journal} {Phys. Rev. B}\ }\textbf {\bibinfo {volume} {89}},\ \bibinfo
  {pages} {075139} (\bibinfo {year} {2014})}\BibitemShut {NoStop}%
\bibitem [{\citenamefont {McClarty}\ \emph {et~al.}(2017)\citenamefont
  {McClarty}, \citenamefont {Krüger}, \citenamefont {Guidi}, \citenamefont
  {Parker}, \citenamefont {Refson}, \citenamefont {Parker}, \citenamefont
  {Prabhakaran},\ and\ \citenamefont {Coldea}}]{McClarty_2017}%
  \BibitemOpen
  \bibfield  {author} {\bibinfo {author} {\bibfnamefont {P.~A.}\ \bibnamefont
  {McClarty}}, \bibinfo {author} {\bibfnamefont {F.}~\bibnamefont {Krüger}},
  \bibinfo {author} {\bibfnamefont {T.}~\bibnamefont {Guidi}}, \bibinfo
  {author} {\bibfnamefont {S.~F.}\ \bibnamefont {Parker}}, \bibinfo {author}
  {\bibfnamefont {K.}~\bibnamefont {Refson}}, \bibinfo {author} {\bibfnamefont
  {A.~W.}\ \bibnamefont {Parker}}, \bibinfo {author} {\bibfnamefont
  {D.}~\bibnamefont {Prabhakaran}},\ and\ \bibinfo {author} {\bibfnamefont
  {R.}~\bibnamefont {Coldea}},\ }\bibfield  {title} {\bibinfo {title}
  {{Topological triplon modes and bound states in a
  Shastry{\textendash}Sutherland magnet}},\ }\href
  {https://doi.org/10.1038/nphys4117} {\bibfield  {journal} {\bibinfo
  {journal} {Nat. Phys.}\ }\textbf {\bibinfo {volume} {13}},\ \bibinfo {pages}
  {736} (\bibinfo {year} {2017})}\BibitemShut {NoStop}%
\bibitem [{\citenamefont {M\"uhlbauer}\ \emph {et~al.}(2019)\citenamefont
  {M\"uhlbauer}, \citenamefont {Honecker}, \citenamefont {P\'erigo},
  \citenamefont {Bergner}, \citenamefont {Disch}, \citenamefont {Heinemann},
  \citenamefont {Erokhin}, \citenamefont {Berkov}, \citenamefont {Leighton},
  \citenamefont {Eskildsen},\ and\ \citenamefont {Michels}}]{Muehlbauer2019}%
  \BibitemOpen
  \bibfield  {author} {\bibinfo {author} {\bibfnamefont {S.}~\bibnamefont
  {M\"uhlbauer}}, \bibinfo {author} {\bibfnamefont {D.}~\bibnamefont
  {Honecker}}, \bibinfo {author} {\bibfnamefont {E.~A.}\ \bibnamefont
  {P\'erigo}}, \bibinfo {author} {\bibfnamefont {F.}~\bibnamefont {Bergner}},
  \bibinfo {author} {\bibfnamefont {S.}~\bibnamefont {Disch}}, \bibinfo
  {author} {\bibfnamefont {A.}~\bibnamefont {Heinemann}}, \bibinfo {author}
  {\bibfnamefont {S.}~\bibnamefont {Erokhin}}, \bibinfo {author} {\bibfnamefont
  {D.}~\bibnamefont {Berkov}}, \bibinfo {author} {\bibfnamefont
  {C.}~\bibnamefont {Leighton}}, \bibinfo {author} {\bibfnamefont {M.~R.}\
  \bibnamefont {Eskildsen}},\ and\ \bibinfo {author} {\bibfnamefont
  {A.}~\bibnamefont {Michels}},\ }\bibfield  {title} {\bibinfo {title}
  {Magnetic small-angle neutron scattering},\ }\href
  {https://doi.org/10.1103/RevModPhys.91.015004} {\bibfield  {journal}
  {\bibinfo  {journal} {Rev. Mod. Phys.}\ }\textbf {\bibinfo {volume} {91}},\
  \bibinfo {pages} {015004} (\bibinfo {year} {2019})}\BibitemShut {NoStop}%
\bibitem [{\citenamefont {Sobota}\ \emph {et~al.}(2021)\citenamefont {Sobota},
  \citenamefont {He},\ and\ \citenamefont {Shen}}]{Sobota2021}%
  \BibitemOpen
  \bibfield  {author} {\bibinfo {author} {\bibfnamefont {J.~A.}\ \bibnamefont
  {Sobota}}, \bibinfo {author} {\bibfnamefont {Y.}~\bibnamefont {He}},\ and\
  \bibinfo {author} {\bibfnamefont {Z.-X.}\ \bibnamefont {Shen}},\ }\bibfield
  {title} {\bibinfo {title} {Angle-resolved photoemission studies of quantum
  materials},\ }\href {https://doi.org/10.1103/RevModPhys.93.025006} {\bibfield
   {journal} {\bibinfo  {journal} {Rev. Mod. Phys.}\ }\textbf {\bibinfo
  {volume} {93}},\ \bibinfo {pages} {025006} (\bibinfo {year}
  {2021})}\BibitemShut {NoStop}%
\bibitem [{\citenamefont {Barthel}\ \emph {et~al.}(2009)\citenamefont
  {Barthel}, \citenamefont {Schollw\"ock},\ and\ \citenamefont
  {White}}]{Barthel2009}%
  \BibitemOpen
  \bibfield  {author} {\bibinfo {author} {\bibfnamefont {T.}~\bibnamefont
  {Barthel}}, \bibinfo {author} {\bibfnamefont {U.}~\bibnamefont
  {Schollw\"ock}},\ and\ \bibinfo {author} {\bibfnamefont {S.~R.}\ \bibnamefont
  {White}},\ }\bibfield  {title} {\bibinfo {title} {Spectral functions in
  one-dimensional quantum systems at finite temperature using the density
  matrix renormalization group},\ }\href
  {https://doi.org/10.1103/PhysRevB.79.245101} {\bibfield  {journal} {\bibinfo
  {journal} {Phys. Rev. B}\ }\textbf {\bibinfo {volume} {79}},\ \bibinfo
  {pages} {245101} (\bibinfo {year} {2009})}\BibitemShut {NoStop}%
\bibitem [{\citenamefont {Karrasch}\ \emph {et~al.}(2012)\citenamefont
  {Karrasch}, \citenamefont {Bardarson},\ and\ \citenamefont
  {Moore}}]{Karrasch2012}%
  \BibitemOpen
  \bibfield  {author} {\bibinfo {author} {\bibfnamefont {C.}~\bibnamefont
  {Karrasch}}, \bibinfo {author} {\bibfnamefont {J.~H.}\ \bibnamefont
  {Bardarson}},\ and\ \bibinfo {author} {\bibfnamefont {J.~E.}\ \bibnamefont
  {Moore}},\ }\bibfield  {title} {\bibinfo {title} {Finite-temperature
  dynamical density matrix renormalization group and the {D}rude weight of
  spin-$1/2$ chains},\ }\href {https://doi.org/10.1103/PhysRevLett.108.227206}
  {\bibfield  {journal} {\bibinfo  {journal} {Phys. Rev. Lett.}\ }\textbf
  {\bibinfo {volume} {108}},\ \bibinfo {pages} {227206} (\bibinfo {year}
  {2012})}\BibitemShut {NoStop}%
\bibitem [{\citenamefont {Karrasch}\ \emph {et~al.}(2013)\citenamefont
  {Karrasch}, \citenamefont {Bardarson},\ and\ \citenamefont
  {Moore}}]{Karrasch2013}%
  \BibitemOpen
  \bibfield  {author} {\bibinfo {author} {\bibfnamefont {C.}~\bibnamefont
  {Karrasch}}, \bibinfo {author} {\bibfnamefont {J.~H.}\ \bibnamefont
  {Bardarson}},\ and\ \bibinfo {author} {\bibfnamefont {J.~E.}\ \bibnamefont
  {Moore}},\ }\bibfield  {title} {\bibinfo {title} {Reducing the numerical
  effort of finite-temperature density matrix renormalization group
  calculations},\ }\href {https://doi.org/10.1088/1367-2630/15/8/083031}
  {\bibfield  {journal} {\bibinfo  {journal} {New J. Phys.}\ }\textbf {\bibinfo
  {volume} {15}},\ \bibinfo {pages} {083031} (\bibinfo {year}
  {2013})}\BibitemShut {NoStop}%
\bibitem [{\citenamefont {Barthel}(2013)}]{Barthel2013}%
  \BibitemOpen
  \bibfield  {author} {\bibinfo {author} {\bibfnamefont {T.}~\bibnamefont
  {Barthel}},\ }\bibfield  {title} {\bibinfo {title} {Precise evaluation of
  thermal response functions by optimized density matrix renormalization group
  schemes},\ }\href {https://doi.org/10.1088/1367-2630/15/7/073010} {\bibfield
  {journal} {\bibinfo  {journal} {New J. Phys.}\ }\textbf {\bibinfo {volume}
  {15}},\ \bibinfo {pages} {073010} (\bibinfo {year} {2013})}\BibitemShut
  {NoStop}%
\bibitem [{\citenamefont {Tiegel}\ \emph {et~al.}(2014)\citenamefont {Tiegel},
  \citenamefont {Manmana}, \citenamefont {Pruschke},\ and\ \citenamefont
  {Honecker}}]{Tiegel2014}%
  \BibitemOpen
  \bibfield  {author} {\bibinfo {author} {\bibfnamefont {A.~C.}\ \bibnamefont
  {Tiegel}}, \bibinfo {author} {\bibfnamefont {S.~R.}\ \bibnamefont {Manmana}},
  \bibinfo {author} {\bibfnamefont {T.}~\bibnamefont {Pruschke}},\ and\
  \bibinfo {author} {\bibfnamefont {A.}~\bibnamefont {Honecker}},\ }\bibfield
  {title} {\bibinfo {title} {Matrix product state formulation of
  frequency-space dynamics at finite temperatures},\ }\href
  {https://doi.org/10.1103/PhysRevB.90.060406} {\bibfield  {journal} {\bibinfo
  {journal} {Phys. Rev. B}\ }\textbf {\bibinfo {volume} {90}},\ \bibinfo
  {pages} {060406} (\bibinfo {year} {2014})}\BibitemShut {NoStop}%
\bibitem [{\citenamefont {Jansen}\ \emph {et~al.}(2022)\citenamefont {Jansen},
  \citenamefont {Bon\v{c}a},\ and\ \citenamefont
  {Heidrich-Meisner}}]{Jansen2022}%
  \BibitemOpen
  \bibfield  {author} {\bibinfo {author} {\bibfnamefont {D.}~\bibnamefont
  {Jansen}}, \bibinfo {author} {\bibfnamefont {J.}~\bibnamefont {Bon\v{c}a}},\
  and\ \bibinfo {author} {\bibfnamefont {F.}~\bibnamefont {Heidrich-Meisner}},\
  }\bibfield  {title} {\bibinfo {title} {Finite-temperature optical
  conductivity with density-matrix renormalization group methods for the
  {H}olstein polaron and bipolaron with dispersive phonons},\ }\href
  {https://doi.org/10.1103/PhysRevB.106.155129} {\bibfield  {journal} {\bibinfo
   {journal} {Phys. Rev. B}\ }\textbf {\bibinfo {volume} {106}},\ \bibinfo
  {pages} {155129} (\bibinfo {year} {2022})}\BibitemShut {NoStop}%
\bibitem [{\citenamefont {Rakovszky}\ \emph {et~al.}(2022)\citenamefont
  {Rakovszky}, \citenamefont {von Keyserlingk},\ and\ \citenamefont
  {Pollmann}}]{Rakovsky2022}%
  \BibitemOpen
  \bibfield  {author} {\bibinfo {author} {\bibfnamefont {T.}~\bibnamefont
  {Rakovszky}}, \bibinfo {author} {\bibfnamefont {C.~W.}\ \bibnamefont {von
  Keyserlingk}},\ and\ \bibinfo {author} {\bibfnamefont {F.}~\bibnamefont
  {Pollmann}},\ }\bibfield  {title} {\bibinfo {title} {Dissipation-assisted
  operator evolution method for capturing hydrodynamic transport},\ }\href
  {https://doi.org/10.1103/PhysRevB.105.075131} {\bibfield  {journal} {\bibinfo
   {journal} {Phys. Rev. B}\ }\textbf {\bibinfo {volume} {105}},\ \bibinfo
  {pages} {075131} (\bibinfo {year} {2022})}\BibitemShut {NoStop}%
\bibitem [{\citenamefont {Sirker}\ and\ \citenamefont
  {Kl\"umper}(2005)}]{PhysRevB.71.241101}%
  \BibitemOpen
  \bibfield  {author} {\bibinfo {author} {\bibfnamefont {J.}~\bibnamefont
  {Sirker}}\ and\ \bibinfo {author} {\bibfnamefont {A.}~\bibnamefont
  {Kl\"umper}},\ }\bibfield  {title} {\bibinfo {title} {Real-time dynamics at
  finite temperature by the density-matrix renormalization group: A
  path-integral approach},\ }\href {https://doi.org/10.1103/PhysRevB.71.241101}
  {\bibfield  {journal} {\bibinfo  {journal} {Phys. Rev. B}\ }\textbf {\bibinfo
  {volume} {71}},\ \bibinfo {pages} {241101} (\bibinfo {year}
  {2005})}\BibitemShut {NoStop}%
\bibitem [{\citenamefont {Kokalj}\ and\ \citenamefont
  {Prelov\v{s}ek}(2009)}]{PhysRevB.80.205117}%
  \BibitemOpen
  \bibfield  {author} {\bibinfo {author} {\bibfnamefont {J.}~\bibnamefont
  {Kokalj}}\ and\ \bibinfo {author} {\bibfnamefont {P.}~\bibnamefont
  {Prelov\v{s}ek}},\ }\bibfield  {title} {\bibinfo {title} {Finite-temperature
  dynamics with the density-matrix renormalization group method},\ }\href
  {https://doi.org/10.1103/PhysRevB.80.205117} {\bibfield  {journal} {\bibinfo
  {journal} {Phys. Rev. B}\ }\textbf {\bibinfo {volume} {80}},\ \bibinfo
  {pages} {205117} (\bibinfo {year} {2009})}\BibitemShut {NoStop}%
\bibitem [{\citenamefont {Suwa}\ and\ \citenamefont {Todo}(2015)}]{Suwa2015}%
  \BibitemOpen
  \bibfield  {author} {\bibinfo {author} {\bibfnamefont {H.}~\bibnamefont
  {Suwa}}\ and\ \bibinfo {author} {\bibfnamefont {S.}~\bibnamefont {Todo}},\
  }\bibfield  {title} {\bibinfo {title} {Generalized moment method for gap
  estimation and quantum {M}onte {C}arlo level spectroscopy},\ }\href
  {https://doi.org/10.1103/PhysRevLett.115.080601} {\bibfield  {journal}
  {\bibinfo  {journal} {Phys. Rev. Lett.}\ }\textbf {\bibinfo {volume} {115}},\
  \bibinfo {pages} {080601} (\bibinfo {year} {2015})}\BibitemShut {NoStop}%
\bibitem [{\citenamefont {Hastings}(2007)}]{Hastings2007}%
  \BibitemOpen
  \bibfield  {author} {\bibinfo {author} {\bibfnamefont {M.~B.}\ \bibnamefont
  {Hastings}},\ }\bibfield  {title} {\bibinfo {title} {An area law for
  one-dimensional quantum systems},\ }\href
  {https://doi.org/10.1088/1742-5468/2007/08/P08024} {\bibfield  {journal}
  {\bibinfo  {journal} {J. Stat. Mech.}\ }\textbf {\bibinfo {volume} {2007}},\
  \bibinfo {pages} {P08024} (\bibinfo {year} {2007})}\BibitemShut {NoStop}%
\bibitem [{\citenamefont {Wolf}\ \emph {et~al.}(2008)\citenamefont {Wolf},
  \citenamefont {Verstraete}, \citenamefont {Hastings},\ and\ \citenamefont
  {Cirac}}]{Wolf2008}%
  \BibitemOpen
  \bibfield  {author} {\bibinfo {author} {\bibfnamefont {M.~M.}\ \bibnamefont
  {Wolf}}, \bibinfo {author} {\bibfnamefont {F.}~\bibnamefont {Verstraete}},
  \bibinfo {author} {\bibfnamefont {M.~B.}\ \bibnamefont {Hastings}},\ and\
  \bibinfo {author} {\bibfnamefont {J.~I.}\ \bibnamefont {Cirac}},\ }\bibfield
  {title} {\bibinfo {title} {Area laws in quantum systems: Mutual information
  and correlations},\ }\href {https://doi.org/10.1103/PhysRevLett.100.070502}
  {\bibfield  {journal} {\bibinfo  {journal} {Phys. Rev. Lett.}\ }\textbf
  {\bibinfo {volume} {100}},\ \bibinfo {pages} {070502} (\bibinfo {year}
  {2008})}\BibitemShut {NoStop}%
\bibitem [{\citenamefont {Krilov}\ \emph {et~al.}(2001)\citenamefont {Krilov},
  \citenamefont {Sim},\ and\ \citenamefont {Berne}}]{Krilov2001}%
  \BibitemOpen
  \bibfield  {author} {\bibinfo {author} {\bibfnamefont {G.}~\bibnamefont
  {Krilov}}, \bibinfo {author} {\bibfnamefont {E.}~\bibnamefont {Sim}},\ and\
  \bibinfo {author} {\bibfnamefont {B.~J.}\ \bibnamefont {Berne}},\ }\bibfield
  {title} {\bibinfo {title} {Quantum time correlation functions from complex
  time {M}onte {C}arlo simulations: A maximum entropy approach},\ }\href
  {https://doi.org/10.1063/1.1331613} {\bibfield  {journal} {\bibinfo
  {journal} {J. Chem. Phys.}\ }\textbf {\bibinfo {volume} {114}},\ \bibinfo
  {pages} {1075–1088} (\bibinfo {year} {2001})}\BibitemShut {NoStop}%
\bibitem [{\citenamefont {Cao}\ \emph {et~al.}(2024)\citenamefont {Cao},
  \citenamefont {Lu}, \citenamefont {Stoudenmire},\ and\ \citenamefont
  {Parcollet}}]{Cao2023}%
  \BibitemOpen
  \bibfield  {author} {\bibinfo {author} {\bibfnamefont {X.}~\bibnamefont
  {Cao}}, \bibinfo {author} {\bibfnamefont {Y.}~\bibnamefont {Lu}}, \bibinfo
  {author} {\bibfnamefont {E.~M.}\ \bibnamefont {Stoudenmire}},\ and\ \bibinfo
  {author} {\bibfnamefont {O.}~\bibnamefont {Parcollet}},\ }\bibfield  {title}
  {\bibinfo {title} {Dynamical correlation functions from complex time
  evolution},\ }\href {https://doi.org/10.1103/PhysRevB.109.235110} {\bibfield
  {journal} {\bibinfo  {journal} {Phys. Rev. B}\ }\textbf {\bibinfo {volume}
  {109}},\ \bibinfo {pages} {235110} (\bibinfo {year} {2024})}\BibitemShut
  {NoStop}%
\bibitem [{\citenamefont {Grundner}\ \emph {et~al.}(2024)\citenamefont
  {Grundner}, \citenamefont {Westhoff}, \citenamefont {Kugler}, \citenamefont
  {Parcollet},\ and\ \citenamefont {Schollw\"ock}}]{Grundner2023}%
  \BibitemOpen
  \bibfield  {author} {\bibinfo {author} {\bibfnamefont {M.}~\bibnamefont
  {Grundner}}, \bibinfo {author} {\bibfnamefont {P.}~\bibnamefont {Westhoff}},
  \bibinfo {author} {\bibfnamefont {F.~B.}\ \bibnamefont {Kugler}}, \bibinfo
  {author} {\bibfnamefont {O.}~\bibnamefont {Parcollet}},\ and\ \bibinfo
  {author} {\bibfnamefont {U.}~\bibnamefont {Schollw\"ock}},\ }\bibfield
  {title} {\bibinfo {title} {Complex time evolution in tensor networks and
  time-dependent {G}reen's functions},\ }\href
  {https://doi.org/10.1103/PhysRevB.109.155124} {\bibfield  {journal} {\bibinfo
   {journal} {Phys. Rev. B}\ }\textbf {\bibinfo {volume} {109}},\ \bibinfo
  {pages} {155124} (\bibinfo {year} {2024})}\BibitemShut {NoStop}%
\bibitem [{\citenamefont {Haegeman}\ \emph {et~al.}(2011)\citenamefont
  {Haegeman}, \citenamefont {Cirac}, \citenamefont {Osborne}, \citenamefont
  {Pi\ifmmode~\check{z}\else \v{z}\fi{}orn}, \citenamefont {Verschelde},\ and\
  \citenamefont {Verstraete}}]{Haegeman2011}%
  \BibitemOpen
  \bibfield  {author} {\bibinfo {author} {\bibfnamefont {J.}~\bibnamefont
  {Haegeman}}, \bibinfo {author} {\bibfnamefont {J.~I.}\ \bibnamefont {Cirac}},
  \bibinfo {author} {\bibfnamefont {T.~J.}\ \bibnamefont {Osborne}}, \bibinfo
  {author} {\bibfnamefont {I.}~\bibnamefont {Pi\ifmmode~\check{z}\else
  \v{z}\fi{}orn}}, \bibinfo {author} {\bibfnamefont {H.}~\bibnamefont
  {Verschelde}},\ and\ \bibinfo {author} {\bibfnamefont {F.}~\bibnamefont
  {Verstraete}},\ }\bibfield  {title} {\bibinfo {title} {{Time-Dependent
  Variational Principle for Quantum Lattices}},\ }\href
  {https://doi.org/10.1103/PhysRevLett.107.070601} {\bibfield  {journal}
  {\bibinfo  {journal} {Phys. Rev. Lett.}\ }\textbf {\bibinfo {volume} {107}},\
  \bibinfo {pages} {070601} (\bibinfo {year} {2011})}\BibitemShut {NoStop}%
\bibitem [{\citenamefont {Haegeman}\ \emph {et~al.}(2016)\citenamefont
  {Haegeman}, \citenamefont {Lubich}, \citenamefont {Oseledets}, \citenamefont
  {Vandereycken},\ and\ \citenamefont {Verstraete}}]{Haegeman2016}%
  \BibitemOpen
  \bibfield  {author} {\bibinfo {author} {\bibfnamefont {J.}~\bibnamefont
  {Haegeman}}, \bibinfo {author} {\bibfnamefont {C.}~\bibnamefont {Lubich}},
  \bibinfo {author} {\bibfnamefont {I.}~\bibnamefont {Oseledets}}, \bibinfo
  {author} {\bibfnamefont {B.}~\bibnamefont {Vandereycken}},\ and\ \bibinfo
  {author} {\bibfnamefont {F.}~\bibnamefont {Verstraete}},\ }\bibfield  {title}
  {\bibinfo {title} {Unifying time evolution and optimization with matrix
  product states},\ }\href {https://doi.org/10.1103/PhysRevB.94.165116}
  {\bibfield  {journal} {\bibinfo  {journal} {Phys. Rev. B}\ }\textbf {\bibinfo
  {volume} {94}},\ \bibinfo {pages} {165116} (\bibinfo {year}
  {2016})}\BibitemShut {NoStop}%
\bibitem [{\citenamefont {Wang}\ \emph {et~al.}(2024)\citenamefont {Wang},
  \citenamefont {McClarty}, \citenamefont {Dankova}, \citenamefont {Honecker},\
  and\ \citenamefont {Wietek}}]{Wang2024a}%
  \BibitemOpen
  \bibfield  {author} {\bibinfo {author} {\bibfnamefont {Z.}~\bibnamefont
  {Wang}}, \bibinfo {author} {\bibfnamefont {P.}~\bibnamefont {McClarty}},
  \bibinfo {author} {\bibfnamefont {D.}~\bibnamefont {Dankova}}, \bibinfo
  {author} {\bibfnamefont {A.}~\bibnamefont {Honecker}},\ and\ \bibinfo
  {author} {\bibfnamefont {A.}~\bibnamefont {Wietek}},\ }\href
  {https://arxiv.org/abs/2405.18475} {\bibinfo {title} {Anomalous thermal
  broadening in the {S}hastry-{S}utherland model and {S}r{C}u$_2(${BO}$_3)_2$}}
  (\bibinfo {year} {2024}),\ \Eprint {https://arxiv.org/abs/2405.18475}
  {arXiv:2405.18475 [cond-mat.str-el]} \BibitemShut {NoStop}%
\bibitem [{\citenamefont {{Sriram Shastry}}\ and\ \citenamefont
  {{Sutherland}}(1981)}]{ShastrySutherland1981}%
  \BibitemOpen
  \bibfield  {author} {\bibinfo {author} {\bibfnamefont {B.}~\bibnamefont
  {{Sriram Shastry}}}\ and\ \bibinfo {author} {\bibfnamefont {B.}~\bibnamefont
  {{Sutherland}}},\ }\bibfield  {title} {\bibinfo {title} {{Exact ground state
  of a quantum mechanical antiferromagnet}},\ }\href
  {https://doi.org/10.1016/0378-4363(81)90838-X} {\bibfield  {journal}
  {\bibinfo  {journal} {Physica B+C}\ }\textbf {\bibinfo {volume} {108}},\
  \bibinfo {pages} {1069} (\bibinfo {year} {1981})}\BibitemShut {NoStop}%
\bibitem [{\citenamefont {{Miyahara}}\ and\ \citenamefont
  {{Ueda}}(2003)}]{miyahara2003}%
  \BibitemOpen
  \bibfield  {author} {\bibinfo {author} {\bibfnamefont {S.}~\bibnamefont
  {{Miyahara}}}\ and\ \bibinfo {author} {\bibfnamefont {K.}~\bibnamefont
  {{Ueda}}},\ }\bibfield  {title} {\bibinfo {title} {{Theory of the orthogonal
  dimer {H}eisenberg spin model for {SrCu}$_{2}$({BO}$_{3}$)$_{2}$}},\ }\href
  {https://doi.org/10.1088/0953-8984/15/9/201} {\bibfield  {journal} {\bibinfo
  {journal} {J. Phys.: Condens. Matter}\ }\textbf {\bibinfo {volume} {15}},\
  \bibinfo {pages} {R327} (\bibinfo {year} {2003})}\BibitemShut {NoStop}%
\bibitem [{\citenamefont {Kageyama}\ \emph {et~al.}(2000)\citenamefont
  {Kageyama}, \citenamefont {Nishi}, \citenamefont {Aso}, \citenamefont
  {Onizuka}, \citenamefont {Yosihama}, \citenamefont {Nukui}, \citenamefont
  {Kodama}, \citenamefont {Kakurai},\ and\ \citenamefont
  {Ueda}}]{kageyama2000}%
  \BibitemOpen
  \bibfield  {author} {\bibinfo {author} {\bibfnamefont {H.}~\bibnamefont
  {Kageyama}}, \bibinfo {author} {\bibfnamefont {M.}~\bibnamefont {Nishi}},
  \bibinfo {author} {\bibfnamefont {N.}~\bibnamefont {Aso}}, \bibinfo {author}
  {\bibfnamefont {K.}~\bibnamefont {Onizuka}}, \bibinfo {author} {\bibfnamefont
  {T.}~\bibnamefont {Yosihama}}, \bibinfo {author} {\bibfnamefont
  {K.}~\bibnamefont {Nukui}}, \bibinfo {author} {\bibfnamefont
  {K.}~\bibnamefont {Kodama}}, \bibinfo {author} {\bibfnamefont
  {K.}~\bibnamefont {Kakurai}},\ and\ \bibinfo {author} {\bibfnamefont
  {Y.}~\bibnamefont {Ueda}},\ }\bibfield  {title} {\bibinfo {title} {Direct
  evidence for the localized single-triplet excitations and the dispersive
  multitriplet excitations in {SrCu}$_{2}$({BO}$_{3}$)$_{2}$},\ }\href
  {https://doi.org/10.1103/PhysRevLett.84.5876} {\bibfield  {journal} {\bibinfo
   {journal} {Phys. Rev. Lett.}\ }\textbf {\bibinfo {volume} {84}},\ \bibinfo
  {pages} {5876} (\bibinfo {year} {2000})}\BibitemShut {NoStop}%
\bibitem [{\citenamefont {Gaulin}\ \emph {et~al.}(2004)\citenamefont {Gaulin},
  \citenamefont {Lee}, \citenamefont {Haravifard}, \citenamefont {Castellan},
  \citenamefont {Berlinsky}, \citenamefont {Dabkowska}, \citenamefont {Qiu},\
  and\ \citenamefont {Copley}}]{gaulin2004}%
  \BibitemOpen
  \bibfield  {author} {\bibinfo {author} {\bibfnamefont {B.~D.}\ \bibnamefont
  {Gaulin}}, \bibinfo {author} {\bibfnamefont {S.~H.}\ \bibnamefont {Lee}},
  \bibinfo {author} {\bibfnamefont {S.}~\bibnamefont {Haravifard}}, \bibinfo
  {author} {\bibfnamefont {J.~P.}\ \bibnamefont {Castellan}}, \bibinfo {author}
  {\bibfnamefont {A.~J.}\ \bibnamefont {Berlinsky}}, \bibinfo {author}
  {\bibfnamefont {H.~A.}\ \bibnamefont {Dabkowska}}, \bibinfo {author}
  {\bibfnamefont {Y.}~\bibnamefont {Qiu}},\ and\ \bibinfo {author}
  {\bibfnamefont {J.~R.~D.}\ \bibnamefont {Copley}},\ }\bibfield  {title}
  {\bibinfo {title} {{High-Resolution Study of Spin Excitations in the Singlet
  Ground State of
  ${\mathrm{S}\mathrm{r}\mathrm{C}\mathrm{u}}_{2}({\mathrm{B}\mathrm{O}}_{3}{)}_{2}$}},\
  }\href {https://doi.org/10.1103/PhysRevLett.93.267202} {\bibfield  {journal}
  {\bibinfo  {journal} {Phys. Rev. Lett.}\ }\textbf {\bibinfo {volume} {93}},\
  \bibinfo {pages} {267202} (\bibinfo {year} {2004})}\BibitemShut {NoStop}%
\bibitem [{\citenamefont {Zayed}\ \emph {et~al.}(2014)\citenamefont {Zayed},
  \citenamefont {R\"uegg}, \citenamefont {Str\"assle}, \citenamefont {Stuhr},
  \citenamefont {Roessli}, \citenamefont {Ay}, \citenamefont {Mesot},
  \citenamefont {Link}, \citenamefont {Pomjakushina}, \citenamefont
  {Stingaciu}, \citenamefont {Conder},\ and\ \citenamefont
  {R\o{}nnow}}]{zayed2014}%
  \BibitemOpen
  \bibfield  {author} {\bibinfo {author} {\bibfnamefont {M.~E.}\ \bibnamefont
  {Zayed}}, \bibinfo {author} {\bibfnamefont {C.}~\bibnamefont {R\"uegg}},
  \bibinfo {author} {\bibfnamefont {T.}~\bibnamefont {Str\"assle}}, \bibinfo
  {author} {\bibfnamefont {U.}~\bibnamefont {Stuhr}}, \bibinfo {author}
  {\bibfnamefont {B.}~\bibnamefont {Roessli}}, \bibinfo {author} {\bibfnamefont
  {M.}~\bibnamefont {Ay}}, \bibinfo {author} {\bibfnamefont {J.}~\bibnamefont
  {Mesot}}, \bibinfo {author} {\bibfnamefont {P.}~\bibnamefont {Link}},
  \bibinfo {author} {\bibfnamefont {E.}~\bibnamefont {Pomjakushina}}, \bibinfo
  {author} {\bibfnamefont {M.}~\bibnamefont {Stingaciu}}, \bibinfo {author}
  {\bibfnamefont {K.}~\bibnamefont {Conder}},\ and\ \bibinfo {author}
  {\bibfnamefont {H.~M.}\ \bibnamefont {R\o{}nnow}},\ }\bibfield  {title}
  {\bibinfo {title} {{Correlated Decay of Triplet Excitations in the
  Shastry-Sutherland Compound SrCu$_2($BO$_3)_2$}},\ }\href
  {https://doi.org/10.1103/PhysRevLett.113.067201} {\bibfield  {journal}
  {\bibinfo  {journal} {Phys. Rev. Lett.}\ }\textbf {\bibinfo {volume} {113}},\
  \bibinfo {pages} {067201} (\bibinfo {year} {2014})}\BibitemShut {NoStop}%
\bibitem [{\citenamefont {Kageyama}\ \emph {et~al.}(1999)\citenamefont
  {Kageyama}, \citenamefont {Yoshimura}, \citenamefont {Stern}, \citenamefont
  {Mushnikov}, \citenamefont {Onizuka}, \citenamefont {Kato}, \citenamefont
  {Kosuge}, \citenamefont {Slichter}, \citenamefont {Goto},\ and\ \citenamefont
  {Ueda}}]{Kageyama1999}%
  \BibitemOpen
  \bibfield  {author} {\bibinfo {author} {\bibfnamefont {H.}~\bibnamefont
  {Kageyama}}, \bibinfo {author} {\bibfnamefont {K.}~\bibnamefont {Yoshimura}},
  \bibinfo {author} {\bibfnamefont {R.}~\bibnamefont {Stern}}, \bibinfo
  {author} {\bibfnamefont {N.~V.}\ \bibnamefont {Mushnikov}}, \bibinfo {author}
  {\bibfnamefont {K.}~\bibnamefont {Onizuka}}, \bibinfo {author} {\bibfnamefont
  {M.}~\bibnamefont {Kato}}, \bibinfo {author} {\bibfnamefont {K.}~\bibnamefont
  {Kosuge}}, \bibinfo {author} {\bibfnamefont {C.~P.}\ \bibnamefont
  {Slichter}}, \bibinfo {author} {\bibfnamefont {T.}~\bibnamefont {Goto}},\
  and\ \bibinfo {author} {\bibfnamefont {Y.}~\bibnamefont {Ueda}},\ }\bibfield
  {title} {\bibinfo {title} {{Exact Dimer Ground State and Quantized
  Magnetization Plateaus in the Two-Dimensional Spin System
  ${\mathrm{SrCu}}_{2}({\mathrm{BO}}_{3}){}_{2}$}},\ }\href
  {https://doi.org/10.1103/PhysRevLett.82.3168} {\bibfield  {journal} {\bibinfo
   {journal} {Phys. Rev. Lett.}\ }\textbf {\bibinfo {volume} {82}},\ \bibinfo
  {pages} {3168} (\bibinfo {year} {1999})}\BibitemShut {NoStop}%
\bibitem [{\citenamefont {Miyahara}\ and\ \citenamefont
  {Ueda}(2000)}]{miyahara69thermodynamic}%
  \BibitemOpen
  \bibfield  {author} {\bibinfo {author} {\bibfnamefont {S.}~\bibnamefont
  {Miyahara}}\ and\ \bibinfo {author} {\bibfnamefont {K.}~\bibnamefont
  {Ueda}},\ }\bibfield  {title} {\bibinfo {title} {Thermodynamic properties of
  three-dimensional orthogonal dimer model for {SrCu$_2$(B0$_3$)$_2$}},\ }\href
  {https://www.jps.or.jp/books/jpsjs/69B/jpsj.69sb.072.pdf} {\bibfield
  {journal} {\bibinfo  {journal} {J. Phys. Soc. Jpn. Suppl B}\ }\textbf
  {\bibinfo {volume} {69}},\ \bibinfo {pages} {72} (\bibinfo {year}
  {2000})}\BibitemShut {NoStop}%
\bibitem [{\citenamefont {Wietek}\ \emph {et~al.}(2019)\citenamefont {Wietek},
  \citenamefont {Corboz}, \citenamefont {Wessel}, \citenamefont {Normand},
  \citenamefont {Mila},\ and\ \citenamefont {Honecker}}]{wietek2019}%
  \BibitemOpen
  \bibfield  {author} {\bibinfo {author} {\bibfnamefont {A.}~\bibnamefont
  {Wietek}}, \bibinfo {author} {\bibfnamefont {P.}~\bibnamefont {Corboz}},
  \bibinfo {author} {\bibfnamefont {S.}~\bibnamefont {Wessel}}, \bibinfo
  {author} {\bibfnamefont {B.}~\bibnamefont {Normand}}, \bibinfo {author}
  {\bibfnamefont {F.}~\bibnamefont {Mila}},\ and\ \bibinfo {author}
  {\bibfnamefont {A.}~\bibnamefont {Honecker}},\ }\bibfield  {title} {\bibinfo
  {title} {{Thermodynamic properties of the {S}hastry-{S}utherland model
  throughout the dimer-product phase}},\ }\href
  {https://doi.org/10.1103/PhysRevResearch.1.033038} {\bibfield  {journal}
  {\bibinfo  {journal} {Phys. Rev. Res.}\ }\textbf {\bibinfo {volume} {1}},\
  \bibinfo {pages} {033038} (\bibinfo {year} {2019})}\BibitemShut {NoStop}%
\bibitem [{\citenamefont {Lemmens}\ \emph {et~al.}(2000)\citenamefont
  {Lemmens}, \citenamefont {Grove}, \citenamefont {Fischer}, \citenamefont
  {G\"untherodt}, \citenamefont {Kotov}, \citenamefont {Kageyama},
  \citenamefont {Onizuka},\ and\ \citenamefont {Ueda}}]{Lemmens2000}%
  \BibitemOpen
  \bibfield  {author} {\bibinfo {author} {\bibfnamefont {P.}~\bibnamefont
  {Lemmens}}, \bibinfo {author} {\bibfnamefont {M.}~\bibnamefont {Grove}},
  \bibinfo {author} {\bibfnamefont {M.}~\bibnamefont {Fischer}}, \bibinfo
  {author} {\bibfnamefont {G.}~\bibnamefont {G\"untherodt}}, \bibinfo {author}
  {\bibfnamefont {V.~N.}\ \bibnamefont {Kotov}}, \bibinfo {author}
  {\bibfnamefont {H.}~\bibnamefont {Kageyama}}, \bibinfo {author}
  {\bibfnamefont {K.}~\bibnamefont {Onizuka}},\ and\ \bibinfo {author}
  {\bibfnamefont {Y.}~\bibnamefont {Ueda}},\ }\bibfield  {title} {\bibinfo
  {title} {{Collective Singlet Excitations and Evolution of Raman Spectral
  Weights in the 2D Spin Dimer Compound
  ${\mathrm{SrCu}}_{2}({\mathrm{BO}}_{3}{)}_{2}$}},\ }\href
  {https://doi.org/10.1103/PhysRevLett.85.2605} {\bibfield  {journal} {\bibinfo
   {journal} {Phys. Rev. Lett.}\ }\textbf {\bibinfo {volume} {85}},\ \bibinfo
  {pages} {2605} (\bibinfo {year} {2000})}\BibitemShut {NoStop}%
\bibitem [{\citenamefont {{Wulferding}}\ \emph {et~al.}(2021)\citenamefont
  {{Wulferding}}, \citenamefont {{Choi}}, \citenamefont {{Lee}}, \citenamefont
  {{Prosnikov}}, \citenamefont {{Gallais}}, \citenamefont {{Lemmens}},
  \citenamefont {{Zhong}}, \citenamefont {{Kageyama}},\ and\ \citenamefont
  {{Choi}}}]{wulferding2021}%
  \BibitemOpen
  \bibfield  {author} {\bibinfo {author} {\bibfnamefont {D.}~\bibnamefont
  {{Wulferding}}}, \bibinfo {author} {\bibfnamefont {Y.}~\bibnamefont
  {{Choi}}}, \bibinfo {author} {\bibfnamefont {S.}~\bibnamefont {{Lee}}},
  \bibinfo {author} {\bibfnamefont {M.~A.}\ \bibnamefont {{Prosnikov}}},
  \bibinfo {author} {\bibfnamefont {Y.}~\bibnamefont {{Gallais}}}, \bibinfo
  {author} {\bibfnamefont {P.}~\bibnamefont {{Lemmens}}}, \bibinfo {author}
  {\bibfnamefont {C.}~\bibnamefont {{Zhong}}}, \bibinfo {author} {\bibfnamefont
  {H.}~\bibnamefont {{Kageyama}}},\ and\ \bibinfo {author} {\bibfnamefont
  {K.-Y.}\ \bibnamefont {{Choi}}},\ }\bibfield  {title} {\bibinfo {title}
  {{{Thermally populated versus field-induced triplon bound states in the
  Shastry-Sutherland lattice SrCu$_{2}$(BO$_{3}$)$_{2}$}}},\ }\href
  {https://doi.org/10.1038/s41535-021-00405-7} {\bibfield  {journal} {\bibinfo
  {journal} {npj Quantum Materials}\ }\textbf {\bibinfo {volume} {6}},\
  \bibinfo {eid} {102} (\bibinfo {year} {2021})}\BibitemShut {NoStop}%
\bibitem [{\citenamefont {Paeckel}\ \emph {et~al.}(2019)\citenamefont
  {Paeckel}, \citenamefont {Köhler}, \citenamefont {Swoboda}, \citenamefont
  {Manmana}, \citenamefont {Schollwöck},\ and\ \citenamefont
  {Hubig}}]{Paeckel2019}%
  \BibitemOpen
  \bibfield  {author} {\bibinfo {author} {\bibfnamefont {S.}~\bibnamefont
  {Paeckel}}, \bibinfo {author} {\bibfnamefont {T.}~\bibnamefont {Köhler}},
  \bibinfo {author} {\bibfnamefont {A.}~\bibnamefont {Swoboda}}, \bibinfo
  {author} {\bibfnamefont {S.~R.}\ \bibnamefont {Manmana}}, \bibinfo {author}
  {\bibfnamefont {U.}~\bibnamefont {Schollwöck}},\ and\ \bibinfo {author}
  {\bibfnamefont {C.}~\bibnamefont {Hubig}},\ }\bibfield  {title} {\bibinfo
  {title} {Time-evolution methods for matrix-product states},\ }\href
  {https://doi.org/https://doi.org/10.1016/j.aop.2019.167998} {\bibfield
  {journal} {\bibinfo  {journal} {Ann. Phys. (N. Y.)}\ }\textbf {\bibinfo
  {volume} {411}},\ \bibinfo {pages} {167998} (\bibinfo {year}
  {2019})}\BibitemShut {NoStop}%
\bibitem [{\citenamefont {Zaletel}\ \emph {et~al.}(2015)\citenamefont
  {Zaletel}, \citenamefont {Mong}, \citenamefont {Karrasch}, \citenamefont
  {Moore},\ and\ \citenamefont {Pollmann}}]{Zaletel2015}%
  \BibitemOpen
  \bibfield  {author} {\bibinfo {author} {\bibfnamefont {M.~P.}\ \bibnamefont
  {Zaletel}}, \bibinfo {author} {\bibfnamefont {R.~S.~K.}\ \bibnamefont
  {Mong}}, \bibinfo {author} {\bibfnamefont {C.}~\bibnamefont {Karrasch}},
  \bibinfo {author} {\bibfnamefont {J.~E.}\ \bibnamefont {Moore}},\ and\
  \bibinfo {author} {\bibfnamefont {F.}~\bibnamefont {Pollmann}},\ }\bibfield
  {title} {\bibinfo {title} {Time-evolving a matrix product state with
  long-ranged interactions},\ }\href
  {https://doi.org/10.1103/PhysRevB.91.165112} {\bibfield  {journal} {\bibinfo
  {journal} {Phys. Rev. B}\ }\textbf {\bibinfo {volume} {91}},\ \bibinfo
  {pages} {165112} (\bibinfo {year} {2015})}\BibitemShut {NoStop}%
\bibitem [{\citenamefont {Vidal}(2003)}]{Vidal2003}%
  \BibitemOpen
  \bibfield  {author} {\bibinfo {author} {\bibfnamefont {G.}~\bibnamefont
  {Vidal}},\ }\bibfield  {title} {\bibinfo {title} {Efficient classical
  simulation of slightly entangled quantum computations},\ }\href
  {https://doi.org/10.1103/PhysRevLett.91.147902} {\bibfield  {journal}
  {\bibinfo  {journal} {Phys. Rev. Lett.}\ }\textbf {\bibinfo {volume} {91}},\
  \bibinfo {pages} {147902} (\bibinfo {year} {2003})}\BibitemShut {NoStop}%
\bibitem [{\citenamefont {Yang}\ and\ \citenamefont {White}(2020)}]{Yang2020}%
  \BibitemOpen
  \bibfield  {author} {\bibinfo {author} {\bibfnamefont {M.}~\bibnamefont
  {Yang}}\ and\ \bibinfo {author} {\bibfnamefont {S.~R.}\ \bibnamefont
  {White}},\ }\bibfield  {title} {\bibinfo {title} {Time-dependent variational
  principle with ancillary {K}rylov subspace},\ }\href
  {https://doi.org/10.1103/PhysRevB.102.094315} {\bibfield  {journal} {\bibinfo
   {journal} {Phys. Rev. B}\ }\textbf {\bibinfo {volume} {102}},\ \bibinfo
  {pages} {094315} (\bibinfo {year} {2020})}\BibitemShut {NoStop}%
\bibitem [{\citenamefont {Fishman}\ \emph {et~al.}(2022)\citenamefont
  {Fishman}, \citenamefont {White},\ and\ \citenamefont
  {Stoudenmire}}]{ITensor}%
  \BibitemOpen
  \bibfield  {author} {\bibinfo {author} {\bibfnamefont {M.}~\bibnamefont
  {Fishman}}, \bibinfo {author} {\bibfnamefont {S.~R.}\ \bibnamefont {White}},\
  and\ \bibinfo {author} {\bibfnamefont {E.~M.}\ \bibnamefont {Stoudenmire}},\
  }\bibfield  {title} {\bibinfo {title} {{The ITensor Software Library for
  Tensor Network Calculations}},\ }\href
  {https://doi.org/10.21468/SciPostPhysCodeb.4} {\bibfield  {journal} {\bibinfo
   {journal} {SciPost Phys. Codebases}\ ,\ \bibinfo {pages} {4}} (\bibinfo
  {year} {2022})}\BibitemShut {NoStop}%
\bibitem [{\citenamefont {Yang}\ \emph {et~al.}(2023)\citenamefont {Yang},
  \citenamefont {Fishman}, \citenamefont {White},\ and\ \citenamefont
  {Stoudenmire}}]{ITensorTDVP}%
  \BibitemOpen
  \bibfield  {author} {\bibinfo {author} {\bibfnamefont {M.}~\bibnamefont
  {Yang}}, \bibinfo {author} {\bibfnamefont {M.}~\bibnamefont {Fishman}},
  \bibinfo {author} {\bibfnamefont {S.~R.}\ \bibnamefont {White}},\ and\
  \bibinfo {author} {\bibfnamefont {E.~M.}\ \bibnamefont {Stoudenmire}},\
  }\href@noop {} {\bibinfo {title} {Itensortdvp}},\ \bibinfo {howpublished}
  {\url{https://github.com/ITensor/ITensorTDVP.jl}} (\bibinfo {year}
  {2023})\BibitemShut {NoStop}%
\bibitem [{\citenamefont {Prabhu}(2018)}]{Prabhu2018}%
  \BibitemOpen
  \bibfield  {author} {\bibinfo {author} {\bibfnamefont {K.~M.~M.}\
  \bibnamefont {Prabhu}},\ }\href {https://doi.org/10.1201/9781315216386}
  {\emph {\bibinfo {title} {Window Functions and Their Applications in Signal
  Processing}}}\ (\bibinfo  {publisher} {CRC Press},\ \bibinfo {year}
  {2018})\BibitemShut {NoStop}%
\bibitem [{\citenamefont {Silver}\ \emph {et~al.}(1990)\citenamefont {Silver},
  \citenamefont {Sivia},\ and\ \citenamefont {Gubernatis}}]{Silver_1990}%
  \BibitemOpen
  \bibfield  {author} {\bibinfo {author} {\bibfnamefont {R.~N.}\ \bibnamefont
  {Silver}}, \bibinfo {author} {\bibfnamefont {D.~S.}\ \bibnamefont {Sivia}},\
  and\ \bibinfo {author} {\bibfnamefont {J.~E.}\ \bibnamefont {Gubernatis}},\
  }\bibfield  {title} {\bibinfo {title} {Maximum-entropy method for analytic
  continuation of quantum {M}onte {C}arlo data},\ }\href
  {https://doi.org/10.1103/PhysRevB.41.2380} {\bibfield  {journal} {\bibinfo
  {journal} {Phys. Rev. B}\ }\textbf {\bibinfo {volume} {41}},\ \bibinfo
  {pages} {2380} (\bibinfo {year} {1990})}\BibitemShut {NoStop}%
\bibitem [{\citenamefont {Sandvik}(1998)}]{Sandvik_SAC_98}%
  \BibitemOpen
  \bibfield  {author} {\bibinfo {author} {\bibfnamefont {A.~W.}\ \bibnamefont
  {Sandvik}},\ }\bibfield  {title} {\bibinfo {title} {Stochastic method for
  analytic continuation of quantum {M}onte {C}arlo data},\ }\href
  {https://doi.org/10.1103/PhysRevB.57.10287} {\bibfield  {journal} {\bibinfo
  {journal} {Phys. Rev. B}\ }\textbf {\bibinfo {volume} {57}},\ \bibinfo
  {pages} {10287} (\bibinfo {year} {1998})}\BibitemShut {NoStop}%
\bibitem [{\citenamefont {Beach}(2004)}]{beach2004identifying}%
  \BibitemOpen
  \bibfield  {author} {\bibinfo {author} {\bibfnamefont {K.~S.~D.}\
  \bibnamefont {Beach}},\ }\href@noop {} {\bibinfo {title} {Identifying the
  maximum entropy method as a special limit of stochastic analytic
  continuation}} (\bibinfo {year} {2004}),\ \Eprint
  {https://arxiv.org/abs/cond-mat/0403055} {arXiv:cond-mat/0403055
  [cond-mat.str-el]} \BibitemShut {NoStop}%
\bibitem [{\citenamefont {Shao}\ and\ \citenamefont
  {Sandvik}(2023)}]{SHAO20231}%
  \BibitemOpen
  \bibfield  {author} {\bibinfo {author} {\bibfnamefont {H.}~\bibnamefont
  {Shao}}\ and\ \bibinfo {author} {\bibfnamefont {A.~W.}\ \bibnamefont
  {Sandvik}},\ }\bibfield  {title} {\bibinfo {title} {Progress on stochastic
  analytic continuation of quantum {M}onte {C}arlo data},\ }\href
  {https://doi.org/https://doi.org/10.1016/j.physrep.2022.11.002} {\bibfield
  {journal} {\bibinfo  {journal} {Phys. Rep.}\ }\textbf {\bibinfo {volume}
  {1003}},\ \bibinfo {pages} {1} (\bibinfo {year} {2023})}\BibitemShut
  {NoStop}%
\bibitem [{\citenamefont {Fei}\ \emph {et~al.}(2021)\citenamefont {Fei},
  \citenamefont {Yeh},\ and\ \citenamefont {Gull}}]{Nevanlinna_PRL}%
  \BibitemOpen
  \bibfield  {author} {\bibinfo {author} {\bibfnamefont {J.}~\bibnamefont
  {Fei}}, \bibinfo {author} {\bibfnamefont {C.-N.}\ \bibnamefont {Yeh}},\ and\
  \bibinfo {author} {\bibfnamefont {E.}~\bibnamefont {Gull}},\ }\bibfield
  {title} {\bibinfo {title} {Nevanlinna analytical continuation},\ }\href
  {https://doi.org/10.1103/PhysRevLett.126.056402} {\bibfield  {journal}
  {\bibinfo  {journal} {Phys. Rev. Lett.}\ }\textbf {\bibinfo {volume} {126}},\
  \bibinfo {pages} {056402} (\bibinfo {year} {2021})}\BibitemShut {NoStop}%
\bibitem [{\citenamefont {White}\ and\ \citenamefont
  {Affleck}(2008)}]{White2008}%
  \BibitemOpen
  \bibfield  {author} {\bibinfo {author} {\bibfnamefont {S.~R.}\ \bibnamefont
  {White}}\ and\ \bibinfo {author} {\bibfnamefont {I.}~\bibnamefont
  {Affleck}},\ }\bibfield  {title} {\bibinfo {title} {Spectral function for the
  {$S=1$} {H}eisenberg antiferromagetic chain},\ }\href
  {https://doi.org/10.1103/PhysRevB.77.134437} {\bibfield  {journal} {\bibinfo
  {journal} {Phys. Rev. B}\ }\textbf {\bibinfo {volume} {77}},\ \bibinfo
  {pages} {134437} (\bibinfo {year} {2008})}\BibitemShut {NoStop}%
\bibitem [{\citenamefont {Wolf}\ \emph {et~al.}(2015)\citenamefont {Wolf},
  \citenamefont {Justiniano}, \citenamefont {McCulloch},\ and\ \citenamefont
  {Schollw\"ock}}]{Wolf2015}%
  \BibitemOpen
  \bibfield  {author} {\bibinfo {author} {\bibfnamefont {F.~A.}\ \bibnamefont
  {Wolf}}, \bibinfo {author} {\bibfnamefont {J.~A.}\ \bibnamefont
  {Justiniano}}, \bibinfo {author} {\bibfnamefont {I.~P.}\ \bibnamefont
  {McCulloch}},\ and\ \bibinfo {author} {\bibfnamefont {U.}~\bibnamefont
  {Schollw\"ock}},\ }\bibfield  {title} {\bibinfo {title} {Spectral functions
  and time evolution from the {C}hebyshev recursion},\ }\href
  {https://doi.org/10.1103/PhysRevB.91.115144} {\bibfield  {journal} {\bibinfo
  {journal} {Phys. Rev. B}\ }\textbf {\bibinfo {volume} {91}},\ \bibinfo
  {pages} {115144} (\bibinfo {year} {2015})}\BibitemShut {NoStop}%
\bibitem [{\citenamefont {Tian}\ and\ \citenamefont {White}(2021)}]{Tian2021}%
  \BibitemOpen
  \bibfield  {author} {\bibinfo {author} {\bibfnamefont {Y.}~\bibnamefont
  {Tian}}\ and\ \bibinfo {author} {\bibfnamefont {S.~R.}\ \bibnamefont
  {White}},\ }\bibfield  {title} {\bibinfo {title} {Matrix product state
  recursion methods for computing spectral functions of strongly correlated
  quantum systems},\ }\href {https://doi.org/10.1103/PhysRevB.103.125142}
  {\bibfield  {journal} {\bibinfo  {journal} {Phys. Rev. B}\ }\textbf {\bibinfo
  {volume} {103}},\ \bibinfo {pages} {125142} (\bibinfo {year}
  {2021})}\BibitemShut {NoStop}%
\end{thebibliography}%

\appendix 
%\section{Appendix}

\section{Projection error of TDVP} 
\label{sec:error_TDVP} 

\begin{figure}[t!]
    \centering
    \includegraphics[width=0.4\textwidth]{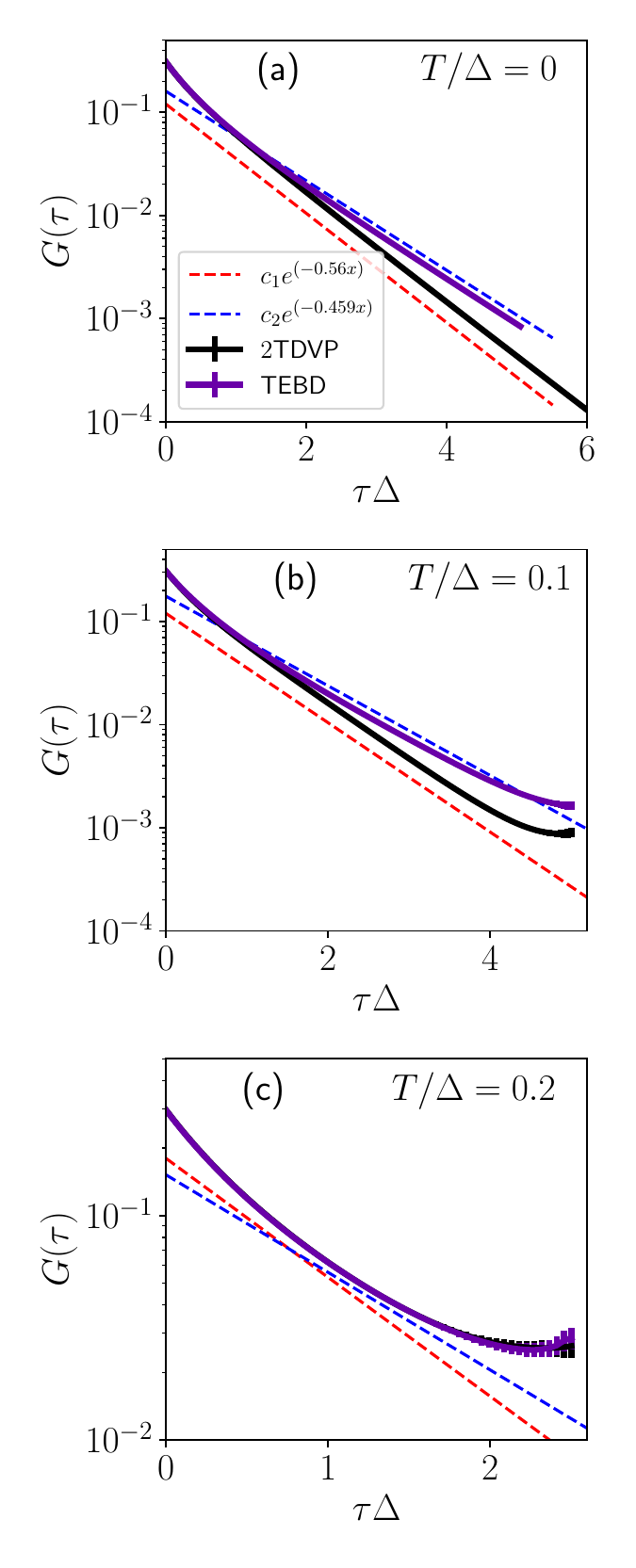}
    \caption{%
    Imaginary time correlation function using TEBD and the 2-site TDVP approach, for 
    $ T /\Delta = 0$ (a), $0.1$ (b) and $0.2$ (c).   Two dashed lines, including $ c_2 e^{ -\Delta \tau } $ ( $  \Delta \approx 0.458$ )  and $ c_1 e^{ - 0.56 \tau } $  are guides  to the eye.  The time evolution approach for measuring correlation function 
    at $T=0$ is based on the ground state via DMRG calculation, where the ones at $T/J_D=0.1$ and $0.2$ are from averaging the METTS configurations.  
    }
    \label{fig:TEBD_TDVP}
\end{figure}

One potential error of the TDVP method exists at the step of projecting the time-dependent Schrödinger equation onto the relevant MPS subspace~\cite{Paeckel2019}.    
This projection error is especially severe when time evolving an MPS with very low bond dimension. It turns out to be a serious issue for the  Shastry-Sutherland model when the temperature is much lower than the first triplon gap $\Delta$.  In this low-energy case, the MPS during the time evolution step does not differ significantly from the ground state, which is known to be a product state of spin singlets living on the off-diagonal bonds.  The projection error leads to a biased estimation of the spin gap, even with the 2-site TDVP approach.   On the other hand, time-evolving block decimation (TEBD)  suffers only from the Trotter error and the MPS truncation error.

As a test, we show a comparison between TDVP and TEBD methods in Fig.~\ref{fig:TEBD_TDVP}.  At low temperatures, one expects the imaginary time correlation function to decay with an exponential factor based on the estimated spin gap ($\Delta \approx 0.458 J_D$) from the perturbation theory.  However, as can be seen from the plot, 
$ G(\tau) $ obtained from the 2-site TDVP method deviates from this expectation with a much steeper slope, for $T/\Delta=0$ and $0.1$.  On the other hand, the long-time correlation from TEBD  resolves the expected spin gap.  
Meanwhile, for temperatures comparable with the spin gap, enough entanglement is built up within the time-evolved MPS and we do not observe a severe projection error in TDVP.  This can be seen from the consistent data between the two methods at $T/\Delta=0.2$.   
For the three temperatures here, we choose the time step in TEBD  as $ d \tau =0.1$ such that the Trotter error is sufficiently small.

\section{Fictitious temperature dependence of stochastic analytic continuation} 
\label{sec:SAC_app}

In the SAC sampling process, one normally expects a characteristic \textit{thermal phase transition temperature} where the specific heat shows a jump. However, 
in the SAC process based on  
our METTS data on  
the $ 16 \times 4 $ Shastry-Sutherland lattice,  
 $\chi^2$ decays toward $\chi^2_{ \text{min} }$ smoothly as a function of the fictitious temperature $\alpha$. Hence we choose to use the weighted average of the spectrum over a wide range of $\alpha$.  It is nevertheless important to check the $\alpha$ dependence of the spectrum, and especially to compare between the imaginary and complex time cases.     

\begin{figure}[t!]
    \centering
    \includegraphics[width=\columnwidth]{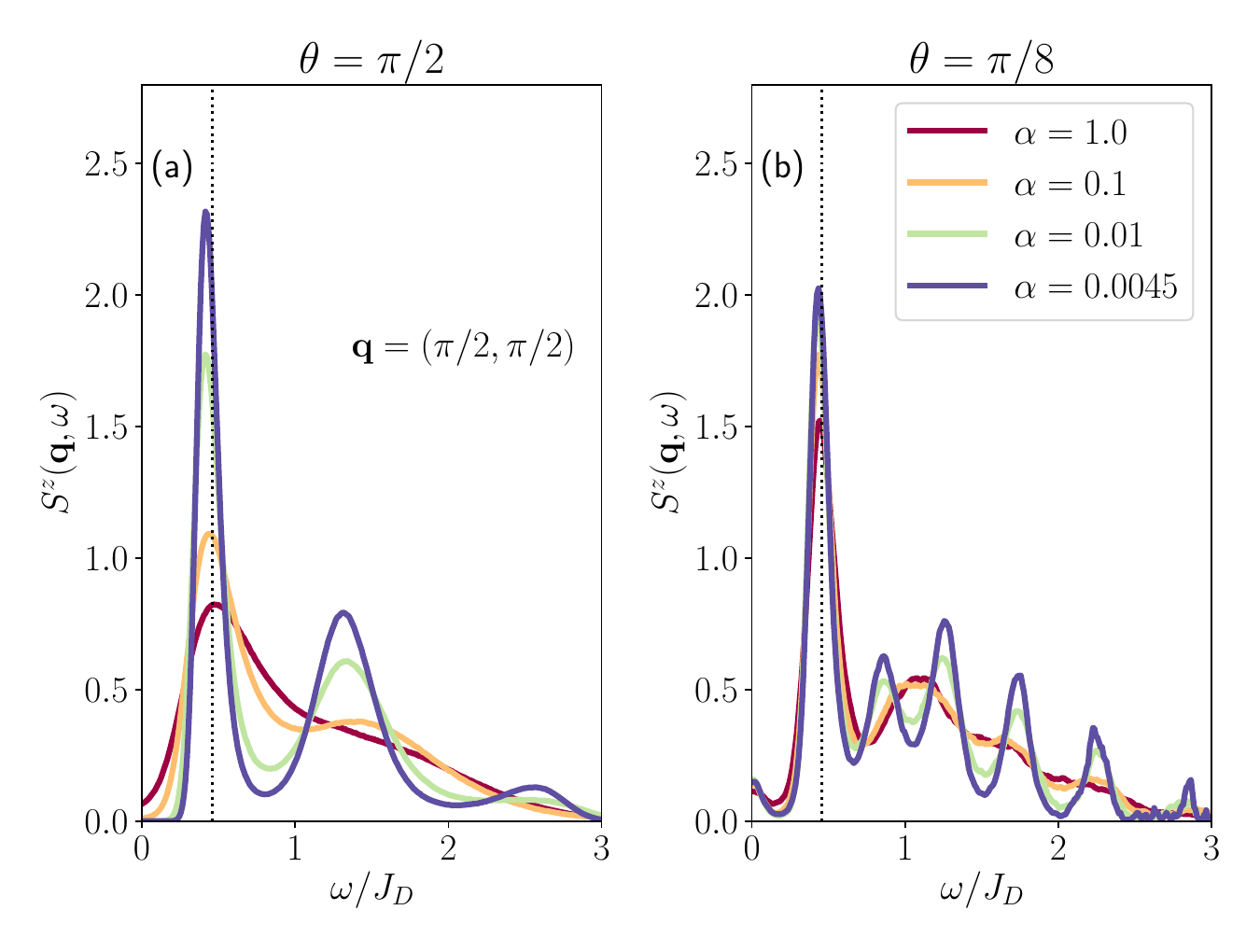}
    \caption{%
    $ {S}(\bm{q}, \omega)$ from stochastic analytic continuation    
     at $\alpha=1.0, 0.1, 0.01$ and $0.0045$, based on $ {S}(\bm{q}, z)$ at $\theta=\pi/2$ (a) and $\pi/8$ (b), for 
    $T/J_D=0.1$.  The simulations at different $\alpha$ are performed via 
    independent annealing processes from high temperatures.   
    }
    \label{fig:SAC_alpha}
\end{figure}

We choose the same set of METTS data that is used in Sec.~\ref{sec:SAC} as showcased here.  
As depicted in Fig.~\ref{fig:SAC_alpha}, the spectral function obtained via the  
imaginary time $\theta=\pi/2$ correlator depends strongly on the choice of  $\alpha$. Taking the spectrum  in  
Fig.~\ref{fig:thetas} as a reference, the weight and position of the second triplon peak are clearly biased for both $\alpha=0.01$ and $0.0045$. On the other hand, the overall shape of the spectrum at $\theta=\pi/8$ is much more stable, although we observe an unphysical splitting of the second peak at small  $\alpha$.  

\section{Comparison of windowing functions} 
\label{sec:windowing}

\begin{figure}[t!]
    \centering
    \includegraphics[width=\columnwidth]{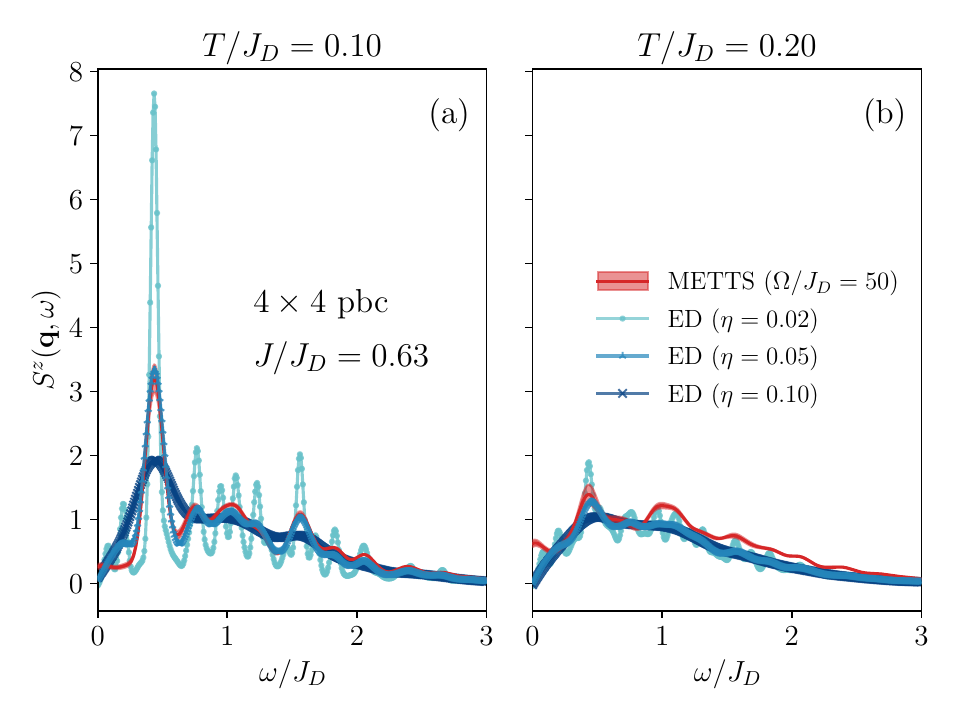}
    \caption{Comparison of spectral functions obtained via ED and the real-time METTS algorithm as in \cref{fig:ed_comparison}, but using a Parzen windowing function \cref{eq:parzen}.}
    \label{fig:ed_comparison_parzen}
\end{figure}

In this appendix we compare the results of computing the spectral functions using different windowing functions to perform the Fourier transform. In the main text in \cref{fig:ed_comparison} we used the Hann window of the form,
\begin{equation}
    w(x) = 0.5 + 0.5 \cos\left(\pi x\right) \quad\text{for}\quad 0 \leq |x| \leq 1.
\end{equation} 
One could ask to what extent the result depends on the windowing function. To answer this question we compare two different windowing functions. First, we show results using the Parzen window of the form,
\begin{equation}
\label{eq:parzen}
w(x) =
\begin{cases}
1 - 6|x|^2 + 6|x|^3 \quad \text{if} \quad 0 \leq |x| \leq 1/2\\
2(1-|x|^3) \quad \text{if} \quad 1/2 \leq |x| \leq 1
\end{cases},
\end{equation}
in \cref{fig:ed_comparison_parzen}. The Parzen windowing function has previously been used in the context of time-dependent DMRG at zero temperature~\cite{Kuehner1999}. 

\begin{figure}[t!]
    \centering
    \includegraphics[width=\columnwidth]{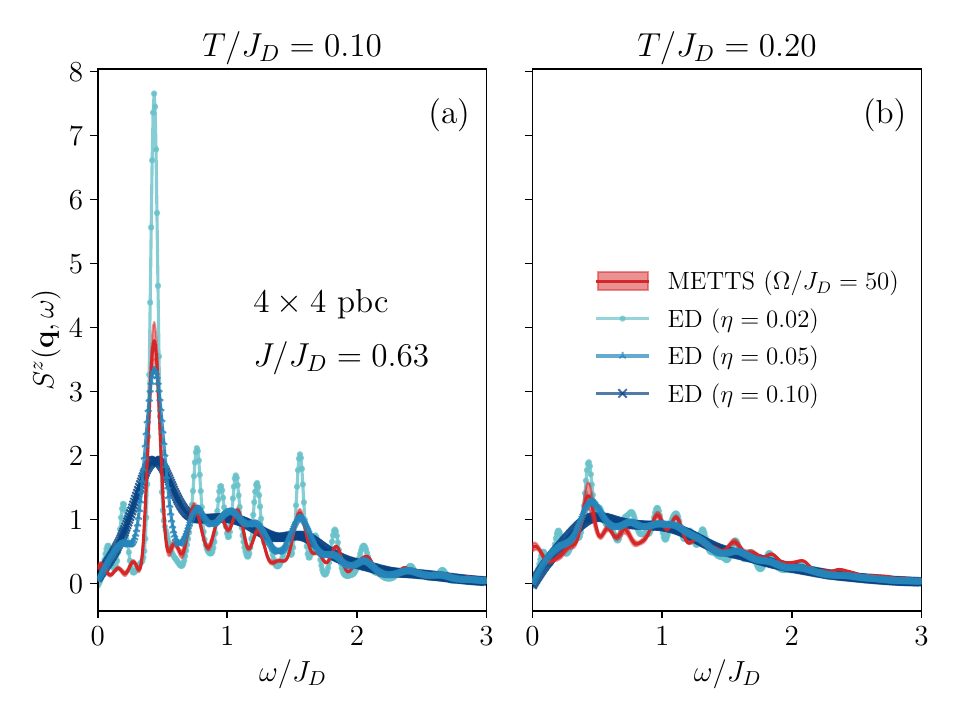}
    \caption{Comparison of spectral functions obtained via ED and the real-time METTS algorithm as in \cref{fig:ed_comparison}, but using a Gaussian windowing function \cref{eq:gaussian}
    of width $\sigma = 1/2$.}
    \label{fig:ed_comparison_gaussian}
\end{figure}

An equally valid choice of windowing function is a Gaussian window of the form, 
\begin{equation}
    \label{eq:gaussian}
    w(x) = \exp\left( - \frac{|x|^2}{2\sigma^2} \right) \quad\text{for}\quad 0 \leq |x| \leq 1,
\end{equation} 
where $\sigma$ controls the width of the Gaussian window. Results from applying the Gaussian windowing function with $\sigma = 1/2$ are shown in \cref{fig:ed_comparison_gaussian}.

Overall, we see that due to the chosen maximal simulation time $\Omega/ J_D = 50$ our results are not strongly affected by the choice of windowing function. The effects of the windowing function are expected to vanish for $\Omega\rightarrow \infty$.

%An issue regarding the SAC process  based on the correlator far from  imaginary time is that the choice of time domain of the correlator used as SAC input,  may have a subtle effect on the real frequency spectrum.  As shown in Fig.~\ref{fig:SAC_start}(b),  unphysical spikes at high frequency exist in the case that all the complex time correlators are taken as input ($ N_{ \text{start}} = 1$) for the SAC, for $\theta=\pi/8$. One notices that these spikes vanish as long as we throw away the first few  time slices; as can be seen from the plot, the spectrum nearly  converges as long as $ N_{ \text{start}} \geq 3$ in this example.      On the other hand, the imaginary-time-based SAC does not suffer from this issue and the spectrum is independent of the choice of $\alpha$, as shown in Fig.~\ref{fig:SAC_start}(a).  

%\begin{figure}[t]
%    \centering
%    \includegraphics[width=\columnwidth]{analytic/Analytic_toss.pdf}
%    \caption{
%    $ N_{ \text{start} }$ dependence of averaged $ {S}(\bm{q}, \omega)$ from stochastic analytic continuation,  
%    based on $ {S}(\bm{q}, z)$ at $\theta=\pi/2$ (a) and $\pi/8$ (b), for 
%    $T=0.1$.  
%    }
%    \label{fig:SAC_start}
%\end{figure}

\end{document}